\documentclass[10pt,journal]{IEEEtran}

\usepackage{graphicx}
\usepackage{amsmath,amsfonts}
\usepackage{algorithmic}
\usepackage{algorithm}
\usepackage{array}
\usepackage[caption=false,font=normalsize,labelfont=sf,textfont=sf]{subfig}
\usepackage{mathtools}
\usepackage{textcomp}
\usepackage{stfloats}
\usepackage{url}
\usepackage{verbatim}
\usepackage{cite}
\usepackage{multirow}
\usepackage{colortbl}
\usepackage{amsthm}
\usepackage{amssymb}
\usepackage{booktabs}
\usepackage{color}
\usepackage{multirow}

\DeclareMathOperator{\prox}{prox}
\DeclareMathOperator{\diag}{diag}

\DeclareMathOperator{\rank}{rank}

\newcommand{\argmin}{\mathop{\mathrm{argmin}}\limits}

\theoremstyle{plain}
\newtheorem{lem}{Lemma}
\newtheorem{thm}{Theorem}

\def\etal{\textit{et al.}}

\def\Order#1{\mathcal{O}(#1)}

\def\InnerProduct<#1>{\langle #1 \rangle}
\def\MatrixBrackets[#1]#2{\lbrack #1 \rbrack_{#2}}
\def\AlgoLfloor#1{\left\lfloor #1 \right}
\def\MatrixIdentity{\mathbf{I}}

\def\OpNormSq#1{\| #1 \|_{\mathrm{op}}^{2}}

\def\SSSTTV{{\text{S}_{3} \text{TTV}}}
\def\llHTV{{l_{0}\text{-}l_{1}\text{HTV}}}

\def\RealSpace#1{\mathbb{R}^{#1}}

\def\SetConvex{C}
\def\FuncIndicator#1{\iota_{#1}}

\def\Projection#1{P_{#1}}

\def\FuncOne{f}
\def\FuncTwo{g}
\def\NumVarOne{N}
\def\NumVarTwo{M}
\def\VarOne{\mathbf{x}}
\def\VarTwo{\mathbf{y}}
\def\MatrixOne{\mathbf{X}}
\def\ElementOne#1{x_{#1}}
\def\IndexOne{i}

\def\HSIClean{\mathbf{u}}
\def\HSIObsv{\mathbf{v}}
\def\NoiseSparse{\mathbf{s}}
\def\NoiseGauss{\mathbf{n}}
\def\NoiseStripe{\mathbf{t}}

\def\ResHSIClean{\HSIClean^{'}}
\def\ResNoiseSparse{\NoiseSparse^{'}}
\def\ResNoiseStripe{\NoiseStripe^{'}}

\def\NumAll{N}
\def\NumVert{N_{1}}
\def\NumHori{N_{2}}
\def\NumBand{N_{3}}

\def\SmallBlock#1{\mathbf{T}_{\HSIClean}^{(#1)}}
\def\NumBlock{L}

\def\IndexBlock{l}

\def\NumBlockVert{N_{1}^{'}}
\def\NumBlockHori{N_{2}^{'}}

\def\NumSingularValue{Q}

\def\IndexAlg{t}

\def\NotationVector{\mathbf{x}}
\def\NotationMatrix{\mathbf{X}}
\def\NotationScalar{x}
\def\NotationNumDim{N}
\def\NotationIndex{n}

\def\FuncProx{\FuncOne}
\def\IndexProx{\gamma}
\def\VarProxOne{\VarOne}
\def\VarProxTwo{\VarTwo}
\def\NumVarProx{\NumVarOne}

\def\Projection#1{P_{#1}}

\def\FuncPrimal#1{\FuncOne_{#1}}
\def\FuncDual#1{\FuncTwo_{#1}}
\def\VarPrimal#1{\VarOne_{#1}}
\def\VarDual#1{\VarTwo_{#1}}
\def\VarDualMatrix#1{\mathbf{Y}_{#1}}

\def\NumVarPrimal{\NumVarOne}
\def\NumVarDual{\NumVarTwo}
\def\DimVarPrimal#1{n_{#1}}
\def\DimVarDual#1{m_{#1}}
\def\IndexPrimal{i}
\def\IndexDual{j}
\def\IndexVarPrimal{k}
\def\IndexVarDual{l}

\def\LinOpPPDS#1{\mathbf{A}_{#1}}
\def\ScalarStepsize#1{\gamma_{#1}}
\def\MatrixStepsize#1{\mathbf{\Gamma}_{#1}}
\def\ParamStepsize#1{\gamma_{#1}}

\def\IndexBand{\IndexPrimal}

\def\DiffOpVert{\mathbf{D}_{v}}
\def\DiffOpHori{\mathbf{D}_{h}}
\def\DiffOpBand{\mathbf{D}_{s}}
\def\DiffOpSp{\mathbf{D}}

\def\DiffOpVertT{\mathbf{D}_{v}^{\top}}
\def\DiffOpHoriT{\mathbf{D}_{h}^{\top}}
\def\DiffOpBandT{\mathbf{D}_{s}^{\top}}
\def\DiffOpSpT{\mathbf{D}^{\top}}

\def\ExpantionOp#1{\mathbf{P}_{#1}}

\def\LeftSingularMatrix{\mathbf{U}}
\def\SingularMatrix#1{\mathbf{\Sigma}_{#1}}
\def\RightSingularMatrixT{\mathbf{V}^{\top}}
\def\SingularValue#1{\rho_{#1}}

\def\RadiusFidel{\varepsilon}
\def\RadiusSparse{\alpha}
\def\RadiusStripe{\beta}

\def\ParamsRadius{\rho}

\def\BallFidel{B_{2, \RadiusFidel}^{\HSIObsv}}
\def\BallSparse{B_{1, \RadiusSparse}}
\def\BallStripe{B_{1, \RadiusStripe}}
\def\SetZero{\{ \mathbf{0} \}}

\def\MinRange{\underline{\mu}}
\def\MaxRange{\bar{\mu}}
\def\SetRange{R_{\MinRange,\MaxRange}}

\def\OperationVec{\VarOne}
\def\OperationMatrix{\MatrixOne}

\def\IndexVert{i}
\def\IndexHori{j}

\def\StanDevGauss{\sigma}
\def\RateSparse{p_{\NoiseSparse}}
\def\RateStripe{p_{\NoiseStripe}}

\def\IndexNoisy{\text{(b)}}
\def\IndexSSTV{\text{(c)}}
\def\IndexHSSTVOne{\text{(d)}}
\def\IndexHSSTVTwo{\text{(e)}}
\def\IndexllHTV{\text{(f)}}
\def\IndexSTV{\text{(g)}}
\def\IndexSSST{\text{(h)}}
\def\IndexLRTDTV{\text{(i)}}
\def\IndexFGSLR{\text{(j)}}
\def\IndexTPTV{\text{(k)}}
\def\IndexQRNND{\text{(l)}}
\def\IndexFastHyMix{\text{(m)}}
\def\IndexSSSTTV{\text{(n)}}

\def\MatHSIClean{\mathbf{U}^{(\IndexBlock)}}

\begin{document}
\bstctlcite{IEEEexample:BSTcontrol}

\title{Spatio-Spectral Structure Tensor Total Variation for Hyperspectral Image Denoising and Destriping}

\author{Shingo~Takemoto,~\IEEEmembership{Student~Member,~IEEE,}
        Kazuki~Naganuma, Shunsuke~Ono,~\IEEEmembership{Member,~IEEE,}
\thanks{S. Takemoto is with the Department of Computer Science, Institute of Science Tokyo, Yokohama, 226-8501, Japan (e-mail: takemoto.s.e908@m.isct.ac.jp).}
\thanks{K. Naganuma is with the Institute of Engineering of Tokyo University of Agriculture and Technology, Tokyo, 184-8588, Japan (e-mail: k-naganuma@go.tuat.ac.jp).}
\thanks{S. Ono is with the Department of Computer Science, Institute of Science Tokyo, Yokohama, 226-8501, Japan (e-mail: ono.s.5af2@m.isct.ac.jp).}
\thanks{This work was supported in part by Grant-in-Aid for JSPS Fellows Grant Number JP24KJ1068 and 23KJ0912, in part by JSPS KAKENHI under Grant 22H03610, 22H00512, 23H01415, 23K17461, 24K03119, and 24K22291, and in part by JST PRESTO under Grant JPMJPR21C4, JST AdCORP under Grant JPMJKB2307, and JST ACT-X Grant JPMJAX23CJ.}}


\markboth{Journal of \LaTeX\ Class Files,~Vol.~14, No.~8, August~2021}%
{Shell \MakeLowercase{\textit{et al.}}: A Sample Article Using IEEEtran.cls for IEEE Journals}


\maketitle

\begin{abstract}
This paper proposes a novel regularization method, named \textit{Spatio-Spectral Structure Tensor Total Variation} ($\SSSTTV$), for denoising and destriping of hyperspectral (HS) images. HS images are inevitably contaminated by various types of noise, during acquisition process, due to the measurement equipment and the environment. For HS image denoising and destriping tasks, Spatio-Spectral Total Variation (SSTV) is widely known as a powerful regularization approach that models the spatio-spectral piecewise smoothness. However, since SSTV refers only to the local differences of pixels/bands, edges and textures that extend beyond adjacent pixels are not preserved during denoising process. To address this problem, we newly introduce $\SSSTTV$, which is designed to preserve two essential physical characteristics of HS images: semi-local spatial structures and spectral correlation across all bands. Specifically, we define $\SSSTTV$ as the sum of the nuclear norms of spatio-spectral structure tensors, which are matrices formed by arranging second-order spatio-spectral difference vectors within semi-local areas. Furthermore, we formulate the HS image denoising and destriping problem as a constrained convex optimization problem involving $\SSSTTV$ and develop an algorithm based on a preconditioned primal-dual splitting method to solve this problem efficiently. Finally, we demonstrate the effectiveness of $\SSSTTV$ by comparing it with existing methods, including state-of-the-art ones through denoising and destriping experiments. The source code is available at \url{https://github.com/MDI-TokyoTech/Spatio-Spectral-Structure-Tensor-Total-Variation}.

\end{abstract}



\section{Introduction}
\IEEEPARstart{H}{yperspectral} (HS) imaging measures a wide spectrum of light ranging from the ultraviolet to the near-infrared.
The rich spectral information of HS images, with more than one hundred bands, can distinguish materials and phenomena that the human eye and existing RGB cameras cannot.
This capability has been applied in diverse fields including agriculture, mineralogy, astronomy, and biotechnology~\cite{Borengasser2007HSIApplications,Grahn2007Techniques, Thenkabail2016VegetationOverview,Lu2020AgricultureOverview}.
However, observed HS images are inevitably contaminated by various types of noise---including thermal noise, quantization noise, shot noise, and stripe noise---due to factors caused by measurement equipment and environment, such as photon effects, atmospheric absorption, dark currents, and sensor disturbances~\cite{Shen2015DenoisingOverview,Rasti2018DenoisingOverview,Shen2022DenoisingOverview}.
Since such noise significantly degrades the performance of subsequent processing such as unmixing~\cite{Bioucas-Dias2012UnmixingOverview,Ma2014UnmixingOverview}, classification~\cite{Ghamisi2017Classification,Li2019Classification,Nicolas2019Classification}, and anomaly detection~\cite{Matteoli2014Anomaly,Su2022Anomaly}, HS image denoising is an essential preprocessing step for the applications.

To obtain desirable HS images from degraded observations, HS recovery methods need to employ strategies to capture the inherent properties of HS images. For such strategies, existing methods adopt deep neural networks (DNN) or regularization functions. DNN approaches capture the properties that are difficult to model mathematically or statistically. Various architectures have been proposed, including Convolutional NN (CNN) based methods~\cite{Yuan2019HSIDCNN, Wang2022NL3DCNN} that extract spatial and spectral correlations; recurrent models~\cite{Wei2021QRNN3D} that leverage quasi-recurrent units to learn the global spectral correlations; transformer-based approaches~\cite{Li2023SST} that capture long-range spatial–spectral relationships via self-attention; and DNNs with subspace framework, where denoising is performed in a low-dimensional coefficient space derived from spectral low-rank decomposition~\cite{Zhuang2023FastHyMix, Peng2024RCILD}. However, they do not separate HS images from noise exhibiting similar spatial structures to HS images. For instance, stripe noise is smooth in one spatial direction, and HS images are often smooth in the direction as well. Therefore, when HS images are contaminated by such types of noise, DNN approaches degrade their performance.

On the other hand, since regularization functions are mathematically designed to accurately separate desirable HS images from various types of noise, they have attracted attention~\cite{Xue2017Robust,Jiang2022Adaptive,Naganuma2024Unmixing}. Over the past decade, non-local similarity methods, including 3D Nonlocal Means (3DNLM)~\cite{Qian20123DNLM}, Block Matching 3D Filtering (BM3D)~\cite{Maggioni2013BM3D}, BM4D~\cite{Chen2014BM4D}, and Non-local Meets Global (NGMeet)~\cite{He2019NGmeet} have been proposed to capture similarities in distant patches, but these methods blur edges. In contrast, total variation (TV) type regularization methods originating from edge-preserving natural image denoising~\cite{Rudin1992TV,Bresson2008TV} have been developed. The pioneering TV-type method for HS image denoising is \textit{Spectral-Spatial Adaptive Hyperspectral TV} (SSAHTV)~\cite{Yuan2012HTV}, which models the spatial piecewise-smoothness of HS images as the sparsity of first-order spatial differences between adjacent pixels. Furthermore, \textit{Anisotropic Spectral Spatial TV} (ASSTV)~\cite{Chang2015ASSTV} extends this model by incorporating first-order spectral differences into the SSAHTV regularization function, capturing the spectral piecewise-smoothness in addition to the spatial piecewise-smoothness of HS images. However, HS images are often severely degraded by various types of noise unlike natural images, and under such conditions, because first-order differences cannot sufficiently distinguish noise components from clean image structures, minimizing the $\ell_{1}$-type norms of first-order differences in SSAHTV and ASSTV causes over-smoothing. In addition, since these methods only refer to adjacent pixels/bands, they corrupt edges and textures that extend beyond adjacent pixels during the denoising process.

One promising TV-type regularization method that overcomes blurring and over-smoothing is Spatio-Spectral TV (SSTV)~\cite{Aggarwal2016SSTV}.
SSTV is defined by the $\ell_{1}$ norm of second-order spatio-spectral differences, i.e., first-order spatial differences of spectral ones to avoid over-smoothing.
In addition, SSTV captures not only the spatial piecewise-smoothness, but also the spatial similarity between adjacent bands by its formulation.
For these reasons, SSTV has been widely used in state-of-the-art HS image denoising methods~\cite{Fan2018SSTV-LRTF,Ince2019GLSSTV,Takeyama2020HSSTV,Wang2021l0l1HTV,Takemoto2022GSSTV}.	
Takeyama $\etal$ proposed \textit{Hybrid Spatio-Spectral TV} (HSSTV)~\cite{Takeyama2020HSSTV}, which enhances SSTV by incorporating SSAHTV.
To recover more detailed spatial structures of HS images, \textit{Graph Spatio-Spectral TV} (GSSTV)~\cite{Takemoto2022GSSTV} was proposed, which weights the spatial difference operator of SSTV based on a graph reflecting the spatial structures of an HS image.
To directly control the degree of the smoothness, Wang $\etal$ proposed the $\ell_{0}\text{-}\ell_{1}$ \textit{hybrid TV} ($\llHTV$)~\cite{Wang2021l0l1HTV}, which incorporates $\ell_{0}$-type constraints of the spatial differences (originally proposed for color image processing~\cite{Ono2017l0Gradient}) into SSTV.
However, SSTV and its extension methods cannot preserve the semi-local spatial structures of HS images during the denoising process because they only refer to adjacent pixels/bands.

As an approach to capture non-local properties and potentially solve the corruption of semi-local spatial structures, we focus on low-rank (LR) regularization methods~\cite{Zhang2014LRMR, Chen2022FGSLR}. HS images have strong correlations between vectors for each pixel across all bands, arising from their composition of a limited variety of material-specific spectra. By promoting spectral low-rankness, LR-type regularization methods capture this non-local property rather than the properties of adjacent pixels or bands. However, since LR modeling does not directly account for spatial structures, several works have proposed combining it with TV~\cite{He2016LRTV, Chen2023TPTV}, or applying tensor decomposition techniques~\cite{Xue2022Tensor1,Xue2022Tensor2,Xue2024Tensor}. Tensor decomposition-based methods~\cite{Wang2018LRTDTV, Chen2020LRTDGS, Sun2022Tensor, Li2024LRTDAHL} aim to extract jointly spatial and spectral global latent structures across the three spatial-spectral modes, but they generally do not explicitly characterize local or semi-local spatial structures of HS images.

By integrating the spectral correlation modeling of LR regularization into the TV framework, Structure Tensor Total Variation (STV)~\cite{Lefkimmiatis2015STV, Wu2017STWNNM, Ono2016ASTV, Kurihara2017SSST} has been developed to preserve semi-local spatial structures. These regularization functions evaluate the spectral correlation in differences instead of the sparsity of differences, as in the standard TV-type methods, via the nuclear norm of \textit{structure tensors}\footnote{Following the original STV paper~\cite{Lefkimmiatis2015STV}, we call a 'matrix' composed of differences as structure 'tensor.' The structure tensor is often used in image processing~\cite{Forstner1987Fast, Weickert1998Anisotropic, Jahne2005Digital}.} consisting of matrices of local differences in small spatial areas and all bands. The original STV~\cite{Lefkimmiatis2015STV} and \textit{Structure Tensor total variation-regularized Weighted Nuclear Norm Minimization} (STWNNM)~\cite{Wu2017STWNNM} capture the semi-local spatial piecewise-smoothness, but do not model the spectral property of HS images due to consisting only of spatial differences. The regularization function of \textit{Arranged Structure tensor TV} (ASTV)~\cite{Ono2016ASTV} characterizes the spectral correlations across all bands by modifying the ordering of spatial differences in the structure tensors. In addition, \textit{Spatio-Spectral Structure Tensor} (SSST)~\cite{Kurihara2017SSST} explicitly exploits the spectral piecewise-smoothness of HS images by including not only spatial differences but also spectral ones in its formulation. However, the existing STV-type regularization methods minimize the nuclear norms of “first-order" differences, leading to spatial or spectral over-smoothing, as in SSAHTV.

From the discussion so far, SSTV and STV-type regularization methods are powerful approaches that capture the underlying spatial and spectral characteristics of HS images.
However, they have their own limitations: SSTV cannot preserve the semi-local spatial structures and STV-type regularization methods cause over-smoothing.
Now a natural question arises: \textit{Can we design a regularization function with the two requirements: avoid over-smoothing, preserve semi-local spatial structures?}

Inspired by both SSTV and the above family of STV, we propose a denoising method for HS images using a newly introduced \textit{spatio-spectral structure tensor total variation} ($\SSSTTV$) model.
The main contributions of this article are summarized as follows.
\begin{enumerate}
	\item We design a novel regularization method, namely $\SSSTTV$. 
	This method includes newly designed function defined as the sum of the nuclear norms of structure tensors (called spatio-spectral structure tensor) that include \textit{second-order spatio-spectral differences}. 
	This function can fully capture the spatial piecewise smoothness, the spatial similarity between adjacent bands, and the spectral correlation across all bands in small spatial areas. This leads to effective noise removal while preserving the semi-local spatial structures of HS images without over-smoothing.
	
	\item We formulate the mixed noise removal problem as a constrained convex optimization problem involving $\SSSTTV$. 
	Inspired by the approach of separating noise components~\cite{Zhang2022Double}, our method incorporates specific hard constraints to characterize Gaussian noise, sparse noise, and stripe noise within the optimization problem, enabling effective removal of these noise types from HS images. 
	By using hard constraints instead of data-fidelity and noise terms, we decouple interdependent hyperparameters into independent ones, making parameter setting easier, as shown in prior studies~\cite{Afonso2011Constraint, Chierchia2015Constraint, Ono2015Constraint, Ono2017Constraint, Ono2019Constraint, Naganuma2022Destriping}.

	\item To solve our optimization problem for the HS image denoising, we develop an efficient algorithm based on a Preconditioned Primal-Dual Splitting method (P-PDS)~\cite{Pock2011PPDS}. 
	Unlike other popular algorithms used in existing HS image denoising methods, such as an alternating direction method of multipliers~\cite{Boyd2011ADMM} and PDS~\cite{Chambolle2011PDS, Condat2013PDS}, P-PDS can automatically determine the appropriate stepsizes based on the problem structure~\cite{Pock2011PPDS,Naganuma2023PPDS}. 
	
\end{enumerate}
Experimental results show the superiority of the proposed method to existing methods including state-of-the-art ones.
The comparison of the features of the existing and proposed methods is summarized in Table.~\ref{tab:ProsCons}.

The paper is organized as follows.
In Sec.~\ref{sec:Preliminaries}, we introduce the mathematical tools required for the proposed method.
Sec.~\ref{sec:proposed_method} provides the proposed HS image denoising method involving $\SSSTTV$.
The experimental results are reported in Sec.~\ref{sec:experiments}.
Finally, we give concluding remarks in Sec.~\ref{sec:conclusion}.
The preliminary version of this paper, without considering stripe noise, mathematical details, comprehensive experimental comparison, or deeper discussion, has appeared in conference proceedings~\cite{Takemoto2023S3TTV}.

\begin{table*}[t]
    \begin{center}
        \caption{Pros and Cons of Existing and Proposed Methods for HS Image Denoising.}
        \label{tab:ProsCons}
        \begin{tabular}{c ccccc}
            \toprule
                Methods & Spatial piecewise smoothness & 
                \begin{tabular}{c}
                    Spatial similarity \\ between adjacent bands
                \end{tabular} & 
                \begin{tabular}{c}
                     Spectral correlation \\ across all bands
                \end{tabular} & Avoiding over-smoothing & Convexity\\
            \cmidrule(lr){1-6}
            \vspace{-0.5mm}
                SSAHTV~\cite{Yuan2012HTV} & $\checkmark$ & -- & -- & -- & $\checkmark$ \\
                SSTV~\cite{Aggarwal2016SSTV} & $\checkmark$ & $\checkmark$ & -- & $\checkmark$ & $\checkmark$\\
                HSSTV~\cite{Takeyama2020HSSTV} & $\checkmark$ & $\checkmark$ & -- & -- & $\checkmark$ \\
                $\llHTV$~\cite{Wang2021l0l1HTV} & $\checkmark$ & $\checkmark$ & -- & $\checkmark$ & -- \\
                GSSTV~\cite{Takemoto2022GSSTV} & $\checkmark$ & $\checkmark$ & -- & $\checkmark$ & $\checkmark$ \\
                LRTDTV~\cite{Wang2018LRTDTV} & $\checkmark$ & -- & $\checkmark$ & $\checkmark$ & -- \\
                FGSLR~\cite{Chen2022FGSLR} & $\checkmark$ & -- & $\checkmark$ & -- & -- \\
                TPTV~\cite{Chen2023TPTV} & $\checkmark$ & -- & $\checkmark$ & $\checkmark$ & -- \\
                STV~\cite{Lefkimmiatis2015STV} & $\checkmark$ & -- & -- & -- & $\checkmark$ \\
                STWNNM~\cite{Wu2017STWNNM} & $\checkmark$ & -- & -- & -- & -- \\
                ASTV~\cite{Ono2016ASTV} & $\checkmark$ & -- & $\checkmark$ & -- & $\checkmark$ \\
                SSST~\cite{Kurihara2017SSST} & $\checkmark$ & -- & $\checkmark$ & -- & $\checkmark$ \\
            \cmidrule(lr){1-6}
                Proposed method & $\checkmark$ & $\checkmark$ & $\checkmark$ & $\checkmark$ & $\checkmark$ \\
            \bottomrule
        \end{tabular}
    \end{center}
    \vspace{-3mm}
\end{table*}
\section{Preliminaries}
\label{sec:Preliminaries}

\subsection{Notations}
\label{subsec:Notations}
Throughout this paper, we denote vectors and matrices by boldface lowercase letters (e.g., $\NotationVector$) and boldface capital letters (e.g., $\NotationMatrix$), respectively.
We treat an HS image, denoted by $\HSIClean$ with $\NumVert$ vertical pixels, $\NumHori$ horizontal pixels, and $\NumBand$ bands.
We denote the total number of elements in the HS image by $\NumAll = \NumVert \NumHori \NumBand$.
For a matrix data $\NotationMatrix \in \RealSpace{\NumVert \times \NumHori}$, the value at the location $(\IndexPrimal, \IndexDual)$ is denoted by $[\NotationMatrix]_{\IndexPrimal,\IndexDual}$.
The $\ell_{1}$-norm and the $\ell_{2}$-norm of a vector $\NotationVector \in \RealSpace{\NotationNumDim}$ are defined as $\| \NotationVector \|_{1} := \sum_{\NotationIndex=1}^{\NotationNumDim} | \NotationScalar_{\NotationIndex} |$ and $\| \NotationVector \|_{2} := \sqrt{\sum_{\NotationIndex=1}^{\NotationNumDim} \NotationScalar_{\NotationIndex}^{2}}$, respectively, where  $\NotationScalar_{\NotationIndex}$ represents the $\NotationIndex$-th entry of $\NotationVector$.
The nuclear norm of a marix, which is the sum of all the singular values, is denoted by $\| \cdot \|_{*}$.
For an HS image $\HSIClean \in \RealSpace{\NumAll}$, let $\DiffOpVert \in \RealSpace{\NumAll \times \NumAll}$, $\DiffOpHori \in \RealSpace{\NumAll \times \NumAll}$, and $\DiffOpBand \in \RealSpace{\NumAll \times \NumAll}$ be the forward difference operators along the horizontal, vertical, and spectral directions, respectively, with the periodic boundary condition.
Here, a spatial difference operator is denoted by $\DiffOpSp := \begin{pmatrix} \DiffOpVertT & \DiffOpHoriT \end{pmatrix}^{\top} \in \RealSpace{2 \NumAll \times \NumAll}$. 
Using $\DiffOpVert$, $\DiffOpHori$, and $\DiffOpBand$, we denote the second-order spatio-spectral differences by $\DiffOpVert \DiffOpBand \HSIClean \in \RealSpace{\NumAll}$ and $\DiffOpHori \DiffOpBand \HSIClean \in \RealSpace{\NumAll}$.
Other notations will be introduced as needed.

\subsection{Proximal Tools}
\label{subsec:Prox}
In this chapter, we introduce basic proximal tools that play a central role in the optimization part of our method.
Let $\FuncProx$ be a \textit{proper lower semi-continuous convex function}.\footnote{A function $\FuncProx \: : \: \RealSpace{\NumVarProx} \rightarrow ( -\infty, \infty ]$ is called a proper lower semi-continuous convex function if $\lbrace \VarProxOne \in \RealSpace{\NumVarProx} | \FuncProx(\VarProxOne) < \infty \rbrace$ is nonempty, $\lbrace \VarProxOne \in \RealSpace{\NumVarProx} | \FuncProx(\VarProxOne) \leq \alpha \rbrace$ is closed for every $\alpha \in \RealSpace{}$, and $\FuncProx(\lambda \VarProxOne + (1-\lambda)\VarProxTwo) \leq \lambda\FuncProx(\VarProxOne) + (1-\lambda)\FuncProx(\VarProxTwo)$ for every $\VarProxOne, \VarProxTwo \in \RealSpace{\NumVarProx}$ and $\lambda \in (0, 1)$.}
Then, for $\IndexProx > 0$, the \textit{proximity operator} of $\FuncProx$ is defined by
\begin{equation}
	\label{eq:Prox}
	\prox_{\IndexProx, \FuncProx}(\VarProxOne) := \arg\min_{\VarProxTwo \in \RealSpace{\NumVarProx}} \FuncProx(\VarProxTwo) + \frac{1}{2 \IndexProx} \| \VarProxOne - \VarProxTwo \|_{2}^{2}.
\end{equation}

The \textit{Fenchel--Rockafellar conjugate function} $\FuncProx^{*}$ of the function $\FuncProx$ is defined by
\begin{equation}
	\label{eq:ConjugateFunction}
	\FuncProx^{*}(\VarProxOne) := \sup_{\VarProxTwo} \InnerProduct<\VarProxOne, \VarProxTwo> - \FuncProx(\VarTwo),
\end{equation}
where $\InnerProduct<\cdot , \cdot>$ is the Euclidean inner product.
Thanks to a generalization of Moreau's identity~\cite{Combettes2013Moreau}, the proximity operator of $\FuncProx^{*}$ is calculated as
\begin{equation}
	\label{eq:ProxConjugate}
	\prox_{\IndexProx, \FuncProx^{*}}(\VarProxOne) = \VarProxOne - \IndexProx \prox_{\frac{1}{\IndexProx} \FuncProx} \left( \frac{1}{\IndexProx} \VarProxOne \right).
\end{equation}

The indicator function of a set $\SetConvex \subset \RealSpace{\NumVarOne}$, denoted by $\FuncIndicator{\SetConvex}$, is defined as 
\begin{equation}
	\label{eq:Indicator_Function}
	\FuncIndicator{\SetConvex} (\VarOne) := 
	\begin{cases}
		0, & \mathrm{if} \: \VarOne \in \SetConvex, \\
		\infty, & \mathrm{otherwise}.
	\end{cases}
\end{equation}
The function $\FuncIndicator{\SetConvex}$ is proper lower semi-continuous convex when $\SetConvex$ is nonempty and closed convex.
The proximity operator of $\FuncIndicator{\SetConvex}$ is equivalent to the projection onto $\SetConvex$, as given by
\begin{equation}
	\label{eq:Projection}
	\prox_{\FuncIndicator{\SetConvex}} (\VarProxOne) = \Projection{\SetConvex}(\VarProxOne) := \argmin_{\VarProxTwo \in \SetConvex} \|\VarProxTwo - \VarProxOne \|_{2}.
\end{equation}

\subsection{Preconditoned Primal-Dual Splitting Method (P-PDS)}
\label{subsec:P-PDS}
The standard PDS~\cite{Chambolle2011PDS, Condat2013PDS} and P-PDS~\cite{Pock2011PPDS}, on which our algorithm is based, are efficient algorithms for solving the following generic form of convex optimization problems:
\begin{align}
	\label{prob:convex_optim_prob}
	\min_{\substack{\VarPrimal{1}, \ldots, \VarPrimal{\NumVarPrimal}, \\ 
			\VarDual{1}, \ldots, \VarDual{\NumVarDual}}} 
	& \sum_{\IndexPrimal=1}^{\NumVarPrimal} \FuncPrimal{\IndexPrimal} (\VarPrimal{\IndexPrimal}) + \sum_{\IndexDual=1}^{\NumVarDual} \FuncDual{\IndexDual} (\VarDual{\IndexDual}) \nonumber \\ 
	& \mathrm{s.t.} \:
	\begin{cases} 
		\VarDual{1} = \sum_{\IndexPrimal=1}^{\NumVarPrimal} \LinOpPPDS{1,\IndexPrimal} \VarPrimal{\IndexPrimal}, \\ 
		\vdots \\ 
		\VarDual{\NumVarDual} = \sum_{\IndexPrimal=1}^{\NumVarPrimal} \LinOpPPDS{\NumVarDual,\IndexPrimal} \VarPrimal{\IndexPrimal}, 
	\end{cases}
\end{align}
where $\FuncPrimal{\IndexPrimal} (\IndexPrimal = 1, \ldots, \NumVarPrimal)$ and $\FuncDual{\IndexDual} (\IndexDual = 1, \ldots, \NumVarDual)$ are lower semi-continuous proper convex functions, $\VarPrimal{\IndexPrimal} \in \RealSpace{\DimVarPrimal{\IndexPrimal}} \: (\IndexPrimal = 1, \dots, \NumVarPrimal)$ are primal variables, $\VarDual{\IndexDual} \in \RealSpace{\DimVarDual{\IndexDual}} \: (\IndexDual = 1, \dots, \NumVarDual)$ are dual variables, and $\LinOpPPDS{\IndexDual,\IndexPrimal} \in \RealSpace{\DimVarDual{\IndexDual} \times  \DimVarPrimal{\IndexPrimal}}$ ($i = 1, \ldots, \NumVarPrimal$, $j = 1, \ldots, \NumVarDual$) are linear operators.

These methods solve Prob.~\eqref{prob:convex_optim_prob} by the following iterative procedures:
\begin{equation}
	\label{algo:P-PDS}
	\AlgoLfloor{
		\begin{array}{l}
			\VarPrimal{1}^{(\IndexAlg+1)} 
			\leftarrow \prox_{\ScalarStepsize{1, 1}, \FuncPrimal{1}}
			\left( \VarPrimal{1}^{(\IndexAlg)} - \ScalarStepsize{1,1} \bigl( \sum_{\IndexDual=1}^{\NumVarDual} \LinOpPPDS{\IndexDual,1}^{\top} \VarDual{\IndexDual}^{(\IndexAlg)} \bigr) \right), \\
			\vdots \\
			\VarPrimal{\NumVarPrimal}^{(\IndexAlg+1)} 
			\leftarrow \prox_{\ScalarStepsize{1, \NumVarPrimal}, \FuncPrimal{\NumVarPrimal}}
			\left( \VarPrimal{\NumVarPrimal}^{(\IndexAlg)} - \ScalarStepsize{1, \NumVarPrimal} \bigl( \sum_{\IndexDual=1}^{\NumVarDual} \LinOpPPDS{\IndexDual,\NumVarPrimal}^{\top} \VarDual{\IndexDual}^{(\IndexAlg)} \bigr) \right), \\
			\VarPrimal{\IndexPrimal}^{'} = 2\VarPrimal{\IndexPrimal}^{(\IndexAlg+1)} - \VarPrimal{\IndexPrimal}^{(\IndexAlg)} \: (\forall \IndexPrimal = 1, \ldots, \NumVarPrimal), \\
			\VarDual{1}^{(\IndexAlg+1)}
			\leftarrow \prox_{\ScalarStepsize{2, 1}, \FuncDual{1}^{*}}
			\left( \VarDual{1}^{(\IndexAlg)} - \ScalarStepsize{2, 1}
			\bigl(\sum_{\IndexPrimal=1}^{\NumVarPrimal} \LinOpPPDS{1,\IndexPrimal} \VarPrimal{\IndexPrimal}^{'} \bigr) \right), \\
			\vdots \\
			\VarDual{\NumVarDual}^{(\IndexAlg+1)}
			\leftarrow \prox_{\ScalarStepsize{2, \NumVarDual}, \FuncDual{\NumVarDual}^{*}}
			\left( \VarDual{\NumVarDual}^{(\IndexAlg)} - \ScalarStepsize{2, \NumVarDual}
			\bigl(\sum_{\IndexPrimal=1}^{\NumVarPrimal} \LinOpPPDS{\NumVarDual,\IndexPrimal}  \VarPrimal{\IndexPrimal}^{'} \bigr) \right), \\
		\end{array}
	}.
\end{equation}
where $\ScalarStepsize{1, \IndexPrimal}(\IndexPrimal = 1, \dots, \NumVarPrimal)$ and $\ScalarStepsize{2, \IndexDual}(\IndexDual = 1, \dots, \NumVarDual)$ are the stepsize parameters.

Here, we introduce the convergence property of P-PDS.
For the convergence analysis, we define the diagonal matrices of the stepsize parameters as follows:
\begin{align}
	\label{eq:StepsizeMatrices}
	\MatrixStepsize{1} & = \diag(\ScalarStepsize{1, 1} \MatrixIdentity_{\DimVarPrimal{1}}, \ldots, \ScalarStepsize{1, \NumVarPrimal} \MatrixIdentity_{\DimVarPrimal{\NumVarPrimal}}), \notag \\
	\MatrixStepsize{2} & = \diag(\ScalarStepsize{2, 1} \MatrixIdentity_{\DimVarDual{1}}, \ldots, \ScalarStepsize{2, \NumVarDual} \MatrixIdentity_{\DimVarDual{\NumVarDual}}),
\end{align}
where $\MatrixIdentity_{\DimVarPrimal{\IndexPrimal}} \in \RealSpace{\DimVarPrimal{\IndexPrimal} \times \DimVarPrimal{\IndexPrimal}}$ and $\MatrixIdentity_{\DimVarDual{\IndexDual}} \in \RealSpace{\DimVarDual{\IndexDual} \times \DimVarDual{\IndexDual}}$ are the identity matrices. We define the linear operator including $\LinOpPPDS{\IndexDual,\IndexPrimal}$ as
\begin{equation}
	\label{eq:LinearOperator}
	\LinOpPPDS{} := 
	\begin{bmatrix}
		\LinOpPPDS{1, 1} & \cdots & \LinOpPPDS{1, \NumVarPrimal} \\
		\vdots & \ddots & \vdots \\
		\LinOpPPDS{\NumVarDual, 1} & \cdots & \LinOpPPDS{\NumVarDual, \NumVarPrimal}
	\end{bmatrix}.
\end{equation}
Then, we state the convergence property of P-PDS.
\begin{thm}
	\label{thm:ConvergenceP-PDS}
	[64, Theorem 1] Let $\MatrixStepsize{1}$ and $\MatrixStepsize{2}$ be symmetric and positive definite matrices satisfying
	\begin{equation}
		\label{ieq:ConvergenceCondition}
		\OpNormSq{\MatrixStepsize{1}^{\frac{1}{2}} \circ \LinOpPPDS{} \circ \MatrixStepsize{2}^{\frac{1}{2}}} < 1.
	\end{equation}
	Then, the sequence $\lbrace \VarPrimal{1}^{(\IndexAlg)}, \ldots, \VarPrimal{\NumVarPrimal}^{(\IndexAlg)}, \VarDual{1}^{(\IndexAlg)}, \ldots, \VarDual{\NumVarDual}^{(\IndexAlg)} \rbrace$ generated by the procedure in \eqref{algo:P-PDS} converges to an optimal solution of Prob.~\eqref{prob:convex_optim_prob}.
\end{thm}

The standard PDS needs to adjust the appropriate stepsize parameters to satisfy the convergence conditions~\eqref{ieq:ConvergenceCondition}. On the other hand, P-PDS can automatically determine the stepsize parameters that guarantee convergence~\cite{Pock2011PPDS,Naganuma2023PPDS}. According to~\cite{Pock2011PPDS}, we summarize the stepsize design and their convergence property.
\begin{lem}
	\label{lem:StepsizeParameters}
	[64, lemma 2] Let the diagonal matrices $\MatrixStepsize{1}, \MatrixStepsize{2}$ and the block matrix $\LinOpPPDS{}$ be set as Eq.~\eqref{eq:StepsizeMatrices} and \eqref{eq:LinearOperator}, respectively. In paticular,
	\begin{align}
	\label{eq:Preconditioners}
	\ScalarStepsize{1, \IndexPrimal} & = \frac{1}{\sum_{\IndexDual = 1}^{\NumVarDual} \sum_{\IndexVarDual = 1}^{\DimVarDual{\IndexDual}} | \MatrixBrackets[\LinOpPPDS{\IndexDual, \IndexPrimal}]{\IndexVarDual, 1} |} , \: (\forall \IndexPrimal = 1, \ldots, \NumVarPrimal), \notag \\
	\ScalarStepsize{2, \IndexDual} & = \frac{1}{\sum_{\IndexPrimal = 1}^{\NumVarPrimal} \sum_{\IndexVarPrimal = 1}^{\DimVarPrimal{\IndexPrimal}} | \MatrixBrackets[\LinOpPPDS{\IndexDual, \IndexPrimal}]{1, \IndexVarPrimal} |}, \: (\forall \IndexDual = 1, \ldots, \NumVarDual),
	\end{align}
	then the inquality in~\eqref{ieq:ConvergenceCondition} holds.
\end{lem}
\section{Proposed Method}
\label{sec:proposed_method}
\begin{figure*}[t]
	\begin{center}
	       \includegraphics[width=\hsize]{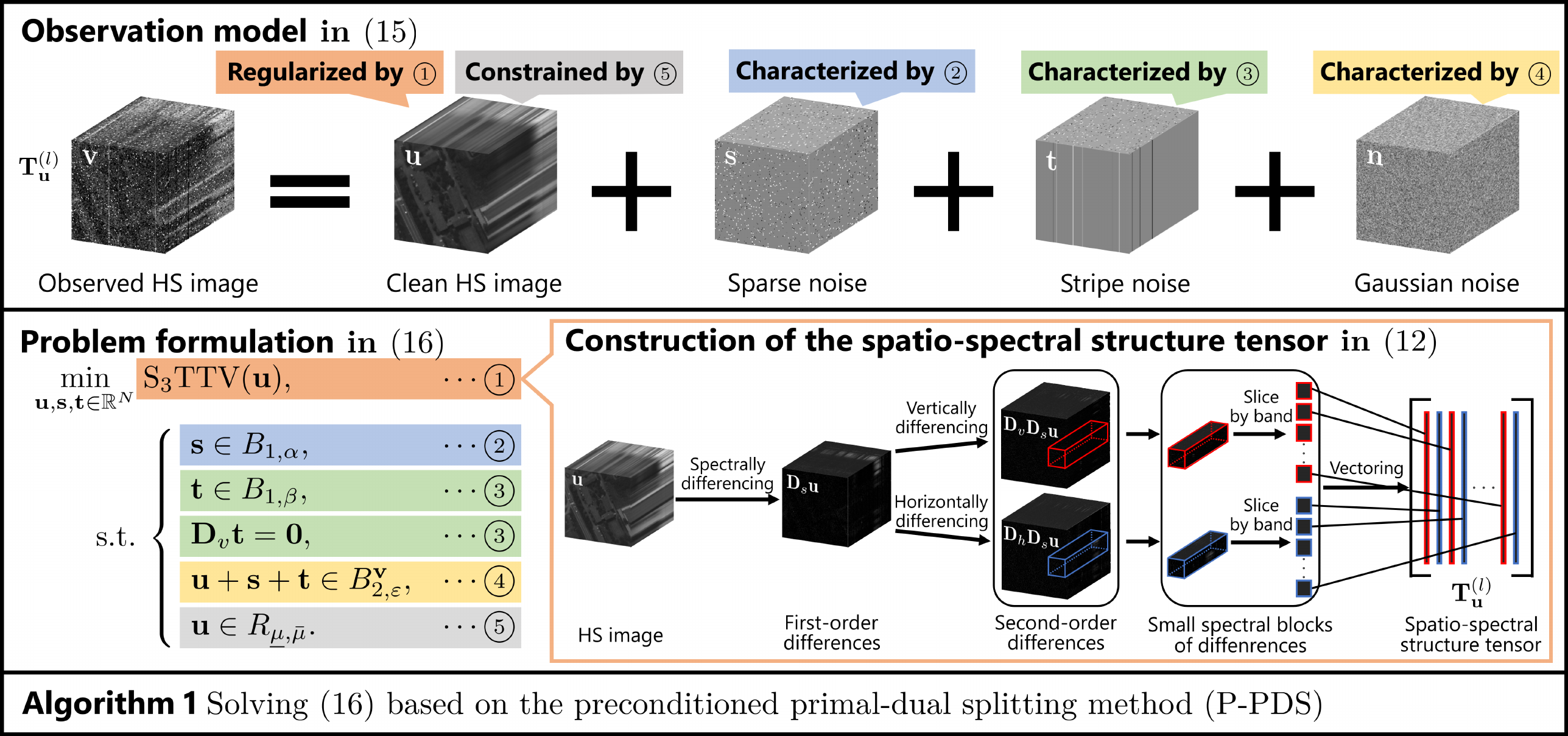}
	\end{center}
	\caption{Illustration of the proposed method, i.e., $\SSSTTV$.}
	\vspace{-2mm}
	\label{fig:schematic_diagram}
\end{figure*}

In the following, we first describe the design of the $\SSSTTV$ regularization function. Next, we consider a situation where an HS image is contaminated with mixed noise and introduce the corresponding observation model. Based on this model, we formulate an HS image denoising problem as a constrained convex optimization problem involving the $\SSSTTV$ regularization function. Finally, we derive an algorithm based on P-PDS to efficiently solve the optimization problem. A schematic diagram of $\SSSTTV$ is shown in Fig.~\ref{fig:schematic_diagram}.

\subsection{Spatio-Spectral Structure Tensor Total Variation ($\SSSTTV$)}
\label{subsec:S3TTV}
Before describing the proposed regularization function, we introduce the notion of \textit{spatio-spectral structure tensor}\footnote{In the SSST paper~\cite{Kurihara2017SSST}, a structure tensor with the same name as the one we proposed (i.e., spatio-spectral structure tensor) is introduced. However, they are essentially different because the structure tensor in SSST consists of first-order differences, whereas that in our regularization function consists of second-order spatio-spectral differences.}.
First, for a given HS image $\HSIClean$, we calculate the second-order spatial-spectral differences $\DiffOpVert \DiffOpBand \HSIClean$ and $\DiffOpHori \DiffOpBand \HSIClean$.
Next, we extract \textit{small spectral blocks} by cropping the second-order spatio-spectral differences to the size $\NumBlockVert \times \NumBlockHori (\NumBlockVert << \NumVert, \NumBlockHori << \NumHori)$ for all bands\footnote{At the boundaries, the block cannot be cropped to an $\NumBlockVert \times \NumBlockHori \times \NumBand$ size. For example, when the difference is cropped to a $3 \times 3 \times \NumBand$ block at a center $(1, 1)$, a $2 \times 2 \times \NumBand$ block is created. In this case, we pad the lacking areas with pixels on the opposite boundaries to make the block $\NumBlockVert \times \NumBlockHori \times \NumBand$.}. Then, the $\IndexBlock$-th spatio-spectral structure tensor $\SmallBlock{\IndexBlock}$ is defined by vectorizing the second-order spatio-spectral differences in the $\IndexBlock$-th small spectral block by band and arranging them in parallel as follows:
\begin{equation}
	\begin{split}
		\label{eq:SSST}
		\SmallBlock{\IndexBlock} :=  
		&\bigl( \lbrack \DiffOpVert \DiffOpBand \HSIClean \rbrack_{1}^{(\IndexBlock)} \: 
		\lbrack\DiffOpHori \DiffOpBand \HSIClean \rbrack_{1}^{(\IndexBlock)} \\ 
		& \: \cdots \lbrack \DiffOpVert \DiffOpBand \HSIClean \rbrack_{\NumBand}^{(\IndexBlock)} \:
		\lbrack\DiffOpHori \DiffOpBand \HSIClean \rbrack_{\NumBand}^{(\IndexBlock)} 
		\bigr) \in \RealSpace{\NumBlockVert \NumBlockHori \times 2\NumBand},
	\end{split}
\end{equation}
where $\lbrack \DiffOpVert \DiffOpBand \HSIClean \rbrack_{\IndexBand}^{(\IndexBlock)} \in \RealSpace{\NumBlockVert \NumBlockHori}$ and $\lbrack \DiffOpHori \DiffOpBand \HSIClean \rbrack_{\IndexBand}^{(\IndexBlock)} \in \RealSpace{\NumBlockVert \NumBlockHori}$ are the second-order spatio-spectral differences of $\IndexBand$-th band in the $\IndexBlock$-th small spectral block. Since HS images have the strong correlation across all bands, $\lbrack \DiffOpVert \DiffOpBand \HSIClean \rbrack_{1}^{(\IndexBlock)}, \ldots, \lbrack \DiffOpVert \DiffOpBand \HSIClean \rbrack_{\NumBand}^{(\IndexBlock)}$ and $\lbrack \DiffOpHori \DiffOpBand \HSIClean \rbrack_{1}^{(\IndexBlock)}, \ldots, \lbrack \DiffOpHori \DiffOpBand \HSIClean \rbrack_{\NumBand}^{(\IndexBlock)}$ are similar vectors, respectively, i.e., the columns of $\SmallBlock{\IndexBlock}$ are approximately linearly dependent. The flow of constructing the spatio-spectral structure tensor is depicted in the middle right of Fig.~\ref{fig:schematic_diagram}.

To capture the spatial piecewise-smoothness, the spatial similarity between adjacent bands, and the spectral correlation of an HS image, we propose a regularization function using the spatio-spectral structure tensors as follows:
\begin{equation}
	\label{eq:S3TTV_Lu}
	\textstyle \SSSTTV (\HSIClean) := \sum_{\IndexBlock=1}^{\NumBlock} \| \SmallBlock{\IndexBlock} \|_{*},
\end{equation}
where $\NumBlock$ is the number of the extracted small spectral blocks.
We call this function as \textit{Spatio-Spectral Structure Tensor Total Variation} ($\SSSTTV$).
Here, $\SmallBlock{\IndexBlock}$ is represented with an operator $\ExpantionOp{\IndexBlock} \in \RealSpace{2 \NumBlockVert \NumBlockHori \NumBand \times 2 \NumVert \NumHori \NumBand}$ that extracts the $\IndexBlock$-th small spectral block as
\begin{equation}
	\label{eq:rewrite_Lu}
	\SmallBlock{\IndexBlock}
	= \ExpantionOp{\IndexBlock} \DiffOpSp \DiffOpBand \HSIClean.
\end{equation}

Minimizing the nuclear norms of the matrices $\SmallBlock{1}, \ldots, \SmallBlock{\NumBlock}$ allows for both the reduction of the energy of second-order differences and the enhancement of the spectral correlation of second-order differences. By reducing the energy of second-order differences instead of first-order differences, the proposed method promotes both the spatial piecewise-smoothness and the spatial similarity between adjacent bands while avoiding over-smoothing, as shown in~\cite{Aggarwal2016SSTV}. Furthermore, experimental analysis also indicates that second-order differences are more effective than first-order differences in distinguishing noise from HS images (see Sec.~\ref{subsec:Discussion}-4) on p. 13 for details). On the other hand, by enhancing the spectral correlation of second-order difference, the proposed method enhances the spectral correlation of HS images (see Appendix~A on p. 16 for proof). Therefore, by solving an optimization problem that incorporates $\SSSTTV$, our method simultaneously captures the above three natures.

\subsection{HS Image Denoising Problem by $\SSSTTV$}
\label{subsec:HSI_Denoising_Problem}
An observed HS image $\HSIObsv \in \RealSpace{\NumAll}$ contaminated by mixed noise is modeled by
\begin{equation}
	\label{eq:Obsevation_model}
	\HSIObsv = \bar{\HSIClean} + \bar{\NoiseSparse} + \bar{\NoiseStripe} + \NoiseGauss,
\end{equation}
where $\bar{\HSIClean}$ is a clean HS image, $\bar{\NoiseSparse}$ is sparse noise, $\bar{\NoiseStripe}$ is stripe noise, and $\NoiseGauss$ is Gaussian noise, respectively.
Modeling different types of noise as separate components is an effective approach for mixed noise removal in HS images~\cite{Zhang2022Double}.

Based on the above observation model, we formulate an HS image denoising problem involving $\SSSTTV$ as a constrained convex optimization problem with the following form:
\begin{equation}
	\label{prob:S3TTV_denoising}
	\min_{\HSIClean, \NoiseSparse, \NoiseStripe \in \RealSpace{\NumAll}} \SSSTTV(\HSIClean) \: \mathrm{s.t.} \:
	\begin{cases} 
		\NoiseSparse \in \BallSparse, \\ 
		\NoiseStripe \in \BallStripe, \\
		\DiffOpVert \NoiseStripe = \mathbf{0}, \\
		\HSIClean + \NoiseSparse + \NoiseStripe \in \BallFidel, \\  
		\HSIClean \in \SetRange,
	\end{cases}
\end{equation}
where
\begin{align}
	\label{eq:constraint_sparse}
	&\BallSparse := \{ \VarOne \in \RealSpace{\NumAll} | \:
	\|\VarOne\|_{1} \leq \RadiusSparse \},  \\
	\label{eq:constraint_stripe}
	&\BallStripe := \{ \VarOne \in \RealSpace{\NumAll} | \:
	\|\VarOne\|_{1} \leq \RadiusStripe \},  \\
	\label{eq:constraint_fidel}
	&\BallFidel := \{ \VarOne \in \RealSpace{\NumAll} | \:
	\|\VarOne - \HSIObsv\|_2 \leq \RadiusFidel \},  \\
	\label{eq:constraint_box}
	&\SetRange := \{ \VarOne \in \RealSpace{\NumAll} | \:
	\MinRange \leq \ElementOne{\IndexOne} \leq \MaxRange  \: (\IndexOne = 1, \dots , \NumAll) \}.
\end{align}

The first constraint characterizes sparse noise $\NoiseSparse$ with the zero-centered $\ell_1$-ball of the radius $\RadiusSparse > 0$. The second constraint controls the intensity of stripe noise $\NoiseStripe$ and the third constraint captures the vertical flatness property by imposing zero to the vertical gradient of $\NoiseStripe$. These constraints effectively characterize stripe noise~\cite{Naganuma2022Destriping}. The fourth constraint serves as data-fidelity with the $\HSIObsv$-centered $\ell_2$-ball of the radius $\RadiusFidel > 0$. The fifth constraint is a box constraint with $\MinRange < \MaxRange$ which represents the dynamic range of $\HSIClean$. For normalized HS images, we can set $\MinRange = 0$ and $\MaxRange = 1$.

Using the first, second, and fourth constraints instead of adding terms to the objective function makes it much easier to adjust the parameters $\RadiusSparse$, $\RadiusStripe$, and $\RadiusFidel$. This is because by expressing multiple terms as constraints, rather than adding them to the objective function, the hyperparameters associated with each term are converted to be independent of each other, and appropriate parameters can be determined without interdependence. Such advantage has been addressed, e.g., in~\cite{Afonso2011Constraint, Chierchia2015Constraint, Ono2015Constraint, Ono2017Constraint, Ono2019Constraint}.

\subsection{Optimization}
\label{subsec:Optim}
To solving Prob.~\eqref{prob:S3TTV_denoising} by an efficient algorithm based on P-PDS~\cite{Pock2011PPDS}, we need to reformulate it into the P-PDS applicable form~\eqref{prob:convex_optim_prob}. Using the indicator functions $\FuncIndicator{\SetZero}$, $\FuncIndicator{\BallFidel}$, $\FuncIndicator{\BallStripe}$, $\FuncIndicator{\BallSparse}$, and $\FuncIndicator{\SetRange}$, we rewrite Prob.~\eqref{prob:S3TTV_denoising} into an equivalent form:
\begin{align}
	\label{prob:denoising2PPDS}
	\min_{
		\substack{
			\HSIClean, \NoiseSparse, \NoiseStripe\\ 
			\VarDualMatrix{1,1}, \ldots , \VarDualMatrix{1,\NumBlock}, 
			\VarDual{2}, \VarDual{3}}} \:
	& \sum_{\IndexBlock=1}^{\NumBlock} \| 
	\VarDualMatrix{1, \IndexBlock} \|_{*}
	+ \FuncIndicator{\SetZero} (\VarDual{2})
	+ \FuncIndicator{\BallFidel} (\VarDual{3}) \nonumber \\
	& + \FuncIndicator{\BallSparse} (\NoiseSparse) 
	+ \FuncIndicator{\BallStripe} (\NoiseStripe)
	+ \FuncIndicator{\SetRange} (\HSIClean),  \nonumber \\
	& \mathrm{s.t.} \:
	\begin{cases} 
		\VarDualMatrix{1,1} = \ExpantionOp{1} \DiffOpSp \DiffOpBand \HSIClean, \\ 
		\vdots \\ 
		\VarDualMatrix{1,\NumBlock} = \ExpantionOp{\NumBlock} \DiffOpSp \DiffOpBand \HSIClean, \\ 
		\VarDual{2} = \DiffOpVert \NoiseStripe, \\
		\VarDual{3} = \HSIClean + \NoiseSparse + \NoiseStripe. 
	\end{cases}
\end{align}
Let $\HSIClean$, $\NoiseSparse$, and $\NoiseStripe$ be the primal variables and $\VarDualMatrix{1,1}, \ldots , \VarDualMatrix{1,\NumBlock}, \VarDual{2}, \VarDual{3}$ be the dual variables. The operators $\ExpantionOp{1}, \ldots, \ExpantionOp{\NumBlock}, \DiffOpSp, \DiffOpBand$, and $\DiffOpVert$ are linear operators. The indicator functions $\FuncIndicator{\SetZero}$, $\FuncIndicator{\BallFidel}$, $\FuncIndicator{\BallStripe}$, $\FuncIndicator{\BallSparse}$, and $\FuncIndicator{\SetRange}$ and the nuclear norm $\| \cdot \|_{*}$ are proper lower semi-continuous convex. Then, by defining,
\begin{align}
	\label{eq:FuncMapping}
	& \FuncPrimal{1} (\HSIClean) := \FuncIndicator{\SetRange} (\HSIClean), \nonumber \\
	& \FuncPrimal{2} (\NoiseSparse) := \FuncIndicator{\BallSparse} (\NoiseSparse), \nonumber \\
	& \FuncPrimal{3} (\NoiseStripe) := \FuncIndicator{\BallStripe} (\NoiseStripe), \nonumber \\
	& \FuncDual{1} (\VarDualMatrix{1, 1}) := \| \VarDualMatrix{1, 1} \|_{*}, \ldots, \FuncDual{\NumBlock}(\VarDualMatrix{1, \NumBlock}) := \| \VarDualMatrix{1, \NumBlock} \|_{*}, \nonumber \\
	& \FuncDual{\NumBlock + 1} (\VarDual{2}) := \FuncIndicator{\SetZero} (\VarDual{2}), \nonumber \\
	& \FuncDual{\NumBlock + 2} (\VarDual{3}) := \FuncIndicator{\BallFidel} (\VarDual{3}),
\end{align}
Prob.~\eqref{prob:denoising2PPDS} is reduced to Prob.~\eqref{prob:convex_optim_prob}. 
Therefore, P-PDS is applicable to Prob.~\eqref{prob:denoising2PPDS}.

\begin{figure}[!t]
	\begin{algorithm}[H]
	    \caption{P-PDS-based solver for (18)}
		\label{algo_DPPDS}
		\begin{algorithmic}[1]
			\REQUIRE $\HSIClean^{(0)}, \NoiseSparse^{(0)}, \NoiseStripe^{(0)},  \VarDualMatrix{1,\IndexBlock}^{(0)}(\IndexBlock = 1, \ldots \NumBlock), \VarDual{2}^{(0)}, \VarDual{3}^{(0)}$
			\ENSURE $\HSIClean^{(\IndexAlg)}$
			\WHILE {A stopping criterion is not satisfied}
    			\STATE $\HSIClean^{(\IndexAlg+1)} \leftarrow$ \\ $ \qquad  \Projection{\SetRange}\left( \HSIClean^{(\IndexAlg)} - \ParamStepsize{\HSIClean}  \bigl(\sum_{\IndexBlock=1}^{\NumBlock} \DiffOpBandT \DiffOpSpT \ExpantionOp{\IndexBlock}^{\top} \VarDualMatrix{1,\IndexBlock}^{(\IndexAlg)} + \VarDual{3}^{(\IndexAlg)} \bigr) \right)$;
    			\STATE $\NoiseSparse^{(\IndexAlg+1)} \leftarrow \prox_{\ParamStepsize{\NoiseSparse}, \FuncIndicator{\BallSparse}} \left( \NoiseSparse^{(\IndexAlg)} - \ParamStepsize{\NoiseSparse} \VarDual{3}^{(\IndexAlg)} \right)$;
                    \STATE $\NoiseStripe^{(\IndexAlg+1)} \leftarrow \prox_{\ParamStepsize{\NoiseStripe}, \FuncIndicator{\BallStripe}} \left( \NoiseStripe^{(\IndexAlg)} - \ParamStepsize{\NoiseStripe} \bigl( \DiffOpVertT \VarDual{2}^{(\IndexAlg)} + \VarDual{3}^{(\IndexAlg)} \bigr) \right)$;
                    \STATE $\ResHSIClean \leftarrow 2\HSIClean^{(\IndexAlg+1)} - \HSIClean^{(\IndexAlg)}$;
                    \STATE $\ResNoiseSparse \leftarrow 2\NoiseSparse^{(\IndexAlg+1)} - \NoiseSparse^{(\IndexAlg)}$;
                    \STATE $\ResNoiseStripe \leftarrow 2\NoiseStripe^{(\IndexAlg+1)} - \NoiseStripe^{(\IndexAlg)}$;
    			\FOR{$\IndexBlock = 1, \ldots, \NumBlock$}
    			    \STATE $\VarDualMatrix{1,\IndexBlock}^{'} \leftarrow \VarDualMatrix{1,\IndexBlock}^{(\IndexAlg)} + \ParamStepsize{\VarDualMatrix{1, \IndexBlock}} \ExpantionOp{\IndexBlock} \DiffOpSp \DiffOpBand \ResHSIClean$;
    			    \STATE $\VarDualMatrix{1,\IndexBlock}^{(\IndexAlg+1)} \leftarrow \VarDualMatrix{1,\IndexBlock}^{'} - \ParamStepsize{\VarDualMatrix{1, \IndexBlock}} \prox_{\ParamStepsize{\VarDualMatrix{1, \IndexBlock}}^{-1}, \| \cdot \|_{*}} \left(\ParamStepsize{\VarDualMatrix{1, \IndexBlock}}^{-1} \VarDualMatrix{1,\IndexBlock}^{'}\right) $;
    		    \ENDFOR
                    \STATE $\VarDual{2}^{(\IndexAlg+1)} \leftarrow \VarDual{2}^{(\IndexAlg)} + \ParamStepsize{\VarDual{2}} \DiffOpVert \ResNoiseStripe$;
    			\STATE $\VarDual{3}^{'} \leftarrow \VarDual{3}^{(\IndexAlg)} + \ParamStepsize{\VarDual{3}} \left( \ResHSIClean + \ResNoiseSparse + \ResNoiseStripe \right)$;
    			\STATE $\VarDual{3}^{(\IndexAlg+1)} \leftarrow \VarDual{3}^{'} - \ParamStepsize{\VarDual{3}} \Projection{\BallFidel} \left(\ParamStepsize{\VarDual{3}}^{-1} \VarDual{3}^{'}  \right)$;
    			\STATE $\IndexAlg \leftarrow \IndexAlg + 1$;
			\ENDWHILE
		\end{algorithmic}
	\end{algorithm}
	\vspace{-2mm}
\end{figure}

We show the detailed algorithm in Alg.~1 based on~\eqref{algo:P-PDS}.
The proximity operators of $\FuncIndicator{\SetRange}$, $\FuncIndicator{\SetZero}$, and  $\FuncIndicator{\BallFidel}$ are calculated by
\begin{align}
	\label{eq:prox_box_constraint}
	\lbrack \prox_{\ParamStepsize{} \FuncIndicator{\SetRange}} (\VarOne) \rbrack_{i} 
	&= \lbrack \Projection{\SetRange} (\VarOne) \rbrack_{i} =
	\begin{cases} 
		\MinRange, & \text{if } \ElementOne{i} < \MinRange, \\ 
		\MaxRange, & \text{if } \ElementOne{i}> \MaxRange, \\ 
		\ElementOne{i}, & \text{otherwise,} 
	\end{cases} \\
	\label{eq:prox_zeroset}
	\prox_{\gamma \FuncIndicator{\SetZero}}(\VarOne) &= \mathbf{0}, \\
	\label{eq:prox_l2ball_constraint}
	\prox_{\gamma \FuncIndicator{\BallFidel}}(\VarOne) &= \Projection{\BallFidel} (\VarOne) = 
	\begin{cases}
		\VarOne, & \text{if } \VarOne \in \BallFidel, \\ 
		\HSIObsv + \frac{\varepsilon (\VarOne - \HSIObsv)}{\| \VarOne - \HSIObsv \|_2}, & \text{otherwise.}
	\end{cases}
\end{align}
The proximity operators of $\FuncIndicator{\BallSparse} (\NoiseSparse)$ and $\FuncIndicator{\BallStripe} (\NoiseStripe)$ can be efficiently computed by a fast $\ell_{1}$-ball projection algorithm~\cite{Condat2016L1ball}.
The proximity operator for the nuclear norm $\| \cdot \|_{*}$ is calculated by
\begin{align}
	\label{eq:prox_nuclear_norm}
	& \prox_{\ParamStepsize{} \| \cdot \|_{*}} (\MatrixOne) 
	= \LeftSingularMatrix \SingularMatrix{\ParamStepsize{}} \RightSingularMatrixT, \notag \\
	& \SingularMatrix{\ParamStepsize{}}
	= \diag \bigl( \max \{ \SingularValue{1} - \ParamStepsize{}, 0\}, \cdots, \max \{ \SingularValue{\NumSingularValue} - \ParamStepsize{}, 0 \} \bigr),
\end{align}
where $\NumSingularValue$ is the number of nonzero singular values, i.e., $\min \{\NumBlockVert \NumBlockHori, 2 \NumBand \}$, and the singular value decomposition of $\MatrixOne$ is $\LeftSingularMatrix \diag \left( \SingularValue{1}, \ldots, \SingularValue{\NumSingularValue} \right) \RightSingularMatrixT$.

Based on Eq.~\eqref{eq:Preconditioners}, the stepsize parameters $\ParamStepsize{\HSIClean}$, $\ParamStepsize{\NoiseSparse}$, $\ParamStepsize{\NoiseStripe}$, $\ParamStepsize{\VarDualMatrix{1, 1}}$,~\ldots,~$\ParamStepsize{\VarDualMatrix{1, \NumBlock}}$, $\ParamStepsize{\VarDual{2}}$, and $\ParamStepsize{\VarDual{3}}$ are given as
\begin{align}
	\label{eq:stepsize_denoising}
	& \ParamStepsize{\HSIClean} = \frac{1}{8 \NumBlock + 1}, \: \ParamStepsize{\NoiseSparse} = 1, \: \ParamStepsize{\NoiseStripe} = \frac{1}{3}, \nonumber \\
	& \ParamStepsize{\VarDualMatrix{1, 1}} = \ldots= \ParamStepsize{\VarDualMatrix{1, \NumBlock}} = \frac{1}{4}, \nonumber \\
	& \ParamStepsize{\VarDual{2}} = \frac{1}{2}, \: \ParamStepsize{\VarDual{3}} = \frac{1}{3}.
\end{align}
From Lemma~\ref{lem:StepsizeParameters}, the above stepsizes satisfy the inequality~\eqref{ieq:ConvergenceCondition}, and the sequence $\{\HSIClean^{(t)}, \NoiseSparse^{(t)}, \NoiseStripe^{(t)}, \VarDualMatrix{1,1}, \ldots , \VarDualMatrix{1,\NumBlock}, \VarDual{2}, \VarDual{3} \}$ generated by Alg.~1 converges to an optimal solution of Prob.~\eqref{prob:denoising2PPDS}.

\subsection{Computational Complexity}
\label{subsec:Comp_Comp}
Table~\ref{tab:ComputationalComplexity} shows the computational complexities of each operator in the proposed algorithm. Based on Table~\ref{tab:ComputationalComplexity}, the computational complexities of each step in Alg.~1 are given as follows:
\begin{itemize}
	\item Step 2: $\mathcal{O} (\NumBlock \NumBlockVert \NumBlockHori \NumBand)$,
	\item Steps 3 and 4: $\mathcal{O} (\NumAll \log{\NumAll})$,
	\item Steps 5, 6, 7, 9, 12, 13, and 14: $\mathcal{O} (\NumAll)$,
	\item Step 10: $\mathcal{O} (\NumBlockVert \NumBlockHori \NumBand \min({\NumBlockVert \NumBlockHori, 2 \NumBand}))$.
\end{itemize}
Thus, the computational complexity for each iteration of Alg.~1 is $\mathcal{O} (\NumBlock \NumBlockVert \NumBlockHori \NumBand \min({\NumBlockVert \NumBlockHori, 2 \NumBand}))$.
\begin{table}[!t]
    \begin{center}
        \caption{Computational Complexity of Each Operation.}
        \label{tab:ComputationalComplexity}
        		\scalebox{0.95}{
        \begin{tabular}{cc}
            \toprule
                Operation & $\mathcal{O}$-notation \\
            \cmidrule(lr){1-2}
            \vspace{-0.5mm}
               $\DiffOpSp \OperationVec, (\OperationVec \in \RealSpace{\NumAll})$ & $\Order{\NumAll}$ \\
               $\DiffOpSpT \OperationVec, (\OperationVec \in \RealSpace{2\NumAll})$ & $\Order{\NumAll}$ \\
               $\DiffOpBand \OperationVec, (\OperationVec \in \RealSpace{\NumAll})$ & $\Order{\NumAll}$ \\
               $\DiffOpBandT \OperationVec, (\OperationVec \in \RealSpace{\NumAll})$ & $\Order{\NumAll}$ \\
               $\DiffOpVert \OperationVec, (\OperationVec \in \RealSpace{\NumAll})$ & $\Order{\NumAll}$ \\
               $\DiffOpVertT \OperationVec, (\OperationVec \in \RealSpace{\NumAll})$ & $\Order{\NumAll}$ \\ 
               $\ExpantionOp{\IndexBlock} \OperationVec, (\OperationVec \in \RealSpace{\NumAll})$ & $\Order{\NumBlockVert \NumBlockHori \NumBand}$ \\
               $\ExpantionOp{\IndexBlock}^{\top} \OperationVec, (\OperationVec \in \RealSpace{\NumBlockVert \NumBlockHori \NumBand})$ & $\Order{\NumBlockVert \NumBlockHori \NumBand}$ \\
               $\Projection{\SetRange}(\OperationVec), (\OperationVec \in \RealSpace{\NumAll})$ & $\Order{\NumAll}$\\
               $\prox_{\ParamStepsize{\NoiseSparse} \FuncIndicator{\BallSparse}} (\OperationVec)$ in~[71], $(\OperationVec \in \RealSpace{\NumAll})$ & $\Order{\NumAll \log{\NumAll}}$ \\
               $\prox_{\ParamStepsize{\NoiseStripe} \FuncIndicator{\BallStripe}} (\OperationVec)$ in~[71] $(\OperationVec \in \RealSpace{\NumAll})$ & $\Order{\NumAll \log{\NumAll}}$ \\
               $\Projection{\BallFidel} (\OperationVec), (\OperationVec \in \RealSpace{\NumAll})$ & $\Order{\NumAll}$\\
               $\prox_{\ParamStepsize{} \| \cdot \|_{*}} (\OperationMatrix) (\OperationMatrix \in \RealSpace{\NumBlockVert \NumBlockHori \times 2\NumBand})$ & $\Order{\NumBlockVert \NumBlockHori \NumBand \min({\NumBlockVert \NumBlockHori, 2 \NumBand})}$ \\
            \bottomrule
        \end{tabular}
        		}
    \end{center}
    \vspace{-3mm}
\end{table}

\subsection{Empirical Validation of the $\SSSTTV$ Design}
\label{subsec:DataDist}
\begin{figure*}[t]
	\begin{center}
		\begin{minipage}{0.325\hsize}
			\centerline{\includegraphics[width=\hsize]{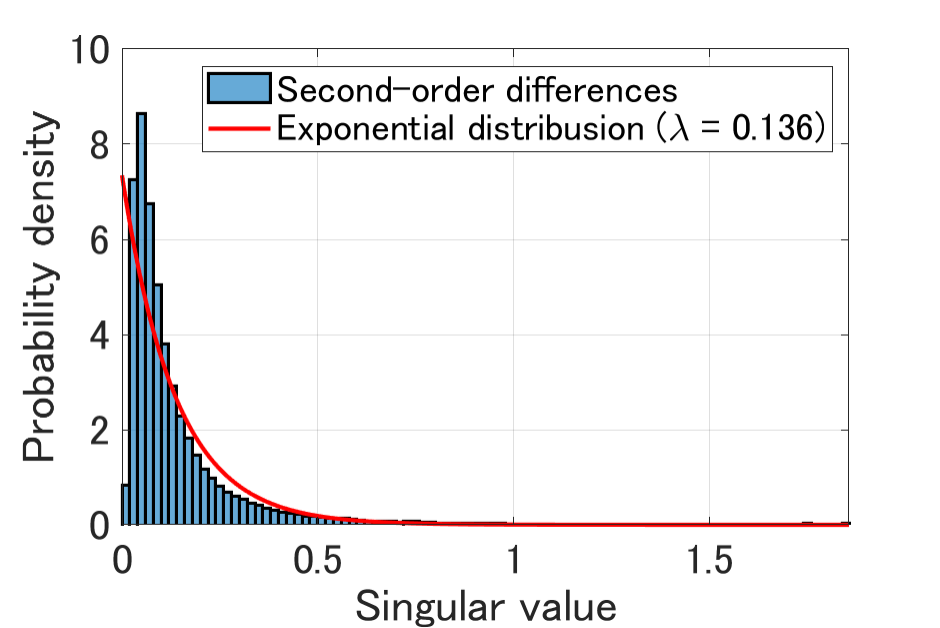}}
		\end{minipage}
		\begin{minipage}{0.325\hsize}
			\centerline{\includegraphics[width=\hsize]{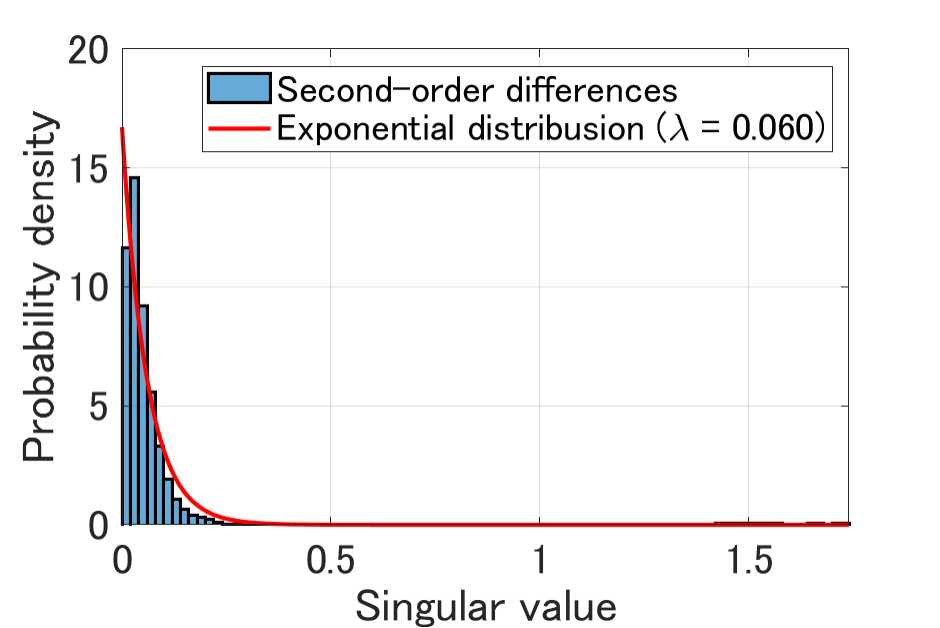}}
		\end{minipage}
		\begin{minipage}{0.325\hsize}
			\centerline{\includegraphics[width=\hsize]{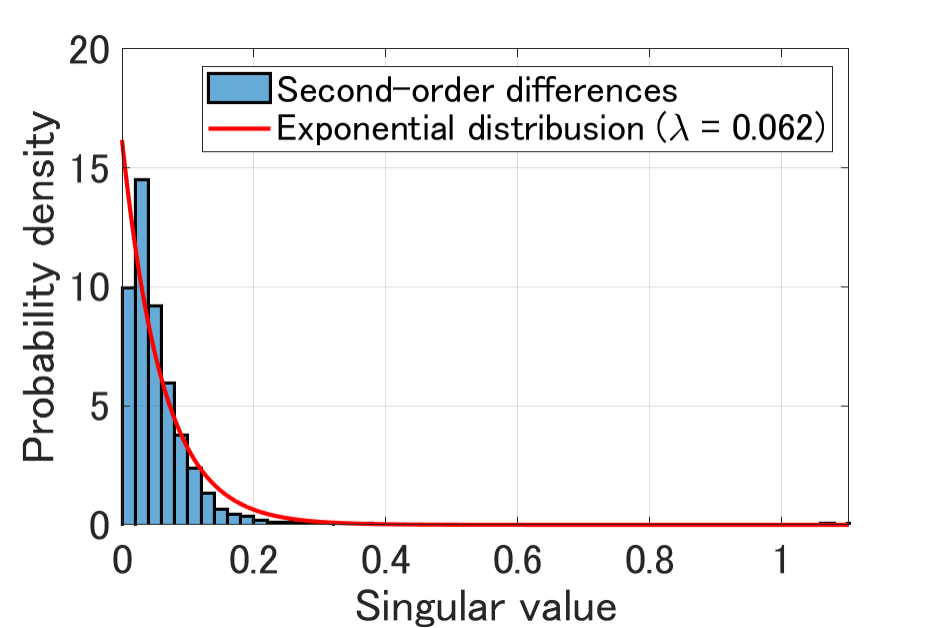}}
		\end{minipage}
		
		\vspace{1mm}
		
		\begin{minipage}{0.325\hsize}
			\centerline{\small{(a)}}
		\end{minipage}
		\begin{minipage}{0.325\hsize}
			\centerline{\small{(b)}}
		\end{minipage}
		\begin{minipage}{0.325\hsize}
			\centerline{\small{(c)}}
		\end{minipage}
		
	\end{center}
	
	\caption{Normalized histograms of singular values computed from all spatio-spectral structure tensors, along with the fitted exponential distributions: (a) Jasper Ridge, (b) Pavia University, and (c) Beltsville.}
	
	\label{fig:data_dist}
\end{figure*}

In this section, we empirically verify the validity of the proposed $\SSSTTV$ design in Eq.~\eqref{eq:S3TTV_Lu} from the perspective of data distribution. Specifically, we plot the distributions of the singular values of the spatio-spectral structure tensors $\SmallBlock{\IndexBlock}$ of HS images. Fig.~\ref{fig:data_dist} shows the normalized histograms of the singular values computed from all spatio-spectral structure tensors extracted from each of the three HS images used in our experiments: \textit{Jasper Ridge}, \textit{Pavia University}, and \textit{Beltsville}. Each histogram is overlaid with a fitted exponential distribution, which corresponds to the positive half of the Laplacian distribution. The exponential distribution is defined as
\begin{equation}
	\label{eq:ExponentialDistribution}
	p(x; \lambda) = \frac{1}{\lambda} \exp\left(-\frac{x}{\lambda} \right), \quad (x \geq 0),
\end{equation}
where $\lambda$ is the scale parameter, which is set as the mean of the singular values. As shown in Fig.~\ref{fig:data_dist}, the singular value distributions closely match the exponential distributions. This indicates that minimizing $\ell_{1}$-norm the singular values of the spatio-spectral structure tensors is reasonable to make observed noisy images closer to HS images of interest. Therefore, the design of $\SSSTTV$ is justified from the perspective of the data distribution.

\begin{table*}[t]
	\begin{center}
		\caption{MPSNRs of the Simulated HS Image Denoising Results.}
		\label{tab:MPSNR}
		\scalebox{0.70}{
			\begin{tabular}{cc cccccccccccc}
				\toprule
				Image & Noise & SSTV~\cite{Aggarwal2016SSTV} & HSSTV1~\cite{Takeyama2020HSSTV} & HSSTV2~\cite{Takeyama2020HSSTV} & $\llHTV$~\cite{Wang2021l0l1HTV} & STV~\cite{Lefkimmiatis2015STV} & SSST~\cite{Kurihara2017SSST} & LRTDTV~\cite{Wang2018LRTDTV} & FGSLR~\cite{Chen2022FGSLR} & TPTV~\cite{Chen2023TPTV} & QRNN3D~\cite{Wei2021QRNN3D} & FastHyMix~\cite{Zhuang2023FastHyMix} & $\SSSTTV$ \\
				\cmidrule(lr){1-14} 

				\multirow{8}{*}{Jasper Ridge} 
				& Case 1 & 
				36.24 & \underline{36.28} & 35.79 & 35.47 & 28.01 & 35.05 & 34.98 & 33.51 & 34.22 & 28.42 & \textbf{38.69} & 35.51 \\ 
				& Case 2 & 
				\underline{39.43} & 39.04 & 39.10 & 38.59 & 29.67 & 36.59 & 37.73 & 35.51 & 39.01 & 31.99 & 27.76 & \textbf{39.99} \\ 
				& Case 3 & 
				34.33 & 34.01 & \underline{34.83} & 34.12 & 27.07 & 32.09 & 34.75 & 33.27 & 33.74 & 27.75 & 25.56 & \textbf{36.15} \\ 
				& Case 4 & 
				42.68 & 41.54 & \underline{44.77} & 43.17 & 38.93 & 41.40 & 36.91 & 41.61 & \textbf{51.57} & 24.56 & 34.88 & 44.26 \\ 
				& Case 5 & 
				39.10 & \underline{39.31} & 38.44 & 38.96 & 30.60 & 37.67 & 35.84 & 35.49 & 37.87 & 23.95 & 35.15 & \textbf{40.04} \\ 
				& Case 6 & 
				34.22 & 34.81 & 33.97 & 34.57 & 27.95 & \underline{34.90} & 33.56 & 33.28 & 33.48 & 23.52 & 34.22 & \textbf{35.50} \\ 
				& Case 7 & 
				\underline{39.40} & 39.12 & 38.71 & 38.47 & 29.40 & 35.72 & 35.95 & 35.32 & 37.96 & 23.71 & 37.22 & \textbf{39.82} \\ 
				& Case 8 & 
				34.68 & 34.22 & \underline{35.15} & 34.27 & 26.79 & 30.99 & 33.49 & 33.07 & 33.15 & 23.22 & 23.17 & \textbf{36.05} \\ 

				\cmidrule(lr){1-14} 

				\multirow{8}{*}{Pavia University} 
				& Case 1 & 
				35.83 & \underline{36.24} & 35.39 & 35.30 & 28.33 & 35.08 & 32.79 & 32.55 & 31.52 & 31.61 & \textbf{37.69} & 35.73 \\ 
				& Case 2 & 
				38.67 & 38.55 & 38.81 & 38.20 & 29.85 & 35.40 & 35.18 & 35.40 & 37.14 & 36.38 & \underline{39.08} & \textbf{39.79} \\ 
				& Case 3 & 
				33.05 & 33.50 & 33.49 & 32.98 & 27.48 & 31.89 & 32.66 & 31.90 & 31.19 & 31.30 & \textbf{36.03} & \underline{34.39} \\ 
				& Case 4 & 
				41.34 & 40.72 & \underline{44.82} & 42.29 & 39.31 & 40.87 & 32.34 & 40.31 & \textbf{48.37} & 35.42 & 34.31 & 43.41 \\ 
				& Case 5 & 
				39.54 & \underline{39.63} & 38.89 & 39.32 & 30.74 & 36.82 & 32.47 & 35.48 & 36.08 & 34.57 & 34.78 & \textbf{40.30} \\ 
				& Case 6 & 
				34.39 & \underline{35.03} & 34.11 & 34.82 & 28.32 & 34.95 & 30.84 & 31.87 & 31.66 & 31.43 & 33.60 & \textbf{35.16} \\ 
				& Case 7 & 
				\underline{38.80} & 38.68 & 38.78 & 38.05 & 29.62 & 34.53 & 32.41 & 35.19 & 35.79 & 33.58 & 31.98 & \textbf{39.36} \\ 
				& Case 8 & 
				33.06 & 33.53 & \underline{33.62} & 32.83 & 27.22 & 30.80 & 30.84 & 31.67 & 31.13 & 30.74 & 31.79 & \textbf{34.33} \\ 

				\cmidrule(lr){1-14} 

				\multirow{8}{*}{Beltsville} 
				& Case 1 & 
				35.20 & \underline{35.89} & 34.88 & 34.71 & 28.95 & 35.79 & 34.15 & 34.58 & 32.24 & 27.91 & \textbf{39.29} & 35.35 \\ 
				& Case 2 & 
				37.87 & 38.19 & 37.66 & 37.34 & 30.66 & 36.74 & \underline{38.67} & 36.65 & 37.78 & 30.34 & 37.85 & \textbf{39.46} \\ 
				& Case 3 & 
				32.86 & 33.87 & 32.87 & 32.68 & 28.20 & 32.26 & 34.05 & \underline{34.34} & 31.75 & 27.52 & \textbf{36.00} & 34.21 \\ 
				& Case 4 & 
				41.21 & 40.88 & 41.48 & 41.56 & 37.73 & 40.18 & 38.79 & \underline{42.89} & \textbf{53.00} & 29.05 & 36.29 & 40.84 \\ 
				& Case 5 & 
				38.46 & \underline{38.75} & 38.04 & 38.16 & 30.57 & 36.98 & 33.31 & 36.55 & 36.83 & 28.88 & 35.14 & \textbf{39.71} \\ 
				& Case 6 & 
				33.92 & 34.64 & 33.74 & 34.21 & 28.47 & \underline{34.72} & 30.44 & 34.15 & 32.92 & 27.93 & 34.12 & \textbf{35.31} \\ 
				& Case 7 & 
				38.02 & \underline{38.22} & 37.66 & 37.23 & 29.87 & 35.35 & 32.76 & 36.37 & 37.09 & 28.44 & 37.13 & \textbf{38.91} \\ 
				& Case 8 & 
				33.07 & 33.92 & 33.20 & 32.65 & 27.65 & 30.82 & 30.44 & \underline{33.94} & 32.31 & 27.29 & 33.23 & \textbf{34.42} \\ 

				\bottomrule
			\end{tabular}
		}
	\end{center}
\end{table*}
\begin{table*}[t]
	\begin{center}
		\caption{MSSIMs of the Simulated HS Image Denoising Results.}
		\label{tab:MSSIM}
		\scalebox{0.70}{
			\begin{tabular}{cc cccccccccccc}
				\toprule
				Image & Noise & SSTV~\cite{Aggarwal2016SSTV} & HSSTV1~\cite{Takeyama2020HSSTV} & HSSTV2~\cite{Takeyama2020HSSTV} & $\llHTV$~\cite{Wang2021l0l1HTV} & STV~\cite{Lefkimmiatis2015STV} & SSST~\cite{Kurihara2017SSST} & LRTDTV~\cite{Wang2018LRTDTV} & FGSLR~\cite{Chen2022FGSLR} & TPTV~\cite{Chen2023TPTV} & QRNN3D~\cite{Wei2021QRNN3D} & FastHyMix~\cite{Zhuang2023FastHyMix} & $\SSSTTV$ \\
				\cmidrule(lr){1-14} 

				\multirow{8}{*}{Jasper Ridge} 
				& Case 1 & 
				0.9266 & \underline{0.9399} & 0.9228 & 0.9063 & 0.7492 & 0.9349 & 0.9152 & 0.9172 & 0.8729 & 0.8427 & \textbf{0.9664} & 0.9121 \\ 
				& Case 2 & 
				0.9631 & \textbf{0.9660} & 0.9606 & 0.9513 & 0.8228 & 0.9518 & 0.9544 & 0.9452 & 0.9521 & 0.8951 & 0.8687 & \underline{0.9647} \\ 
				& Case 3 & 
				0.9086 & 0.9099 & \underline{0.9151} & 0.8959 & 0.7156 & 0.8779 & 0.9123 & 0.9142 & 0.8676 & 0.8235 & 0.7833 & \textbf{0.9266} \\ 
				& Case 4 & 
				0.9823 & 0.9783 & 0.9823 & 0.9826 & 0.9740 & 0.9803 & 0.9572 & 0.9836 & \textbf{0.9895} & 0.8587 & 0.9106 & \underline{0.9838} \\ 
				& Case 5 & 
				0.9570 & \underline{0.9638} & 0.9522 & 0.9525 & 0.8433 & 0.9613 & 0.9316 & 0.9436 & 0.9370 & 0.8389 & 0.9051 & \textbf{0.9648} \\ 
				& Case 6 & 
				0.8854 & 0.9090 & 0.8829 & 0.8879 & 0.7448 & \textbf{0.9328} & 0.8787 & 0.9106 & 0.8574 & 0.7946 & 0.8900 & \underline{0.9131} \\ 
				& Case 7 & 
				0.9625 & \textbf{0.9662} & 0.9569 & 0.9494 & 0.8146 & 0.9426 & 0.9326 & 0.9423 & 0.9451 & 0.8290 & 0.9509 & \underline{0.9641} \\ 
				& Case 8 & 
				0.9129 & 0.9136 & \underline{0.9183} & 0.8958 & 0.7050 & 0.8484 & 0.8807 & 0.9084 & 0.8530 & 0.7787 & 0.6000 & \textbf{0.9257} \\ 

				\cmidrule(lr){1-14} 

				\multirow{8}{*}{Pavia University} 
				& Case 1 & 
				0.9254 & \underline{0.9389} & 0.9171 & 0.9145 & 0.7182 & 0.9269 & 0.8650 & 0.9034 & 0.8120 & 0.9111 & \textbf{0.9558} & 0.9204 \\ 
				& Case 2 & 
				0.9599 & 0.9616 & 0.9586 & 0.9544 & 0.7902 & 0.9345 & 0.9196 & 0.9436 & 0.9399 & 0.9599 & \textbf{0.9723} & \underline{0.9669} \\ 
				& Case 3 & 
				0.8784 & 0.8915 & 0.8848 & 0.8761 & 0.6766 & 0.8582 & 0.8624 & 0.8846 & 0.8071 & 0.9027 & \textbf{0.9470} & \underline{0.9041} \\ 
				& Case 4 & 
				0.9762 & 0.9742 & \textbf{0.9853} & 0.9784 & 0.9727 & 0.9771 & 0.9077 & 0.9822 & 0.9754 & 0.9570 & 0.9337 & \underline{0.9850} \\ 
				& Case 5 & 
				0.9657 & \underline{0.9689} & 0.9596 & 0.9626 & 0.8202 & 0.9477 & 0.8808 & 0.9445 & 0.9216 & 0.9475 & 0.9180 & \textbf{0.9704} \\ 
				& Case 6 & 
				0.8980 & \underline{0.9177} & 0.8915 & 0.9049 & 0.7192 & \textbf{0.9254} & 0.8137 & 0.8976 & 0.8193 & 0.9068 & 0.8968 & 0.9112 \\ 
				& Case 7 & 
				0.9612 & \underline{0.9629} & 0.9587 & 0.9532 & 0.7826 & 0.9207 & 0.8778 & 0.9414 & 0.9221 & 0.9343 & 0.9132 & \textbf{0.9647} \\ 
				& Case 8 & 
				0.8788 & \underline{0.8923} & 0.8869 & 0.8731 & 0.6649 & 0.8210 & 0.8127 & 0.8807 & 0.8068 & 0.8900 & 0.8726 & \textbf{0.9029} \\ 

				\cmidrule(lr){1-14} 

				\multirow{8}{*}{Beltsville} 
				& Case 1 & 
				0.9121 & 0.9310 & 0.9058 & 0.9004 & 0.7256 & \underline{0.9328} & 0.8776 & 0.9131 & 0.8172 & 0.8404 & \textbf{0.9641} & 0.9075 \\ 
				& Case 2 & 
				0.9510 & 0.9566 & 0.9474 & 0.9442 & 0.7994 & 0.9412 & 0.9512 & 0.9483 & 0.9425 & 0.8905 & \textbf{0.9665} & \underline{0.9618} \\ 
				& Case 3 & 
				0.8712 & 0.8941 & 0.8696 & 0.8700 & 0.6870 & 0.8633 & 0.8772 & \underline{0.9090} & 0.8064 & 0.8270 & \textbf{0.9455} & 0.8986 \\ 
				& Case 4 & 
				0.9728 & 0.9729 & 0.9728 & 0.9730 & 0.9556 & 0.9713 & 0.9579 & \textbf{0.9832} & \underline{0.9823} & 0.8837 & 0.9440 & 0.9721 \\ 
				& Case 5 & 
				0.9540 & \underline{0.9605} & 0.9491 & 0.9508 & 0.7987 & 0.9462 & 0.8889 & 0.9484 & 0.9327 & 0.8755 & 0.9165 & \textbf{0.9648} \\ 
				& Case 6 & 
				0.8807 & 0.9056 & 0.8759 & 0.8878 & 0.7044 & \textbf{0.9187} & 0.7900 & 0.9094 & 0.8432 & 0.8471 & 0.8956 & \underline{0.9101} \\ 
				& Case 7 & 
				0.9526 & \underline{0.9582} & 0.9475 & 0.9435 & 0.7717 & 0.9249 & 0.8862 & 0.9462 & 0.9325 & 0.8655 & 0.9542 & \textbf{0.9592} \\ 
				& Case 8 & 
				0.8757 & 0.8967 & 0.8767 & 0.8687 & 0.6586 & 0.8255 & 0.7891 & \textbf{0.9053} & 0.8268 & 0.8250 & 0.9017 & \underline{0.9021} \\ 

				\bottomrule
				
			\end{tabular}
		}
	\end{center}
\end{table*}

\section{Experiments}
\label{sec:experiments}
\begin{figure*}[t]
	\begin{center}
		\begin{minipage}{0.125\hsize}
			\centerline{\includegraphics[width=\hsize]{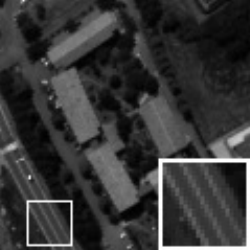}} 
		\end{minipage}
		\begin{minipage}{0.125\hsize}
			\centerline{\includegraphics[width=\hsize]{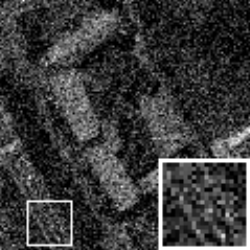}} 
		\end{minipage}
		\begin{minipage}{0.125\hsize}
			\centerline{\includegraphics[width=\hsize]{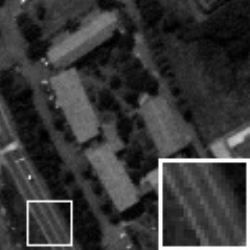}} 
		\end{minipage}
		\begin{minipage}{0.125\hsize}
			\centerline{\includegraphics[width=\hsize]{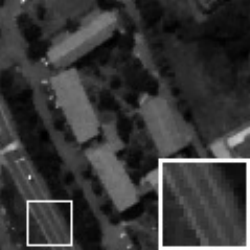}} 
		\end{minipage}
		\begin{minipage}{0.125\hsize}
			\centerline{\includegraphics[width=\hsize]{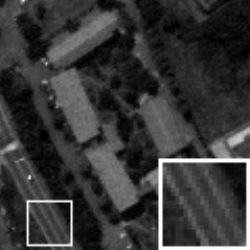}} 
		\end{minipage}
		\begin{minipage}{0.125\hsize}
			\centerline{\includegraphics[width=\hsize]{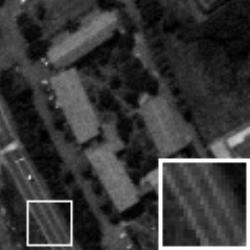}} 
		\end{minipage}
		\begin{minipage}{0.125\hsize}
			\centerline{\includegraphics[width=\hsize]{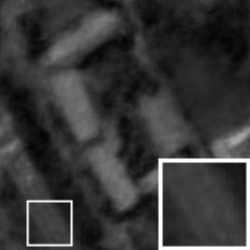}} 
		\end{minipage}
		\begin{minipage}{0.050\hsize}
			\centerline{\hspace{\hsize}} 
		\end{minipage}
		
		\vspace{1mm}
		
		\begin{minipage}{0.125\hsize}
			\centerline{\hspace{\hsize}} 
		\end{minipage}
		\begin{minipage}{0.125\hsize}
			\centerline{\includegraphics[width=\hsize]{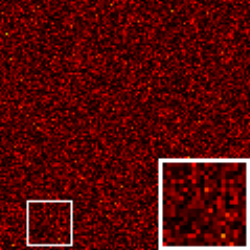}} 
		\end{minipage}
		\begin{minipage}{0.125\hsize}
			\centerline{\includegraphics[width=\hsize]{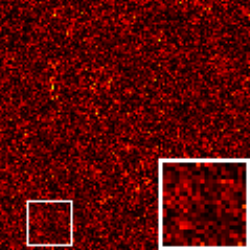}} 
		\end{minipage}
		\begin{minipage}{0.125\hsize}
			\centerline{\includegraphics[width=\hsize]{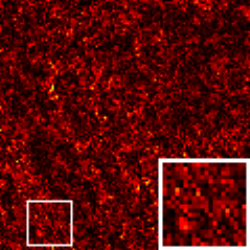}} 
		\end{minipage}
		\begin{minipage}{0.125\hsize}
			\centerline{\includegraphics[width=\hsize]{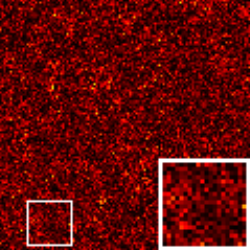}} 
		\end{minipage}
		\begin{minipage}{0.125\hsize}
			\centerline{\includegraphics[width=\hsize]{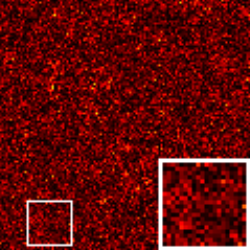}} 
		\end{minipage}
		\begin{minipage}{0.125\hsize}
			\centerline{\includegraphics[width=\hsize]{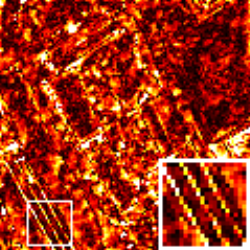}} 
		\end{minipage}
		\begin{minipage}{0.050\hsize}
			\centerline{\includegraphics[width=\hsize]{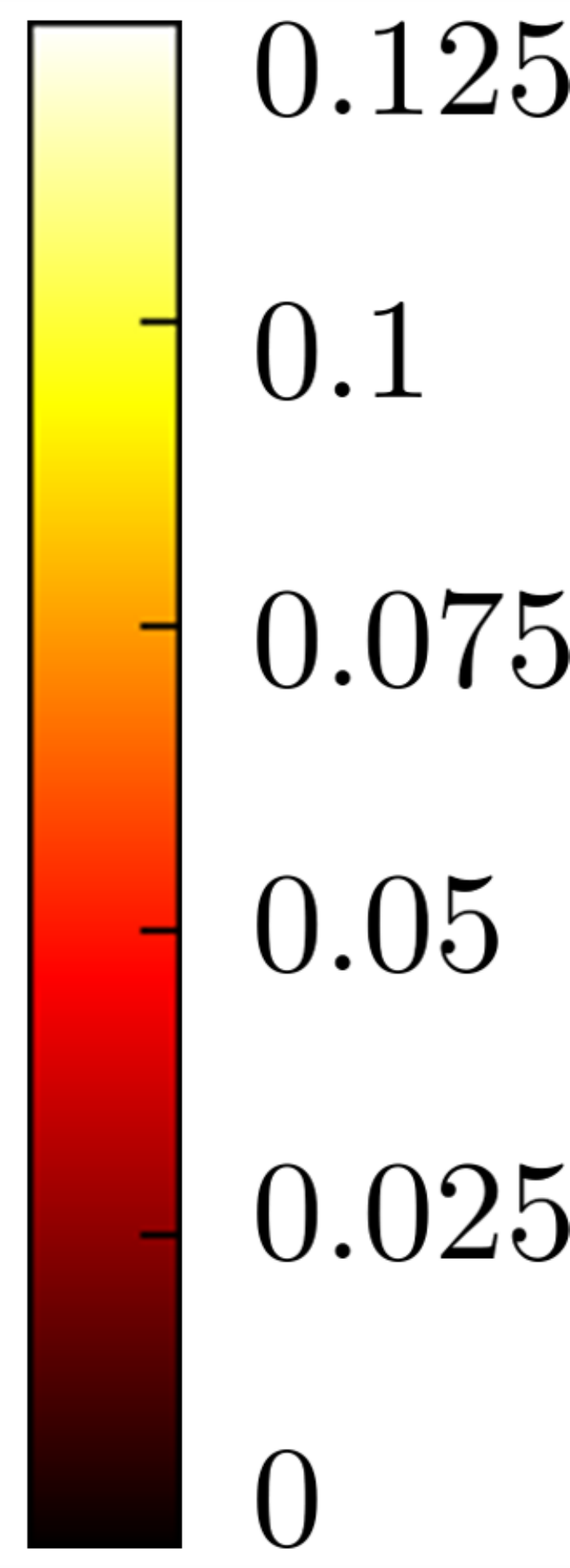}} 
		\end{minipage}
		
		\vspace{1mm}
		
		\begin{minipage}{0.125\hsize}
			\centerline{\small{(a)}}
		\end{minipage}
		\begin{minipage}{0.125\hsize}
			\centerline{\small{(b)}}
		\end{minipage}
		\begin{minipage}{0.125\hsize}
			\centerline{\small{(c)}}
		\end{minipage}
		\begin{minipage}{0.125\hsize}
			\centerline{\small{(d)}}
		\end{minipage}
		\begin{minipage}{0.125\hsize}
			\centerline{\small{(e)}}
		\end{minipage}
		\begin{minipage}{0.125\hsize}
			\centerline{\small{(f)}}
		\end{minipage}
		\begin{minipage}{0.125\hsize}
			\centerline{\small{(g)}}
		\end{minipage}
		\begin{minipage}{0.050\hsize}
			\centerline{\hspace{\hsize}} 
		\end{minipage}
		
		
		\vspace{2mm}
		
		\begin{minipage}{0.125\hsize}
			\centerline{\includegraphics[width=\hsize]{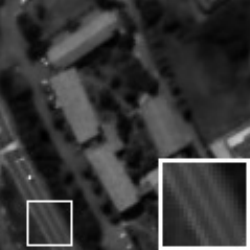}} 
		\end{minipage}
		\begin{minipage}{0.125\hsize}
			\centerline{\includegraphics[width=\hsize]{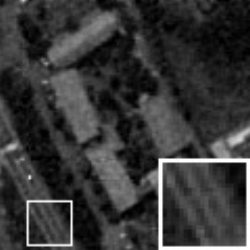}} 
		\end{minipage}
		\begin{minipage}{0.125\hsize}
			\centerline{\includegraphics[width=\hsize]{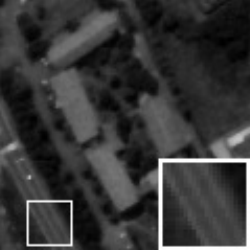}} 
		\end{minipage}
		\begin{minipage}{0.125\hsize}
			\centerline{\includegraphics[width=\hsize]{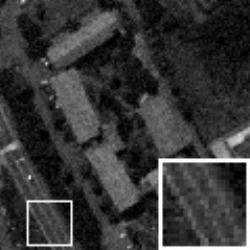}} 
		\end{minipage}
		\begin{minipage}{0.125\hsize}
			\centerline{\includegraphics[width=\hsize]{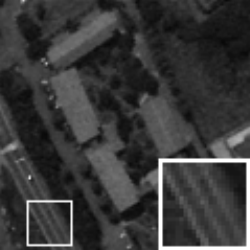}} 
		\end{minipage}
		\begin{minipage}{0.125\hsize}
			\centerline{\includegraphics[width=\hsize]{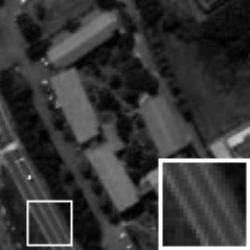}} 
		\end{minipage}
		\begin{minipage}{0.125\hsize}
			\centerline{\includegraphics[width=\hsize]{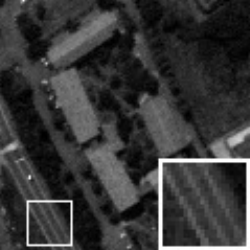}} 
		\end{minipage}
		\begin{minipage}{0.050\hsize}
			\centerline{\hspace{\hsize}} 
		\end{minipage}
		
		\vspace{1mm}
		
		\begin{minipage}{0.125\hsize}
			\centerline{\includegraphics[width=\hsize]{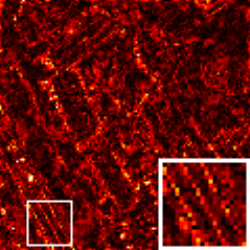}} 
		\end{minipage}
		\begin{minipage}{0.125\hsize}
			\centerline{\includegraphics[width=\hsize]{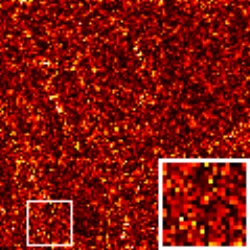}} 
		\end{minipage}
		\begin{minipage}{0.125\hsize}
			\centerline{\includegraphics[width=\hsize]{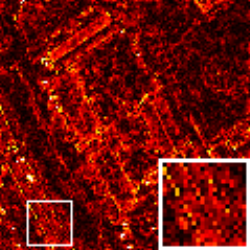}} 
		\end{minipage}
		\begin{minipage}{0.125\hsize}
			\centerline{\includegraphics[width=\hsize]{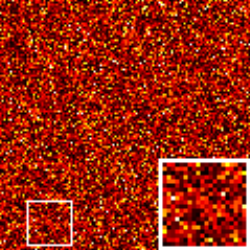}} 
		\end{minipage}
		\begin{minipage}{0.125\hsize}
			\centerline{\includegraphics[width=\hsize]{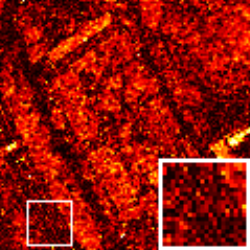}} 
		\end{minipage}
		\begin{minipage}{0.125\hsize}
			\centerline{\includegraphics[width=\hsize]{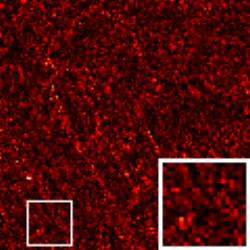}} 
		\end{minipage}
		\begin{minipage}{0.125\hsize}
			\centerline{\includegraphics[width=\hsize]{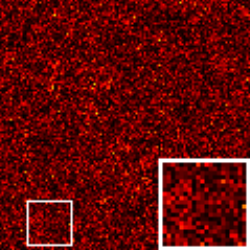}} 
		\end{minipage}
		\begin{minipage}{0.050\hsize}
			\centerline{\includegraphics[width=\hsize]{./result_image/colorbar_hot-eps-converted-to.pdf}} 
		\end{minipage}
		
		\vspace{1mm}
		
		\begin{minipage}{0.125\hsize}
			\centerline{\small{(h)}}
		\end{minipage}
		\begin{minipage}{0.125\hsize}
			\centerline{\small{(i)}}
		\end{minipage}
		\begin{minipage}{0.125\hsize}
			\centerline{\small{(j)}}
		\end{minipage}
		\begin{minipage}{0.125\hsize}
			\centerline{\small{(k)}}
		\end{minipage}
		\begin{minipage}{0.125\hsize}
			\centerline{\small{(l)}}
		\end{minipage}
		\begin{minipage}{0.125\hsize}
			\centerline{\small{(m)}}
		\end{minipage}
		\begin{minipage}{0.125\hsize}
			\centerline{\small{(n)}}
		\end{minipage}
		\begin{minipage}{0.050\hsize}
			\centerline{\hspace{\hsize}} 
		\end{minipage}

	\end{center}
	
	\vspace{-3mm}
	\caption{Denoising results for Pavia University with the 44th band in Case 1. Upper row images are the restored images by each method. Lower row images are the absolute difference between the original image and each restored image (the range of the only observed noisy image in $\IndexNoisy$ is $[0,1]$). (a) Ground-truth. (b) Observed noisy image. (c) SSTV. (d) HSSTV1. (e) HSSTV2. (f) $\llHTV$. (g) STV. (h) SSST. (i) LRTDTV. (j) FGSLR. (k) TPTV. (l) QRNN3D. (m) FastHyMix. (n) $\SSSTTV$ (ours).}
	\label{fig:result_image_Case1_PU}
\end{figure*}
To demonstrate the effectiveness of $\SSSTTV$, we conducted mixed noise removal experiments on HS image contaminated with simulated or real noise.
We compared $\SSSTTV$ with four types of methods; SSTV-based methods, i.e., SSTV~\cite{Aggarwal2016SSTV}, HSSTV~\cite{Takeyama2020HSSTV}, and $\llHTV$~\cite{Wang2021l0l1HTV}; STV-based methods, i.e., STV~\cite{Lefkimmiatis2015STV} and SSST~\cite{Kurihara2017SSST}; LR-based methods, i.e., LRTDTV~\cite{Wang2018LRTDTV}, FGSLR~\cite{Chen2022FGSLR}, and TPTV~\cite{Chen2023TPTV}; and DNN-based methods, i.e., QRNN3D~\cite{Wei2021QRNN3D} and FastHyMix~\cite{Zhuang2023FastHyMix}. 
Here, HSSTV with $\ell_{1}$-norm and $\ell_{1,2}$-norm are denoted by HSSTV1 and HSSTV2, respectively.
For a fair comparison, the regularization functions of the P-PDS applicable methods, i.e., SSTV, HSSTV1, HSSTV2, $\llHTV$, STV, and SSST, were replaced with the $\SSSTTV$ regularization function in Prob.~\eqref{prob:S3TTV_denoising}, and we solve each problem by P-PDS.
For LRTDTV, FGSLR, TPTV, QRNN3D, and FastHyMix, we used implementation codes published by the authors\footnote{The LRTDTV, FGSLR, TPTV, QRNN3D, and FastHyMix implementation codes are available at 
\\ https://github.com/zhaoxile/Hyperspectral-Image-Restoration-via-Total-Variation-Regularized-Low-rank-Tensor-Decomposition, \\https://chenyong1993.github.io/yongchen.github.io/, \\https://github.com/chuchulyf/ETPTV, \\https://github.com/Vandermode/QRNN3D?tab=readme-ov-file, \\and https://github.com/LinaZhuang/HSI-MixedNoiseRemoval-FastHyMix, respectively.}. For QRNN3D, we performed fine-tuning using Pavia Centre\footnote{\url{https://www.ehu/ccwintco/index/php/Hyperspectral_Remote_Sensing_Scenes}}to improve noise removal performance.

\subsection{Simulated HS Image Experiments}
\begin{figure*}[t]
	\begin{center}
		\begin{minipage}{0.125\hsize}
			\centerline{\includegraphics[width=\hsize]{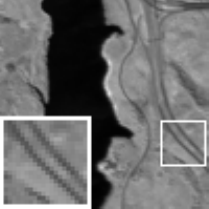}} 
		\end{minipage}
		\begin{minipage}{0.125\hsize}
			\centerline{\includegraphics[width=\hsize]{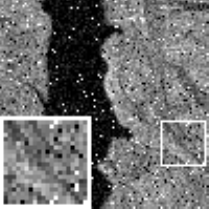}} 
		\end{minipage}
		\begin{minipage}{0.125\hsize}
			\centerline{\includegraphics[width=\hsize]{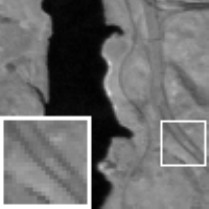}} 
		\end{minipage}
		\begin{minipage}{0.125\hsize}
			\centerline{\includegraphics[width=\hsize]{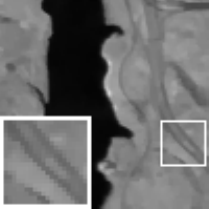}} 
		\end{minipage}
		\begin{minipage}{0.125\hsize}
			\centerline{\includegraphics[width=\hsize]{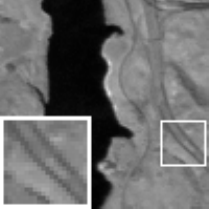}} 
		\end{minipage}
		\begin{minipage}{0.125\hsize}
			\centerline{\includegraphics[width=\hsize]{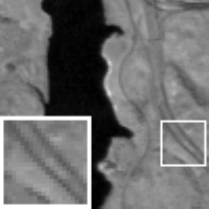}} 
		\end{minipage}
		\begin{minipage}{0.125\hsize}
			\centerline{\includegraphics[width=\hsize]{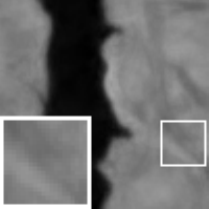}} 
		\end{minipage}
		\begin{minipage}{0.050\hsize}
			\centerline{\hspace{\hsize}} 
		\end{minipage}
		
		\vspace{1mm}
		
		\begin{minipage}{0.125\hsize}
			\centerline{\hspace{\hsize}} 
		\end{minipage}
		\begin{minipage}{0.125\hsize}
			\centerline{\includegraphics[width=\hsize]{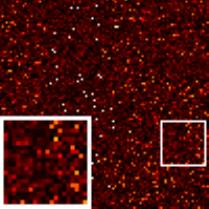}} 
		\end{minipage}
		\begin{minipage}{0.125\hsize}
			\centerline{\includegraphics[width=\hsize]{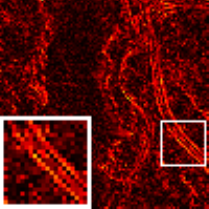}} 
		\end{minipage}
		\begin{minipage}{0.125\hsize}
			\centerline{\includegraphics[width=\hsize]{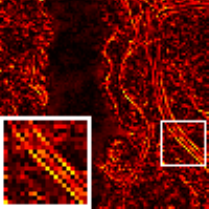}} 
		\end{minipage}
		\begin{minipage}{0.125\hsize}
			\centerline{\includegraphics[width=\hsize]{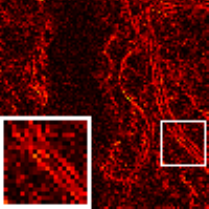}} 
		\end{minipage}
		\begin{minipage}{0.125\hsize}
			\centerline{\includegraphics[width=\hsize]{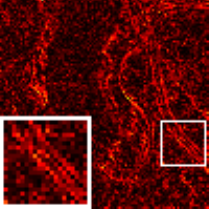}} 
		\end{minipage}
		\begin{minipage}{0.125\hsize}
			\centerline{\includegraphics[width=\hsize]{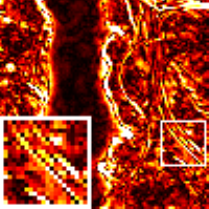}} 
		\end{minipage}
		\begin{minipage}{0.050\hsize}
			\centerline{\includegraphics[width=\hsize]{./result_image/colorbar_hot-eps-converted-to.pdf}} 
		\end{minipage}
		
		\vspace{1mm}
		
		\begin{minipage}{0.125\hsize}
			\centerline{\small{(a)}}
		\end{minipage}
		\begin{minipage}{0.125\hsize}
			\centerline{\small{(b)}}
		\end{minipage}
		\begin{minipage}{0.125\hsize}
			\centerline{\small{(c)}}
		\end{minipage}
		\begin{minipage}{0.125\hsize}
			\centerline{\small{(d)}}
		\end{minipage}
		\begin{minipage}{0.125\hsize}
			\centerline{\small{(e)}}
		\end{minipage}
		\begin{minipage}{0.125\hsize}
			\centerline{\small{(f)}}
		\end{minipage}
		\begin{minipage}{0.125\hsize}
			\centerline{\small{(g)}}
		\end{minipage}
		\begin{minipage}{0.050\hsize}
			\centerline{\hspace{\hsize}} 
		\end{minipage}
		
		
		\vspace{2mm}
		
		\begin{minipage}{0.125\hsize}
			\centerline{\includegraphics[width=\hsize]{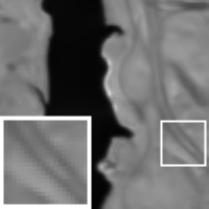}} 
		\end{minipage}
		\begin{minipage}{0.125\hsize}
			\centerline{\includegraphics[width=\hsize]{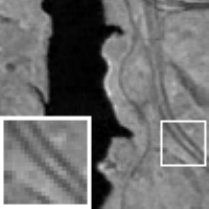}} 
		\end{minipage}
		\begin{minipage}{0.125\hsize}
			\centerline{\includegraphics[width=\hsize]{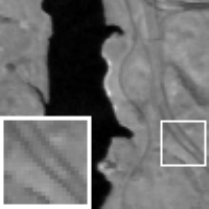}} 
		\end{minipage}
		\begin{minipage}{0.125\hsize}
			\centerline{\includegraphics[width=\hsize]{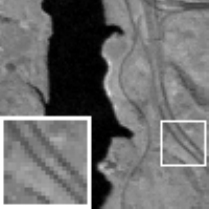}} 
		\end{minipage}
		\begin{minipage}{0.125\hsize}
			\centerline{\includegraphics[width=\hsize]{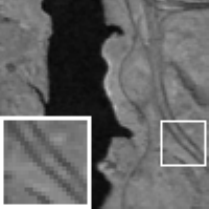}} 
		\end{minipage}
		\begin{minipage}{0.125\hsize}
			\centerline{\includegraphics[width=\hsize]{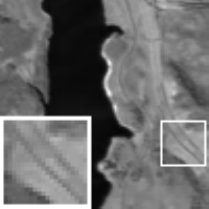}} 
		\end{minipage}
		\begin{minipage}{0.125\hsize}
			\centerline{\includegraphics[width=\hsize]{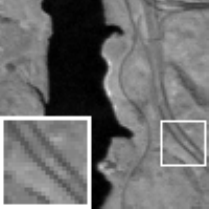}} 
		\end{minipage}
		\begin{minipage}{0.050\hsize}
			\centerline{\hspace{\hsize}} 
		\end{minipage}
		
		\vspace{1mm}
		
		\begin{minipage}{0.125\hsize}
			\centerline{\includegraphics[width=\hsize]{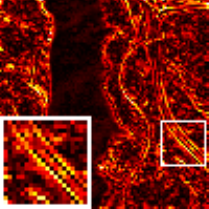}} 
		\end{minipage}
		\begin{minipage}{0.125\hsize}
			\centerline{\includegraphics[width=\hsize]{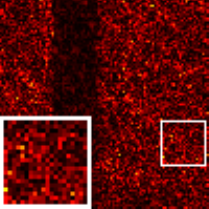}} 
		\end{minipage}
		\begin{minipage}{0.125\hsize}
			\centerline{\includegraphics[width=\hsize]{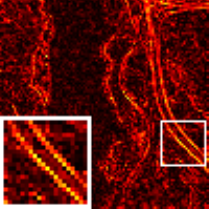}} 
		\end{minipage}
		\begin{minipage}{0.125\hsize}
			\centerline{\includegraphics[width=\hsize]{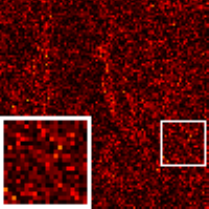}} 
		\end{minipage}
		\begin{minipage}{0.125\hsize}
			\centerline{\includegraphics[width=\hsize]{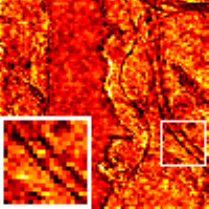}} 
		\end{minipage}
		\begin{minipage}{0.125\hsize}
			\centerline{\includegraphics[width=\hsize]{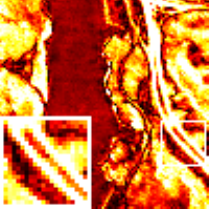}} 
		\end{minipage}
		\begin{minipage}{0.125\hsize}
			\centerline{\includegraphics[width=\hsize]{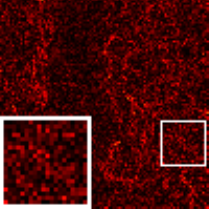}} 
		\end{minipage}
		\begin{minipage}{0.050\hsize}
			\centerline{\includegraphics[width=\hsize]{./result_image/colorbar_hot-eps-converted-to.pdf}} 
		\end{minipage}
		
		\vspace{1mm}
		
		\begin{minipage}{0.125\hsize}
			\centerline{\small{(h)}}
		\end{minipage}
		\begin{minipage}{0.125\hsize}
			\centerline{\small{(i)}}
		\end{minipage}
		\begin{minipage}{0.125\hsize}
			\centerline{\small{(j)}}
		\end{minipage}
		\begin{minipage}{0.125\hsize}
			\centerline{\small{(k)}}
		\end{minipage}
		\begin{minipage}{0.125\hsize}
			\centerline{\small{(l)}}
		\end{minipage}
		\begin{minipage}{0.125\hsize}
			\centerline{\small{(m)}}
		\end{minipage}
		\begin{minipage}{0.125\hsize}
			\centerline{\small{(n)}}
		\end{minipage}
		\begin{minipage}{0.050\hsize}
			\centerline{\hspace{\hsize}} 
		\end{minipage}

	\end{center}
	
	\vspace{-3mm}
	\caption{Denoising results for Jasper Ridge with the 91st band in Case 3. Upper row images are the restored images by each method. Lower row images are the absolute difference between the original image and each restored image (the range of the only observed noisy image in $\IndexNoisy$ is $[0,1]$). (a) Ground-truth. (b) Observed noisy image. (c) SSTV. (d) HSSTV1. (e) HSSTV2. (f) $\llHTV$. (g) STV. (h) SSST. (i) LRTDTV. (j) FGSLR. (k) TPTV. (l) QRNN3D. (m) FastHyMix. (n) $\SSSTTV$ (ours).}
	\label{fig:result_image_Case3_JR}
\end{figure*}
\begin{figure*}[t]
	\begin{center}
		\begin{minipage}{0.125\hsize}
			\centerline{\includegraphics[width=\hsize]{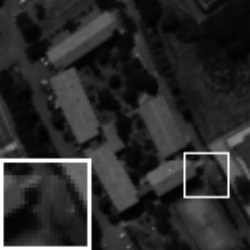}} 
		\end{minipage}
		\begin{minipage}{0.125\hsize}
			\centerline{\includegraphics[width=\hsize]{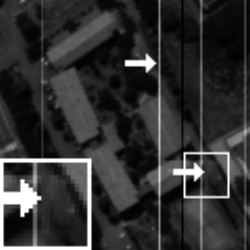}} 
		\end{minipage}
		\begin{minipage}{0.125\hsize}
			\centerline{\includegraphics[width=\hsize]{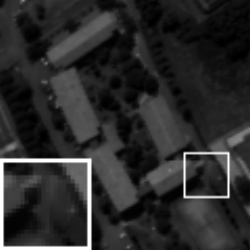}} 
		\end{minipage}
		\begin{minipage}{0.125\hsize}
			\centerline{\includegraphics[width=\hsize]{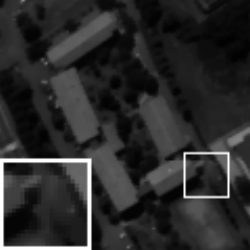}} 
		\end{minipage}
		\begin{minipage}{0.125\hsize}
			\centerline{\includegraphics[width=\hsize]{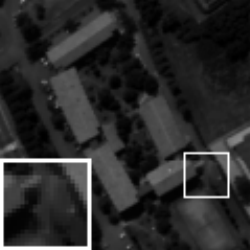}} 
		\end{minipage}
		\begin{minipage}{0.125\hsize}
			\centerline{\includegraphics[width=\hsize]{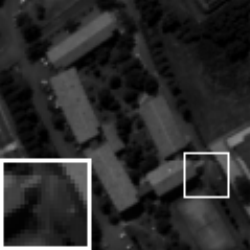}} 
		\end{minipage}
		\begin{minipage}{0.125\hsize}
			\centerline{\includegraphics[width=\hsize]{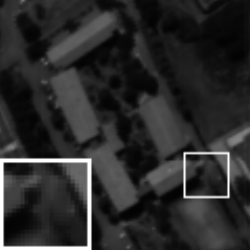}} 
		\end{minipage}
		\begin{minipage}{0.050\hsize}
			\centerline{\hspace{\hsize}} 
		\end{minipage}
		
		\vspace{1mm}
		
		\begin{minipage}{0.125\hsize}
			\centerline{\hspace{\hsize}} 
		\end{minipage}
		\begin{minipage}{0.125\hsize}
			\centerline{\includegraphics[width=\hsize]{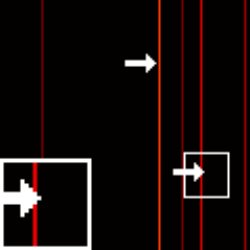}} 
		\end{minipage}
		\begin{minipage}{0.125\hsize}
			\centerline{\includegraphics[width=\hsize]{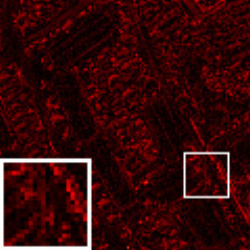}} 
		\end{minipage}
		\begin{minipage}{0.125\hsize}
			\centerline{\includegraphics[width=\hsize]{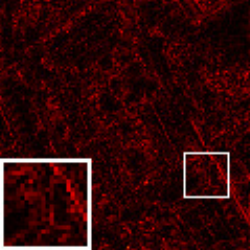}} 
		\end{minipage}
		\begin{minipage}{0.125\hsize}
			\centerline{\includegraphics[width=\hsize]{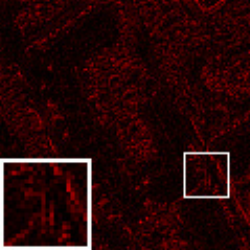}} 
		\end{minipage}
		\begin{minipage}{0.125\hsize}
			\centerline{\includegraphics[width=\hsize]{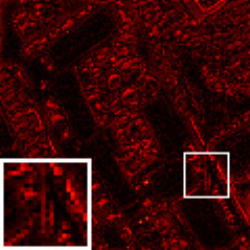}} 
		\end{minipage}
		\begin{minipage}{0.125\hsize}
			\centerline{\includegraphics[width=\hsize]{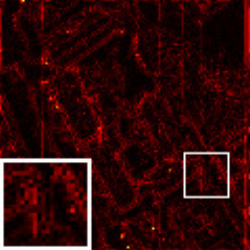}} 
		\end{minipage}
		\begin{minipage}{0.050\hsize}
			\centerline{\includegraphics[width=\hsize]{./result_image/colorbar_hot-eps-converted-to.pdf}} 
		\end{minipage}
		
		\vspace{1mm}
		
		\begin{minipage}{0.125\hsize}
			\centerline{\small{(a)}}
		\end{minipage}
		\begin{minipage}{0.125\hsize}
			\centerline{\small{(b)}}
		\end{minipage}
		\begin{minipage}{0.125\hsize}
			\centerline{\small{(c)}}
		\end{minipage}
		\begin{minipage}{0.125\hsize}
			\centerline{\small{(d)}}
		\end{minipage}
		\begin{minipage}{0.125\hsize}
			\centerline{\small{(e)}}
		\end{minipage}
		\begin{minipage}{0.125\hsize}
			\centerline{\small{(f)}}
		\end{minipage}
		\begin{minipage}{0.125\hsize}
			\centerline{\small{(g)}}
		\end{minipage}
		\begin{minipage}{0.050\hsize}
			\centerline{\hspace{\hsize}} 
		\end{minipage}
		
		
		\vspace{2mm}
		
		\begin{minipage}{0.125\hsize}
			\centerline{\includegraphics[width=\hsize]{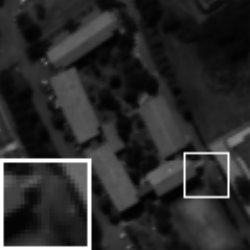}} 
		\end{minipage}
		\begin{minipage}{0.125\hsize}
			\centerline{\includegraphics[width=\hsize]{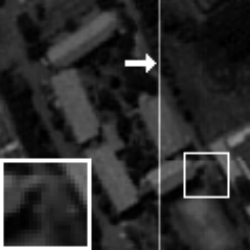}} 
		\end{minipage}
		\begin{minipage}{0.125\hsize}
			\centerline{\includegraphics[width=\hsize]{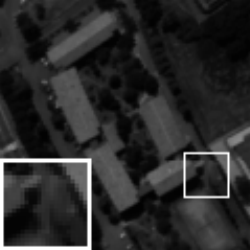}} 
		\end{minipage}
		\begin{minipage}{0.125\hsize}
			\centerline{\includegraphics[width=\hsize]{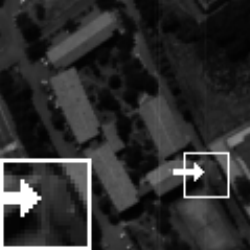}} 
		\end{minipage}
		\begin{minipage}{0.125\hsize}
			\centerline{\includegraphics[width=\hsize]{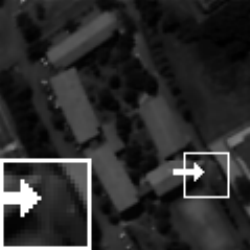}} 
		\end{minipage}
		\begin{minipage}{0.125\hsize}
			\centerline{\includegraphics[width=\hsize]{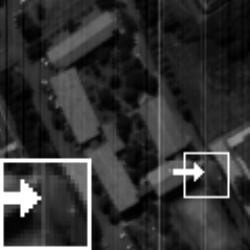}} 
		\end{minipage}
		\begin{minipage}{0.125\hsize}
			\centerline{\includegraphics[width=\hsize]{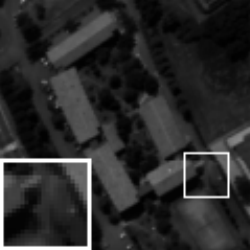}} 
		\end{minipage}
		\begin{minipage}{0.050\hsize}
			\centerline{\hspace{\hsize}} 
		\end{minipage}
		
		\vspace{1mm}
		
		\begin{minipage}{0.125\hsize}
			\centerline{\includegraphics[width=\hsize]{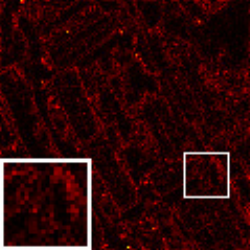}} 
		\end{minipage}
		\begin{minipage}{0.125\hsize}
			\centerline{\includegraphics[width=\hsize]{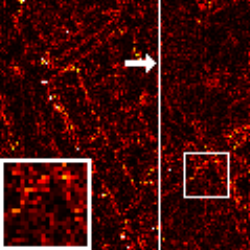}} 
		\end{minipage}
		\begin{minipage}{0.125\hsize}
			\centerline{\includegraphics[width=\hsize]{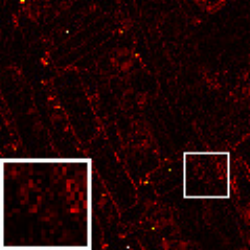}} 
		\end{minipage}
		\begin{minipage}{0.125\hsize}
			\centerline{\includegraphics[width=\hsize]{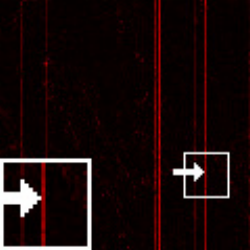}} 
		\end{minipage}
		\begin{minipage}{0.125\hsize}
			\centerline{\includegraphics[width=\hsize]{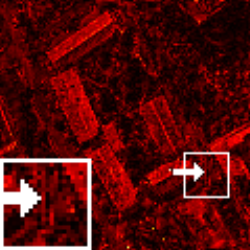}} 
		\end{minipage}
		\begin{minipage}{0.125\hsize}
			\centerline{\includegraphics[width=\hsize]{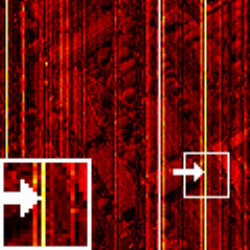}} 
		\end{minipage}
		\begin{minipage}{0.125\hsize}
			\centerline{\includegraphics[width=\hsize]{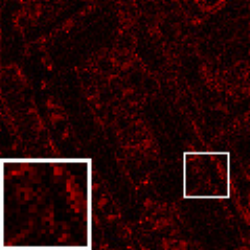}} 
		\end{minipage}
		\begin{minipage}{0.050\hsize}
			\centerline{\includegraphics[width=\hsize]{./result_image/colorbar_hot-eps-converted-to.pdf}} 
		\end{minipage}
		
		\vspace{1mm}
		
		\begin{minipage}{0.125\hsize}
			\centerline{\small{(h)}}
		\end{minipage}
		\begin{minipage}{0.125\hsize}
			\centerline{\small{(i)}}
		\end{minipage}
		\begin{minipage}{0.125\hsize}
			\centerline{\small{(j)}}
		\end{minipage}
		\begin{minipage}{0.125\hsize}
			\centerline{\small{(k)}}
		\end{minipage}
		\begin{minipage}{0.125\hsize}
			\centerline{\small{(l)}}
		\end{minipage}
		\begin{minipage}{0.125\hsize}
			\centerline{\small{(m)}}
		\end{minipage}
		\begin{minipage}{0.125\hsize}
			\centerline{\small{(n)}}
		\end{minipage}
		\begin{minipage}{0.050\hsize}
			\centerline{\hspace{\hsize}} 
		\end{minipage}

	\end{center}
	
	\vspace{-3mm}
	\caption{Denoising results for Pavia University with the 61st band in Case 4. Upper row images are the restored images by each method. Lower row images are the absolute difference between the original image and each restored image (the range of the only observed noisy image in $\IndexNoisy$ is $[0,1]$). (a) Ground-truth. (b) Observed noisy image. (c) SSTV. (d) HSSTV1. (e) HSSTV2. (f) $\llHTV$. (g) STV. (h) SSST. (i) LRTDTV. (j) FGSLR. (k) TPTV. (l) QRNN3D. (m) FastHyMix. (n) $\SSSTTV$ (ours).}
	\label{fig:result_image_Case4_PU}
\end{figure*}
\begin{figure*}[t]
	\begin{center}
		\begin{minipage}{0.125\hsize}
			\centerline{\includegraphics[width=\hsize]{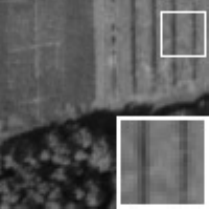}} 
		\end{minipage}
		\begin{minipage}{0.125\hsize}
			\centerline{\includegraphics[width=\hsize]{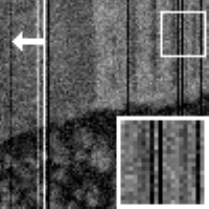}} 
		\end{minipage}
		\begin{minipage}{0.125\hsize}
			\centerline{\includegraphics[width=\hsize]{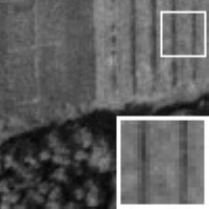}} 
		\end{minipage}
		\begin{minipage}{0.125\hsize}
			\centerline{\includegraphics[width=\hsize]{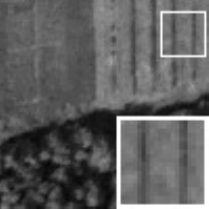}} 
		\end{minipage}
		\begin{minipage}{0.125\hsize}
			\centerline{\includegraphics[width=\hsize]{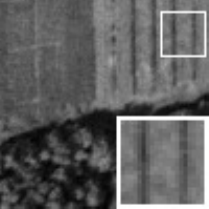}} 
		\end{minipage}
		\begin{minipage}{0.125\hsize}
			\centerline{\includegraphics[width=\hsize]{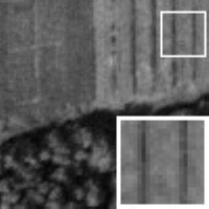}} 
		\end{minipage}
		\begin{minipage}{0.125\hsize}
			\centerline{\includegraphics[width=\hsize]{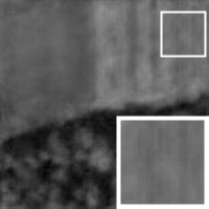}} 
		\end{minipage}
		\begin{minipage}{0.050\hsize}
			\centerline{\hspace{\hsize}} 
		\end{minipage}
		
		\vspace{1mm}
		
		\begin{minipage}{0.125\hsize}
			\centerline{\hspace{\hsize}} 
		\end{minipage}
		\begin{minipage}{0.125\hsize}
			\centerline{\includegraphics[width=\hsize]{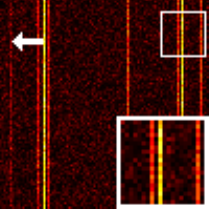}} 
		\end{minipage}
		\begin{minipage}{0.125\hsize}
			\centerline{\includegraphics[width=\hsize]{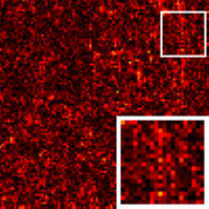}} 
		\end{minipage}
		\begin{minipage}{0.125\hsize}
			\centerline{\includegraphics[width=\hsize]{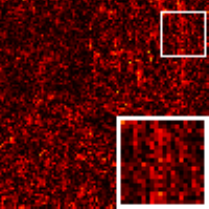}} 
		\end{minipage}
		\begin{minipage}{0.125\hsize}
			\centerline{\includegraphics[width=\hsize]{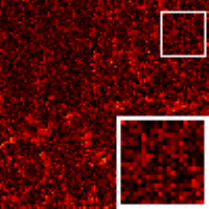}} 
		\end{minipage}
		\begin{minipage}{0.125\hsize}
			\centerline{\includegraphics[width=\hsize]{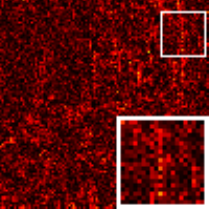}} 
		\end{minipage}
		\begin{minipage}{0.125\hsize}
			\centerline{\includegraphics[width=\hsize]{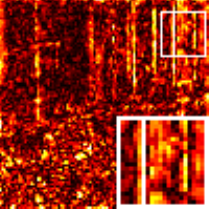}} 
		\end{minipage}
		\begin{minipage}{0.050\hsize}
			\centerline{\includegraphics[width=\hsize]{./result_image/colorbar_hot-eps-converted-to.pdf}} 
		\end{minipage}
		
		\vspace{1mm}
		
		\begin{minipage}{0.125\hsize}
			\centerline{\small{(a)}}
		\end{minipage}
		\begin{minipage}{0.125\hsize}
			\centerline{\small{(b)}}
		\end{minipage}
		\begin{minipage}{0.125\hsize}
			\centerline{\small{(c)}}
		\end{minipage}
		\begin{minipage}{0.125\hsize}
			\centerline{\small{(d)}}
		\end{minipage}
		\begin{minipage}{0.125\hsize}
			\centerline{\small{(e)}}
		\end{minipage}
		\begin{minipage}{0.125\hsize}
			\centerline{\small{(f)}}
		\end{minipage}
		\begin{minipage}{0.125\hsize}
			\centerline{\small{(g)}}
		\end{minipage}
		\begin{minipage}{0.050\hsize}
			\centerline{\hspace{\hsize}} 
		\end{minipage}
		
		
		\vspace{2mm}
		
		\begin{minipage}{0.125\hsize}
			\centerline{\includegraphics[width=\hsize]{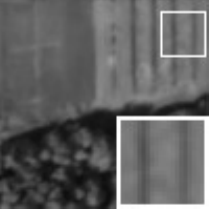}} 
		\end{minipage}
		\begin{minipage}{0.125\hsize}
			\centerline{\includegraphics[width=\hsize]{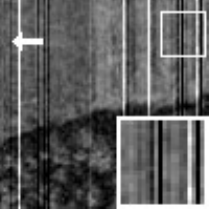}} 
		\end{minipage}
		\begin{minipage}{0.125\hsize}
			\centerline{\includegraphics[width=\hsize]{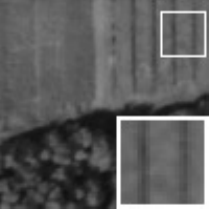}} 
		\end{minipage}
		\begin{minipage}{0.125\hsize}
			\centerline{\includegraphics[width=\hsize]{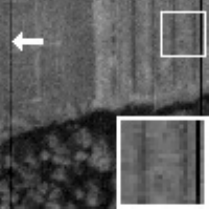}} 
		\end{minipage}
		\begin{minipage}{0.125\hsize}
			\centerline{\includegraphics[width=\hsize]{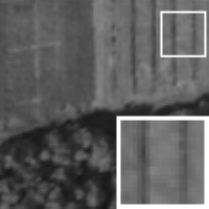}} 
		\end{minipage}
		\begin{minipage}{0.125\hsize}
			\centerline{\includegraphics[width=\hsize]{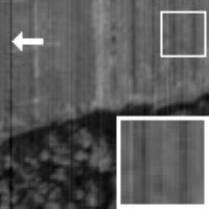}} 
		\end{minipage}
		\begin{minipage}{0.125\hsize}
			\centerline{\includegraphics[width=\hsize]{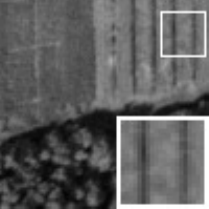}} 
		\end{minipage}
		\begin{minipage}{0.050\hsize}
			\centerline{\hspace{\hsize}} 
		\end{minipage}
		
		\vspace{1mm}
		
		\begin{minipage}{0.125\hsize}
			\centerline{\includegraphics[width=\hsize]{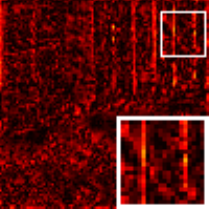}} 
		\end{minipage}
		\begin{minipage}{0.125\hsize}
			\centerline{\includegraphics[width=\hsize]{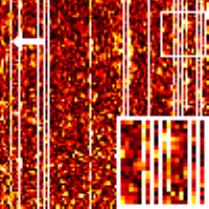}} 
		\end{minipage}
		\begin{minipage}{0.125\hsize}
			\centerline{\includegraphics[width=\hsize]{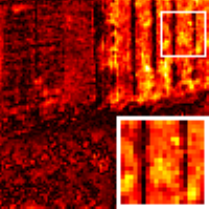}} 
		\end{minipage}
		\begin{minipage}{0.125\hsize}
			\centerline{\includegraphics[width=\hsize]{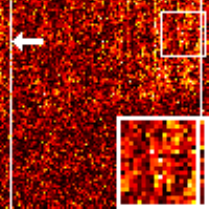}} 
		\end{minipage}
		\begin{minipage}{0.125\hsize}
			\centerline{\includegraphics[width=\hsize]{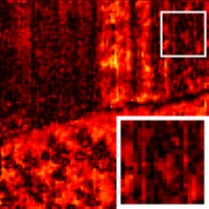}} 
		\end{minipage}
		\begin{minipage}{0.125\hsize}
			\centerline{\includegraphics[width=\hsize]{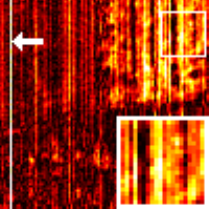}} 
		\end{minipage}
		\begin{minipage}{0.125\hsize}
			\centerline{\includegraphics[width=\hsize]{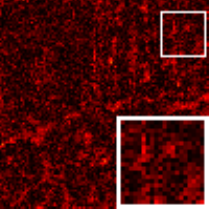}} 
		\end{minipage}
		\begin{minipage}{0.050\hsize}
			\centerline{\includegraphics[width=\hsize]{./result_image/colorbar_hot-eps-converted-to.pdf}} 
		\end{minipage}
		
		\vspace{1mm}
		
		\begin{minipage}{0.125\hsize}
			\centerline{\small{(h)}}
		\end{minipage}
		\begin{minipage}{0.125\hsize}
			\centerline{\small{(i)}}
		\end{minipage}
		\begin{minipage}{0.125\hsize}
			\centerline{\small{(j)}}
		\end{minipage}
		\begin{minipage}{0.125\hsize}
			\centerline{\small{(k)}}
		\end{minipage}
		\begin{minipage}{0.125\hsize}
			\centerline{\small{(l)}}
		\end{minipage}
		\begin{minipage}{0.125\hsize}
			\centerline{\small{(m)}}
		\end{minipage}
		\begin{minipage}{0.125\hsize}
			\centerline{\small{(n)}}
		\end{minipage}
		\begin{minipage}{0.050\hsize}
			\centerline{\hspace{\hsize}} 
		\end{minipage}

	\end{center}
	
	\vspace{-3mm}
	\caption{Denoising results for Beltsville with the 44th band in Case 5. Upper row images are the restored images by each method. Lower row images are the absolute difference between the original image and each restored image (the range of the only observed noisy image in $\IndexNoisy$ is $[0,1]$). (a) Ground-truth. (b) Observed noisy image. (c) SSTV. (d) HSSTV1. (e) HSSTV2. (f) $\llHTV$. (g) STV. (h) SSST. (i) LRTDTV. (j) FGSLR. (k) TPTV. (l) QRNN3D. (m) FastHyMix. (n) $\SSSTTV$ (ours).}
	\label{fig:result_image_Case5_BV}
\end{figure*}
\begin{figure*}[t]
	\begin{center}
		\begin{minipage}{0.125\hsize}
			\centerline{\includegraphics[width=\hsize]{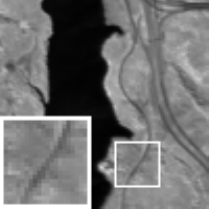}} 
		\end{minipage}
		\begin{minipage}{0.125\hsize}
			\centerline{\includegraphics[width=\hsize]{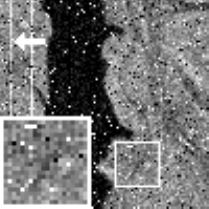}} 
		\end{minipage}
		\begin{minipage}{0.125\hsize}
			\centerline{\includegraphics[width=\hsize]{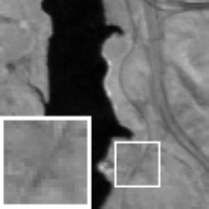}} 
		\end{minipage}
		\begin{minipage}{0.125\hsize}
			\centerline{\includegraphics[width=\hsize]{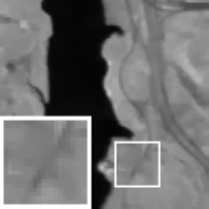}} 
		\end{minipage}
		\begin{minipage}{0.125\hsize}
			\centerline{\includegraphics[width=\hsize]{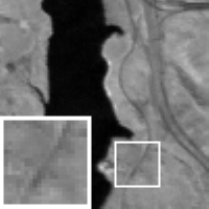}} 
		\end{minipage}
		\begin{minipage}{0.125\hsize}
			\centerline{\includegraphics[width=\hsize]{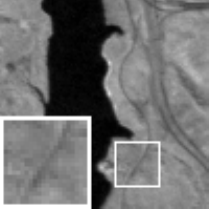}} 
		\end{minipage}
		\begin{minipage}{0.125\hsize}
			\centerline{\includegraphics[width=\hsize]{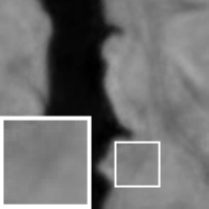}} 
		\end{minipage}
		\begin{minipage}{0.050\hsize}
			\centerline{\hspace{\hsize}} 
		\end{minipage}
		
		\vspace{1mm}
		
		\begin{minipage}{0.125\hsize}
			\centerline{\hspace{\hsize}} 
		\end{minipage}
		\begin{minipage}{0.125\hsize}
			\centerline{\includegraphics[width=\hsize]{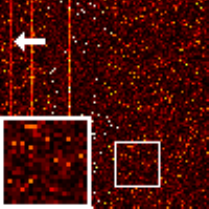}} 
		\end{minipage}
		\begin{minipage}{0.125\hsize}
			\centerline{\includegraphics[width=\hsize]{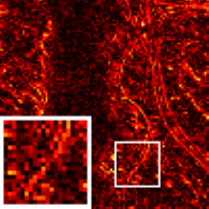}} 
		\end{minipage}
		\begin{minipage}{0.125\hsize}
			\centerline{\includegraphics[width=\hsize]{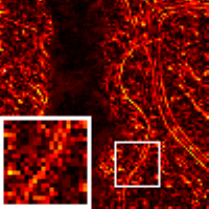}} 
		\end{minipage}
		\begin{minipage}{0.125\hsize}
			\centerline{\includegraphics[width=\hsize]{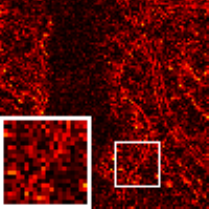}} 
		\end{minipage}
		\begin{minipage}{0.125\hsize}
			\centerline{\includegraphics[width=\hsize]{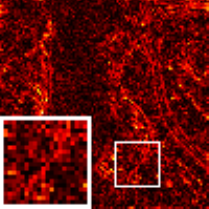}} 
		\end{minipage}
		\begin{minipage}{0.125\hsize}
			\centerline{\includegraphics[width=\hsize]{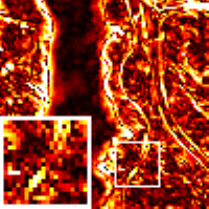}} 
		\end{minipage}
		\begin{minipage}{0.050\hsize}
			\centerline{\includegraphics[width=\hsize]{./result_image/colorbar_hot-eps-converted-to.pdf}} 
		\end{minipage}
		
		\vspace{1mm}
		
		\begin{minipage}{0.125\hsize}
			\centerline{\small{(a)}}
		\end{minipage}
		\begin{minipage}{0.125\hsize}
			\centerline{\small{(b)}}
		\end{minipage}
		\begin{minipage}{0.125\hsize}
			\centerline{\small{(c)}}
		\end{minipage}
		\begin{minipage}{0.125\hsize}
			\centerline{\small{(d)}}
		\end{minipage}
		\begin{minipage}{0.125\hsize}
			\centerline{\small{(e)}}
		\end{minipage}
		\begin{minipage}{0.125\hsize}
			\centerline{\small{(f)}}
		\end{minipage}
		\begin{minipage}{0.125\hsize}
			\centerline{\small{(g)}}
		\end{minipage}
		\begin{minipage}{0.050\hsize}
			\centerline{\hspace{\hsize}} 
		\end{minipage}
		
		
		\vspace{2mm}
		
		\begin{minipage}{0.125\hsize}
			\centerline{\includegraphics[width=\hsize]{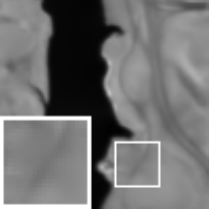}} 
		\end{minipage}
		\begin{minipage}{0.125\hsize}
			\centerline{\includegraphics[width=\hsize]{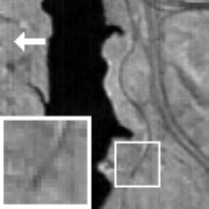}} 
		\end{minipage}
		\begin{minipage}{0.125\hsize}
			\centerline{\includegraphics[width=\hsize]{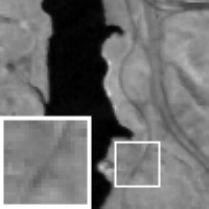}} 
		\end{minipage}
		\begin{minipage}{0.125\hsize}
			\centerline{\includegraphics[width=\hsize]{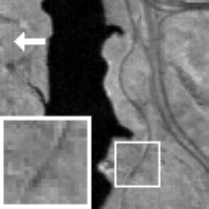}} 
		\end{minipage}
		\begin{minipage}{0.125\hsize}
			\centerline{\includegraphics[width=\hsize]{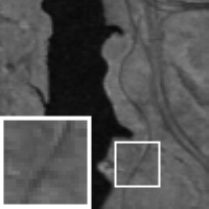}} 
		\end{minipage}
		\begin{minipage}{0.125\hsize}
			\centerline{\includegraphics[width=\hsize]{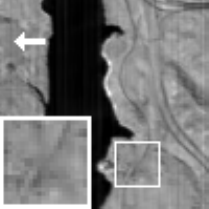}} 
		\end{minipage}
		\begin{minipage}{0.125\hsize}
			\centerline{\includegraphics[width=\hsize]{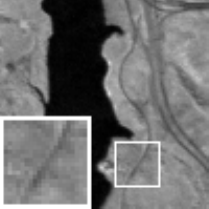}} 
		\end{minipage}
		\begin{minipage}{0.050\hsize}
			\centerline{\hspace{\hsize}} 
		\end{minipage}
		
		\vspace{1mm}
		
		\begin{minipage}{0.125\hsize}
			\centerline{\includegraphics[width=\hsize]{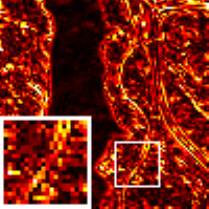}} 
		\end{minipage}
		\begin{minipage}{0.125\hsize}
			\centerline{\includegraphics[width=\hsize]{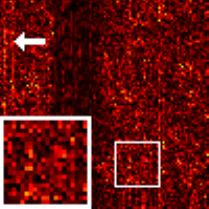}} 
		\end{minipage}
		\begin{minipage}{0.125\hsize}
			\centerline{\includegraphics[width=\hsize]{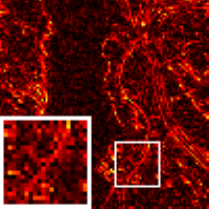}} 
		\end{minipage}
		\begin{minipage}{0.125\hsize}
			\centerline{\includegraphics[width=\hsize]{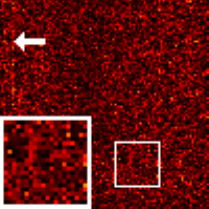}} 
		\end{minipage}
		\begin{minipage}{0.125\hsize}
			\centerline{\includegraphics[width=\hsize]{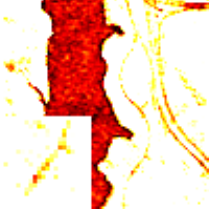}} 
		\end{minipage}
		\begin{minipage}{0.125\hsize}
			\centerline{\includegraphics[width=\hsize]{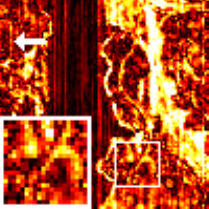}} 
		\end{minipage}
		\begin{minipage}{0.125\hsize}
			\centerline{\includegraphics[width=\hsize]{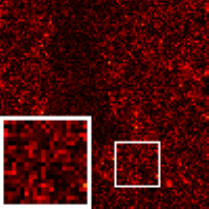}} 
		\end{minipage}
		\begin{minipage}{0.050\hsize}
			\centerline{\includegraphics[width=\hsize]{./result_image/colorbar_hot-eps-converted-to.pdf}} 
		\end{minipage}
		
		\vspace{1mm}
		
		\begin{minipage}{0.125\hsize}
			\centerline{\small{(h)}}
		\end{minipage}
		\begin{minipage}{0.125\hsize}
			\centerline{\small{(i)}}
		\end{minipage}
		\begin{minipage}{0.125\hsize}
			\centerline{\small{(j)}}
		\end{minipage}
		\begin{minipage}{0.125\hsize}
			\centerline{\small{(k)}}
		\end{minipage}
		\begin{minipage}{0.125\hsize}
			\centerline{\small{(l)}}
		\end{minipage}
		\begin{minipage}{0.125\hsize}
			\centerline{\small{(m)}}
		\end{minipage}
		\begin{minipage}{0.125\hsize}
			\centerline{\small{(n)}}
		\end{minipage}
		\begin{minipage}{0.050\hsize}
			\centerline{\hspace{\hsize}} 
		\end{minipage}

	\end{center}
	
	\vspace{-3mm}
	\caption{Denoising results for Jasper Ridge with the 131st band in Case 8. Upper row images are the restored images by each method. Lower row images are the absolute difference between the original image and each restored image (the range of the only observed noisy image in $\IndexNoisy$ is $[0,1]$). (a) Ground-truth. (b) Observed noisy image. (c) SSTV. (d) HSSTV1. (e) HSSTV2. (f) $\llHTV$. (g) STV. (h) SSST. (i) LRTDTV. (j) FGSLR. (k) TPTV. (l) QRNN3D. (m) FastHyMix. (n) $\SSSTTV$ (ours).}
	\label{fig:result_image_Case8_JR}
\end{figure*}

As ground-truth HS images, we adopt three HS image dataset: \textit{Jasper Ridge}\footnote{\url{https://rslab.ut.ac.ir/data}} cropped to size $100 \times 100 \times 198$, \textit{Pavia University}\footnote{\url{https://www.ehu/ccwintco/index/php/Hyperspectral_Remote_Sensing_Scenes}} cropped to size $120 \times 120 \times 98$, and \textit{Beltsville}\footnote{\url{https://www.spectir.com/contact#free-data-samples}} cropped to size $100 \times 100 \times 128$.
All the intensities of three HS images were normalized within the range $[0, 1]$.

HS images are often degraded by a mixture of various types of noise in real-world scenarios.
Thus, in the experiments, we consider the following eight cases of noise contamination:
\begin{itemize}
	\setlength{\leftskip}{18pt}
	\item [Case 1:] The observed HS image is contaminated by only white Gaussian noise with the standard deviation $\StanDevGauss = 0.05$.
	\item [Case 2:] The observed HS image is contaminated by white Gaussian noise with the standard deviation $\StanDevGauss = 0.05$ and salt-and-pepper noise with the rate $\RateSparse = 0.05$.
	\item [Case 3:] The observed HS image is contaminated by white Gaussian noise with the standard deviation $\StanDevGauss = 0.1$ and salt-and-pepper noise with the rate $\RateSparse = 0.05$.
	\item [Case 4:] {The observed HS image is contaminated by only vertical stripe noise whose intensity is uniformly random in the range $[-0.5, 0.5]$ with the rate $\RateStripe = 0.05$.}
	\item [Case 5:] The observed HS image is contaminated by white Gaussian noise with the standard deviation $\StanDevGauss = 0.05$ and vertical stripe noise whose intensity is uniformly random in the range $[-0.5, 0.5]$ with the rate $\RateStripe = 0.05$.
	\item [Case 6:] The observed HS image is contaminated by white Gaussian noise with the standard deviation $\StanDevGauss = 0.1$ and vertical stripe noise whose intensity is uniformly random in the range $[-0.5, 0.5]$ with the rate $\RateStripe = 0.05$.
	\item [Case 7:] The observed HS image is contaminated by white Gaussian noise with the standard deviation $\StanDevGauss = 0.05$, salt-and-pepper noise with the rate $\RateSparse = 0.05$, and vertical stripe noise whose intensity is uniformly random in the range $[-0.5, 0.5]$ with the rate $\RateStripe = 0.05$.
	\item [Case 8:] The observed HS image is contaminated by white Gaussian noise with the standard deviation $\StanDevGauss = 0.1$, salt-and-pepper noise with the rates $\RateSparse = 0.05$, and vertical stripe noise whose intensity is uniformly random in the range $[-0.5, 0.5]$ with the rates $\RateStripe = 0.05$.
\end{itemize}

\begin{table}[!t]
	\begin{center}
		\caption{Hyperparameter Settings in Each Method.}
		\label{tab:HyperParam}
			\begin{tabular}{cc}
				\toprule
				Methods & Parameters \\
				\cmidrule(lr){1-2}
				SSTV~\cite{Aggarwal2016SSTV} 
				& $\ParamsRadius = 0.95$
				\vspace{2mm} \\

				\multirow{2}{*}{HSSTV~\cite{Takeyama2020HSSTV}} 
				& $\ParamsRadius = 0.95$ \\
				& $\omega = 0.05$ 
				\vspace{2mm} \\
				
				\multirow{3}{*}{$\llHTV$~\cite{Wang2021l0l1HTV}} 
				& $\ParamsRadius = 0.95$ \\
				& Threshold of $\ell_{1,0}\text{-norm} = 0.02, 0.03, 0.04$ \\
				& Stepsize reduction factor = $0.999$
		 		\vspace{2mm} \\

				\multirow{2}{*}{STV~\cite{Lefkimmiatis2015STV}} 
				& $\ParamsRadius = 0.95$ \\
				& Spatial blocksize = $[10, 10]$ 
				\vspace{2mm} \\

				\multirow{2}{*}{SSST~\cite{Kurihara2017SSST}} 
				& $\ParamsRadius = 0.95$ \\
				& Spatial blocksize = $[10, 10]$ 
				\vspace{2mm} \\

				\multirow{4}{*}{LRTDTV~\cite{Wang2018LRTDTV}} 
				& $\tau = 1$ \\
				& $C = 10, 15, 20, 25$ \\
				& $\lambda = 100 * C / \sqrt{\NumVert \NumHori}$ \\
				& $\text{Rank} = (0.8\NumVert, 0.8\NumHori, 10)$ 
				\vspace{2mm} \\

				\multirow{4}{*}{FGSLR~\cite{Chen2022FGSLR}} 
				& $\beta = 0.1, 0.5$ \\
				& $\mu = 5, 10$ \\
				& $\delta = 0.5, 50$\\
				& Norm of $B$ is $\ell_{2,1}$-norm or Frobenius norm
				\vspace{2mm} \\

				\multirow{4}{*}{TPTV~\cite{Chen2023TPTV}} 
				& $\lambda = 5e-4, 1e-4, 1e-3, 1e-2, 1, 5e-2$ \\
				& $\text{Max iteration} = 50, 100$ \\
				& $\text{Initial rank} = 2$ \\
				& $\text{Rank} = (7, 7, 5)$
				\vspace{2mm} \\

				\multirow{4}{*}{QRNN3D~\cite{Wei2021QRNN3D}}
				& pretraining: $\text{epoch} = 100, \text{Batch size} = {16, 64}$ \\
				& fine-tuning: $\text{epoch} = 50, \text{Batch size} = 64$ \\
				& common setting: $\text{learning rate} = [1e-3, 1e-5]$ \\
				& (see [17, Table I\hspace{-0.2mm}I] for detail)\\
				\vspace{2mm} \\
				
				FastHyMix~\cite{Zhuang2023FastHyMix} 
				& $k$ subspace $= 4, 8, 12$
				\vspace{2mm} \\

				\multirow{2}{*}{$\SSSTTV$ (ours)} & $\ParamsRadius = 0.95$ \\
				& Spatial blocksize = $[10, 10]$ \\

				\bottomrule
			\end{tabular}
	\end{center}
	\vspace{-3mm}
\end{table}

The block size of $\SSSTTV$ was set to $10 \times 10 \times \NumBand$.
The radii $\RadiusSparse$, $\RadiusStripe$, and $\RadiusFidel$ were set as follows:
\begin{equation}
	\label{eq:RadiusSet}
	\RadiusSparse = \ParamsRadius \tfrac{\NumAll \RateSparse}{2}, \:
	\RadiusStripe = \ParamsRadius \tfrac{0.5 \NumAll \RateStripe (1 - \RateSparse)}{2}, \: \RadiusFidel = \ParamsRadius \sqrt{\StanDevGauss^2 \NumAll (1 - \RateSparse)},
\end{equation}
where the parameter $\ParamsRadius$ was set to $0.95$.
The hyperparameters, including the comparison methods, are shown in Table~\ref{tab:HyperParam}.
For specific cases, adjustments were made to improve the accuracy of the parameter settings. In Case 1, where only Gaussian noise is present, the noise concentrates more on the corresponding term compared to mixed noise cases. To reflect this, $\ParamsRadius$ was set to $0.98$. Similarly, in Case 4, where only Stripe noise is present, $\ParamsRadius$ was also set to $0.98$ for the same reason. Furthermore, in Case 4, the fidelity term $\| v - u - t \|_{2}$ becomes zero, which can lead to instability in the solution. To address this, $\RadiusFidel$ was fixed to $0.01$ to ensure stability.
The stopping criterion of Alg.~1 were set as follows:
\begin{equation}
	\label{eq:StopCri_simulated}
	\frac{\| \HSIClean^{(\IndexAlg+1)} - \HSIClean^{(\IndexAlg)} \|_2}{\| \HSIClean^{(\IndexAlg)}\|_{2}} < 1.0 \times 10^{-5}.
\end{equation}

For the quantitative evaluation, we employed the mean peak signal-to-noise ratio (MPSNR):
\begin{equation}
	\label{eq:MPSNR}
	\mathrm{MPSNR} = \frac{1}{\NumBand} \sum_{\IndexBand=1}^{\NumBand} 10\log_{10}\frac{\NumVert \NumHori}{\|\HSIClean_{\IndexBand} - \bar{\HSIClean}_{\IndexBand}\|_{2}^{2}},
\end{equation}
and the mean structural similarity index (MSSIM)~\cite{Wang2004SSIM}:
\begin{equation}
	\label{eq:MSSIM}
	\mathrm{MSSIM} = \frac{1}{\NumBand} \sum_{\IndexBand=1}^{\NumBand} \mathrm{SSIM}(\HSIClean_{\IndexBand}, \bar{\HSIClean}_{\IndexBand}),
\end{equation}
where $\HSIClean_{\IndexBand}$ and $\bar{\HSIClean}_{\IndexBand}$ are the $\IndexBand$-th band of the ground true HS image $\HSIClean$ and the estimated HS image $\bar{\HSIClean}$, respectively.
Generally, higher MPSNR and MSSIM values are corresponding to better denoising performances. Because the boundary conditions are circulant, we evaluate them by cutting off the first and last three bands. 

\setcounter{subsubsection}{0}
\subsubsection{Quantitative Comparison}
\label{subsubsec:QuantitativeComparison}
Tables~\ref{tab:MPSNR} and~\ref{tab:MSSIM} respectively show MPSNRs and MSSIMs in the experiments on the HS image contaminated with simulated noise. The best and second best results are highlighted in bold and underlined, respectively.
STV is worse in all cases. The SSST results show high MSSIM results in Case 6 (stripe noise only) for MSSIMs. However, its effectiveness declines when the HS image is contaminated with sparse noise. Similarly, the performance of LRTDTV drops when affected by stripe noise. TPTV performs well in Case 4, achieving the highest MPSNRs under contamination by only stripe noise, but shows a decline in cases involving higher levels of Gaussian noise. QRNN3D performs well for Pavia University, but shows limited performance for Jasper Ridge and Beltsville. FastHyMix generally achieves high performance in Gaussian and sparse noise removal experiments, but exhibits significant performance drops under contamination by stripe noise. In contrast, SSTV, HSSTV1, HSSTV2, $\llHTV$, and FGSLR exhibit more stable performance across different noise types. Among these methods, HSSTV1 shows superior MSSIM performance for Jasper Ridge and Pavia University, while FGSLR shows superior MSSIM performance for Beltsville. On the other hand, $\SSSTTV$ achieves the best MPSNRs in most cases, and in the two exceptions, it still ranks second highest. Moreover, $\SSSTTV$ shows a high overall performance independent of the HS images.

\subsubsection{Visual Quality Comparison for restored images}
Figs.~\ref{fig:result_image_Case1_PU}-\ref{fig:result_image_Case8_JR} show the results of HS image denoising and destriping.
The lower row images are the absolute difference between the original image and each restored image.

Fig.~\ref{fig:result_image_Case1_PU} shows the denoising results for Pavia University in Case 1, i.e. under contamination by only Gaussian noise, and Fig.~\ref{fig:result_image_Case3_JR} shows the results for Jasper Ridge in Case 3, which adds sparse noise to the Gaussian noise condition Case 1.
In both cases, the restored images by STV, SSST, and FGSLR in $\IndexSTV$, $\IndexSSST$, and $\IndexFGSLR$ exhibit significant spatial over-smoothing, losing fine spatial details. The restored images by the SSTV-based methods (SSTV, HSSTV1, HSSTV2, and l0-l1HTV) in $\IndexSSTV$, $\IndexHSSTVOne$, $\IndexHSSTVTwo$, and $\IndexllHTV$ have more structure than those of STV, SSST, and FGSLR. However, in Case 3, edges and textures appear in these difference images between the ground-truth and these restored images. Especially in the enlarged images, the edges of the road are clearly visible. In other words, these detailed structures are lost in the restored images. LRTDTV and TPTV manage to maintain structural details, but residual noise remains in both cases. Compared to these methods, QRNN3D removes noise more effectively while preserving structures, but the overall brightness is noticeably shifted in both cases. Similarly, the restored image by FastHyMix in Case 3 appears overly bright, likely due to outliers from sparse noise. In contrast, FastHyMix performs best in Gaussian noise removal in Case 1. On the other hand, as shown in $\IndexSSSTTV$, the proposed method, $\SSSTTV$, consistently removes noise while preserving structural details in both cases. In particular, the edges of the road in the enlarged area of both restored images are effectively reconstructed.


Fig.~\ref{fig:result_image_Case4_PU} shows the denoising results for Pavia University in Case 4, i.e. under contamination by only stripe noise, and Fig.~\ref{fig:result_image_Case5_BV} shows the results for Beltsville in Case 5, which adds Gaussian noise to the stripe noise condition Case 4.
TPTV restores structural details most effectively in Case 4. However, including Case 5, TPTV fails to remove the stripe noise. LRTDTV also leaves even more residual stripe noise. These arise from the mischaracterization of stripe noise. Another TV-LR hybrid method, FGSLR, effectively removes stripe noise, although brightness values are shifted in Case 5. The restored image by the deep learning-based method FastHyMix in $\IndexFastHyMix$ shows residual stripe noise, likely because the spatial correlation of stripe noise led the model to misinterpret the noise as significant features. Similarly, in Case 4, the restored image by QRNN3D in $\IndexQRNND$ also retains stripe noise at the locations indicated by arrows, but QRNN3D appears more robust than FastHyMix. As for other methods, including our method, stripe noise is adequately removed. This is thanks to the second and third constraints in Eq.~\eqref{prob:S3TTV_denoising} that characterize stripe noise. Furthermore, the methods that mainly use the second-order spatio-spectral differences, i.e., SSTV, HSSTV1, HSSTV2, $\llHTV$, and our method fully recover HS images without over-smoothing.

Fig.~\ref{fig:result_image_Case8_JR} shows the denoising results for Jasper Ridge in Case 8, which is the most contaminated case with Gaussian, sparse, and stripe noise. 
For STV and SSST in $\IndexSTV$ and $\IndexSSST$, no stripe noise remains in the restored images, but edges and textures are lost along with the noise, because the nuclear norm of the ``first-order'' differences is directly suppressed. SSTV, HSSTV1, HSSTV2, and $\llHTV$ in $\IndexSSTV$, $\IndexHSSTVOne$, $\IndexHSSTVTwo$, and $\IndexllHTV$ restore edges and textures more clearly than STV and SSST, but still remove some edges and textures (see the difference images in $\IndexSSTV$, $\IndexHSSTVOne$, $\IndexHSSTVTwo$, and $\IndexllHTV$ of Fig.~\ref{fig:result_image_Case8_JR}). This would be due to the fact that these methods suffer more from the limitation of referring only to adjacent pixels/bands. LRTDTV and TPTV almost recover edges and textures in the restored images of $\IndexLRTDTV$ and $\IndexTPTV$, but as in Case 4 and Case 5, stripe noise remains at the positions indicated by the arrows. Unlike LRTDTV and TPTV, FGSLR successfully removes stripe noise, however, edges and textures remain in the difference image, similar to the SSTV-based methods. In the results of FastHyMix, stripe noise remains, and the overall brightness values are shifted. For QRNN3D, the restored image in $\IndexQRNND$ shows even greater changes in brightness across the entire image. This is likely because, under high-intensity mixed noise conditions, it is challenging to automatically extract features such as spectral low-rankness and spatial correlations. On the other hand, the difference image of $\SSSTTV$ in $\IndexSSSTTV$ is closer to black than the other methods. Furthermore, no edges or textures are visible in the enlarged image. These suggest that $\SSSTTV$ removes Gaussian, sparse, and stripe noise most effectively and recovers edges and textures with the highest accuracy.

\subsubsection{Visual Quality Comparison for restored spectra}
\begin{figure*}[t]
	\begin{center}
		\begin{minipage}{0.15\hsize}
			\centerline{\includegraphics[width=\hsize]{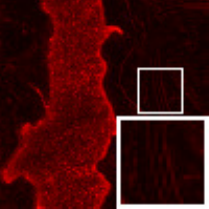}} 
		\end{minipage}
		\begin{minipage}{0.15\hsize}
			\centerline{\includegraphics[width=\hsize]{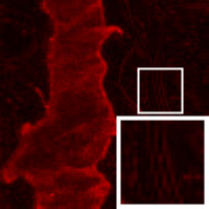}} 
		\end{minipage}
		\begin{minipage}{0.15\hsize}
			\centerline{\includegraphics[width=\hsize]{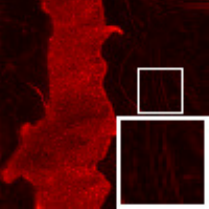}} 
		\end{minipage}
		\begin{minipage}{0.15\hsize}
			\centerline{\includegraphics[width=\hsize]{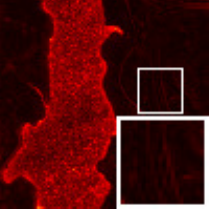}} 
		\end{minipage}
		\begin{minipage}{0.15\hsize}
			\centerline{\includegraphics[width=\hsize]{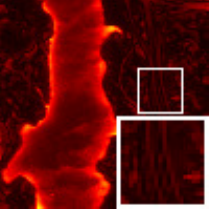}} 
		\end{minipage}
		\begin{minipage}{0.15\hsize}
			\centerline{\includegraphics[width=\hsize]{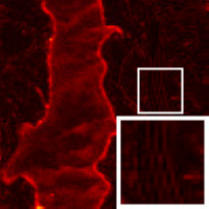}} 
		\end{minipage}
		\begin{minipage}{0.050\hsize}
			\centerline{\hspace{\hsize}} 
		\end{minipage}
		
		\vspace{1mm}
		
		\begin{minipage}{0.15\hsize}
			\centerline{\includegraphics[width=\hsize]{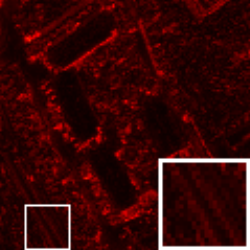}} 
		\end{minipage}
		\begin{minipage}{0.15\hsize}
			\centerline{\includegraphics[width=\hsize]{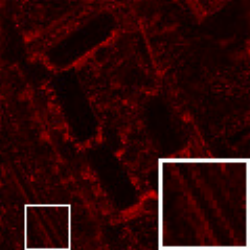}} 
		\end{minipage}
		\begin{minipage}{0.15\hsize}
			\centerline{\includegraphics[width=\hsize]{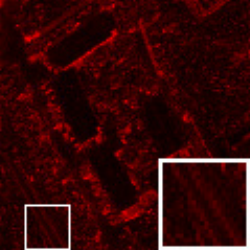}} 
		\end{minipage}
		\begin{minipage}{0.15\hsize}
			\centerline{\includegraphics[width=\hsize]{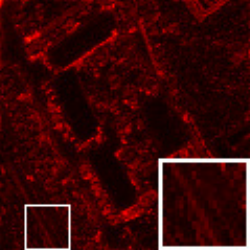}} 
		\end{minipage}
		\begin{minipage}{0.15\hsize}
			\centerline{\includegraphics[width=\hsize]{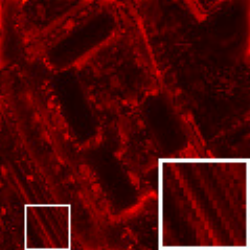}} 
		\end{minipage}
		\begin{minipage}{0.15\hsize}
			\centerline{\includegraphics[width=\hsize]{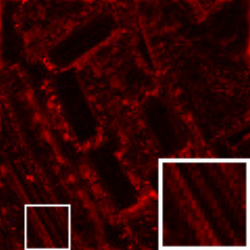}} 
		\end{minipage}
		\begin{minipage}{0.050\hsize}
			\centerline{\hspace{\hsize}} 
		\end{minipage}

		\vspace{1mm}
		
		\begin{minipage}{0.15\hsize}
			\centerline{\includegraphics[width=\hsize]{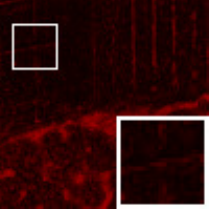}} 
		\end{minipage}
		\begin{minipage}{0.15\hsize}
			\centerline{\includegraphics[width=\hsize]{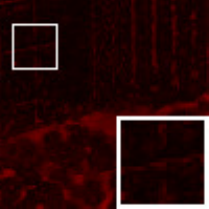}} 
		\end{minipage}
		\begin{minipage}{0.15\hsize}
			\centerline{\includegraphics[width=\hsize]{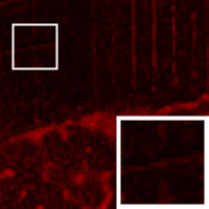}} 
		\end{minipage}
		\begin{minipage}{0.15\hsize}
			\centerline{\includegraphics[width=\hsize]{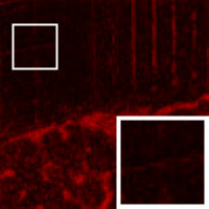}} 
		\end{minipage}
		\begin{minipage}{0.15\hsize}
			\centerline{\includegraphics[width=\hsize]{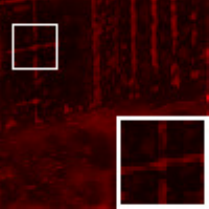}} 
		\end{minipage}
		\begin{minipage}{0.15\hsize}
			\centerline{\includegraphics[width=\hsize]{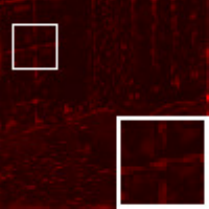}} 
		\end{minipage}
		\begin{minipage}{0.050\hsize}
			\centerline{\includegraphics[width=\hsize]{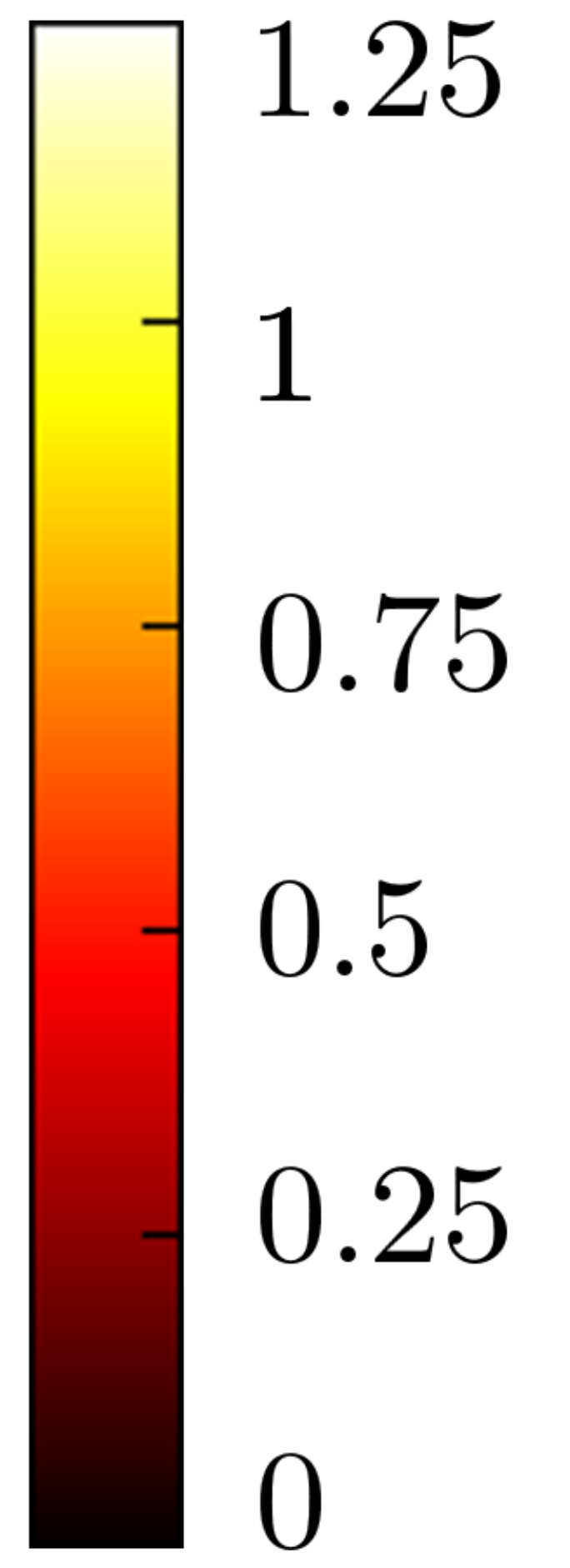}} 
		\end{minipage}
		
		\vspace{1mm}
		
		\begin{minipage}{0.15\hsize}
			\centerline{\small{(c): SSTV}}
		\end{minipage}
		\begin{minipage}{0.15\hsize}
			\centerline{\small{(d): HSSTV1}}
		\end{minipage}
		\begin{minipage}{0.15\hsize}
			\centerline{\small{(e): HSSTV2}}
		\end{minipage}
		\begin{minipage}{0.15\hsize}
			\centerline{\small{(f): $\llHTV$}}
		\end{minipage}
		\begin{minipage}{0.15\hsize}
			\centerline{\small{(g): STV}}
		\end{minipage}
		\begin{minipage}{0.15\hsize}
			\centerline{\small{(h): SSST}}
		\end{minipage}
		\begin{minipage}{0.050\hsize}
			\centerline{\hspace{\hsize}} 
		\end{minipage}
		
		\vspace{2mm}
		
		\begin{minipage}{0.15\hsize}
			\centerline{\includegraphics[width=\hsize]{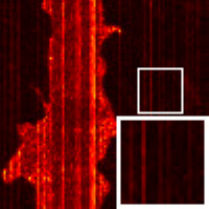}} 
		\end{minipage}
		\begin{minipage}{0.15\hsize}
			\centerline{\includegraphics[width=\hsize]{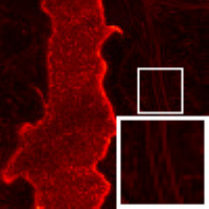}} 
		\end{minipage}
		\begin{minipage}{0.15\hsize}
			\centerline{\includegraphics[width=\hsize]{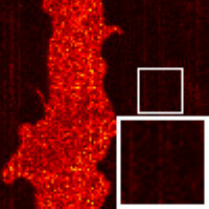}} 
		\end{minipage}
		\begin{minipage}{0.15\hsize}
			\centerline{\includegraphics[width=\hsize]{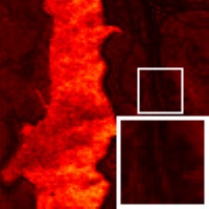}} 
		\end{minipage}
		\begin{minipage}{0.15\hsize}
			\centerline{\includegraphics[width=\hsize]{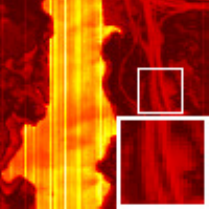}} 
		\end{minipage}
		\begin{minipage}{0.15\hsize}
			\centerline{\includegraphics[width=\hsize]{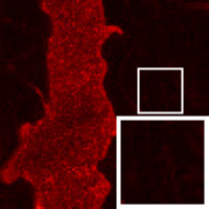}} 
		\end{minipage}
		\begin{minipage}{0.050\hsize}
			\centerline{\hspace{\hsize}} 
		\end{minipage}
		
		\vspace{1mm}
		
		\begin{minipage}{0.15\hsize}
			\centerline{\includegraphics[width=\hsize]{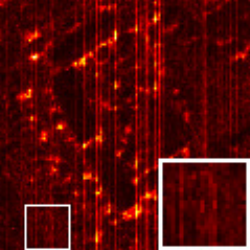}} 
		\end{minipage}
		\begin{minipage}{0.15\hsize}
			\centerline{\includegraphics[width=\hsize]{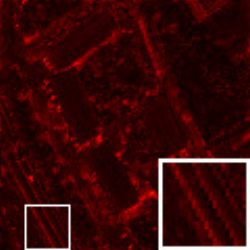}} 
		\end{minipage}
		\begin{minipage}{0.15\hsize}
			\centerline{\includegraphics[width=\hsize]{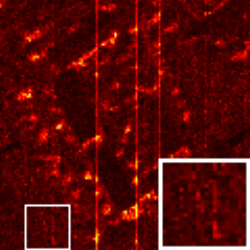}} 
		\end{minipage}
		\begin{minipage}{0.15\hsize}
			\centerline{\includegraphics[width=\hsize]{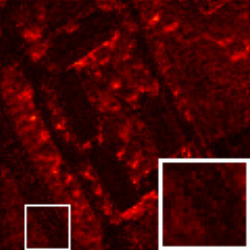}} 
		\end{minipage}
		\begin{minipage}{0.15\hsize}
			\centerline{\includegraphics[width=\hsize]{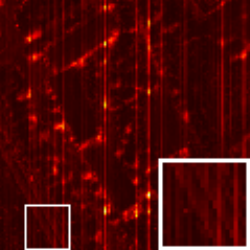}} 
		\end{minipage}
		\begin{minipage}{0.15\hsize}
			\centerline{\includegraphics[width=\hsize]{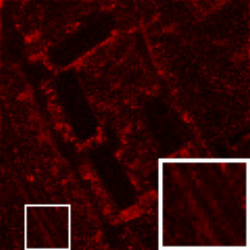}} 
		\end{minipage}
		\begin{minipage}{0.050\hsize}
			\centerline{\hspace{\hsize}} 
		\end{minipage}

		\vspace{1mm}
		
		\begin{minipage}{0.15\hsize}
			\centerline{\includegraphics[width=\hsize]{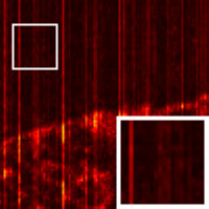}} 
		\end{minipage}
		\begin{minipage}{0.15\hsize}
			\centerline{\includegraphics[width=\hsize]{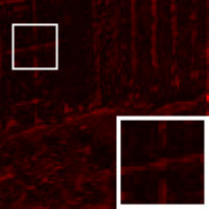}} 
		\end{minipage}
		\begin{minipage}{0.15\hsize}
			\centerline{\includegraphics[width=\hsize]{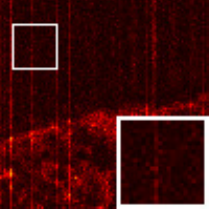}} 
		\end{minipage}
		\begin{minipage}{0.15\hsize}
			\centerline{\includegraphics[width=\hsize]{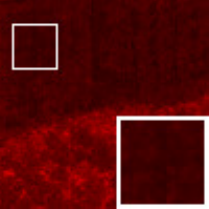}} 
		\end{minipage}
		\begin{minipage}{0.15\hsize}
			\centerline{\includegraphics[width=\hsize]{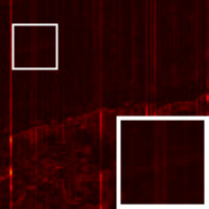}} 
		\end{minipage}
		\begin{minipage}{0.15\hsize}
			\centerline{\includegraphics[width=\hsize]{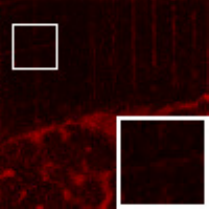}} 
		\end{minipage}
		\begin{minipage}{0.050\hsize}
			\centerline{\includegraphics[width=\hsize]{./fig_supplement/SAM_map_color_woboundary/colorbar_hot_r1.25-eps-converted-to.pdf}} 
		\end{minipage}
		
		\vspace{1mm}
		
		\begin{minipage}{0.15\hsize}
			\centerline{\small{(i): LRTDTV}}
		\end{minipage}
		\begin{minipage}{0.15\hsize}
			\centerline{\small{(j): FGSLR}}
		\end{minipage}
		\begin{minipage}{0.15\hsize}
			\centerline{\small{(k): TPTV}}
		\end{minipage}
		\begin{minipage}{0.15\hsize}
			\centerline{\small{(l): FastHyMix}}
		\end{minipage}
		\begin{minipage}{0.15\hsize}
			\centerline{\small{(m): QRNN3D}}
		\end{minipage}
		\begin{minipage}{0.15\hsize}
			\centerline{\small{(n): $\SSSTTV$}}
		\end{minipage}
		\begin{minipage}{0.050\hsize}
			\centerline{\hspace{\hsize}} 
		\end{minipage}
		
	\end{center}
	
	\vspace{-3mm}
	\caption{2-D SAMs for Jasper Ridge (top), Pavia University (middle), and Beltsville (bottom).}
	
	\label{fig:2D_SAM}
\end{figure*}
%

To assess the spectral restoration accuracy, we analyzed the 2-D Spectral Angle Mapper (SAM) for Jasper Ridge, Pavia University, and Beltsville under Case 8 in Fig.~\ref{fig:2D_SAM}.
The 2-D SAM at the $(\IndexVert, \IndexHori)$-th pixel is computed as:
\begin{equation}
	\label{eq:2D_SAM}
	\mathrm{SAM} (\IndexVert, \IndexHori) = \arccos \left( \frac{\HSIClean_{\IndexVert, \IndexHori}^{\top} \bar{{\HSIClean}}_{\IndexVert, \IndexHori}}{\|\HSIClean_{\IndexVert, \IndexHori} \|_{2} \|\bar{{\HSIClean}}_{\IndexVert, \IndexHori} \|_{2}} \right),
\end{equation}
where $\HSIClean_{\IndexVert, \IndexHori} \in \RealSpace{\NumBand}$ and $\bar{{\HSIClean}}_{\IndexVert, \IndexHori} \in \RealSpace{\NumBand}$ are the spectral vectors at pixel $(\IndexVert, \IndexHori)$ in the clean HS image and the restored HS image, respectively. Since the boundary conditions are circulant, we evaluate them by cutting off the first and last three bands.
QRNN3D in $\IndexQRNND$ yields relatively high SAM values across the entire image, indicating spectral distortion.
The SAM maps of LRTDTV, TPTV, and FastHyMix in $\IndexLRTDTV$, $\IndexTPTV$, and $\IndexFastHyMix$ retain stripe noise.
Across all methods, high SAMs are observed in regions with low signal intensity, such as the verticallly extended red area in Jasper Ridge and the horizontally extended red area in Beltsville. This is due to the higher sensitivity of low-intensity signals to noise contamination. On the other hand, HSSTV1 in Jasper Ridge and FGSLR in Beltsville achieve lower SAMs. This indicates that these methods perform well in regions where the same material is continuously present.
However, the existing methods, especially STV, SSST, and FGSLR, exhibit worse SAMs in material boundaries and fine structures, such as roads and trenches. This effect is caused by spectral values being overly influenced by neighboring different spectra.
In contrast, $\SSSTTV$ excels at restoring spectral details in these complex regions with mixed materials. The enlarged areas in Fig.~\ref{fig:2D_SAM} show that $\SSSTTV$ achieves the lowest SAMs at the roads and the material boundaries.
Such high-precision restoration in mixed-material regions is valuable for applications where accurate spectral differences between materials are critical for analysis, such as mineral exploration and landmine detection.

\subsection{Real HS Image Experiment}
\label{subsec:RealHSIExpt}
\begin{figure*}[t]
    \begin{center}
        \begin{minipage}{0.125\hsize}
            \centerline{\includegraphics[width=\hsize]{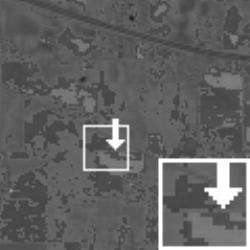}} 
        \end{minipage}
        \begin{minipage}{0.125\hsize}
            \centerline{\includegraphics[width=\hsize]{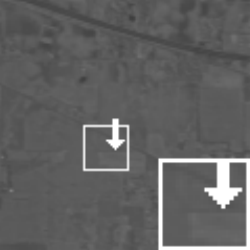}} 
        \end{minipage}
        \begin{minipage}{0.125\hsize}
            \centerline{\includegraphics[width=\hsize]{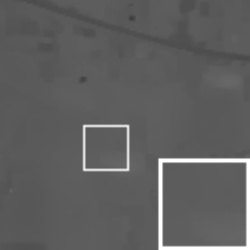}} 
        \end{minipage}
        \begin{minipage}{0.125\hsize}
            \centerline{\includegraphics[width=\hsize]{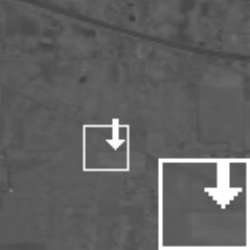}} 
        \end{minipage}
        \begin{minipage}{0.125\hsize}
            \centerline{\includegraphics[width=\hsize]{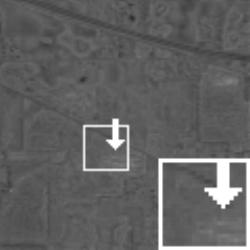}} 
        \end{minipage}
        \begin{minipage}{0.125\hsize}
        	\centerline{\includegraphics[width=\hsize]{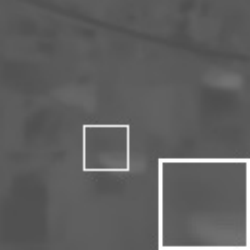}} 
        \end{minipage}
        \begin{minipage}{0.125\hsize}
            \centerline{\includegraphics[width=\hsize]{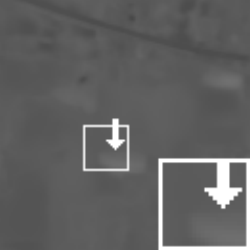}} 
        \end{minipage}

        \vspace{1mm}

		\begin{minipage}{0.125\hsize}
			\centerline{\small{(b)}}
		\end{minipage}
		\begin{minipage}{0.125\hsize}
			\centerline{\small{(c)}}
		\end{minipage}
		\begin{minipage}{0.125\hsize}
			\centerline{\small{(d)}}
		\end{minipage}
		\begin{minipage}{0.125\hsize}
			\centerline{\small{(e)}}
		\end{minipage}
        \begin{minipage}{0.125\hsize}
			\centerline{\small{(f)}}
		\end{minipage}
		\begin{minipage}{0.125\hsize}
			\centerline{\small{(g)}}
		\end{minipage}
        \begin{minipage}{0.125\hsize}
			\centerline{\small{(h)}}
		\end{minipage}
  
        \vspace{2mm}

        \begin{minipage}{0.125\hsize}
			\centerline{\hspace{\hsize}} 
		\end{minipage}
        \begin{minipage}{0.125\hsize}
            \centerline{\includegraphics[width=\hsize]{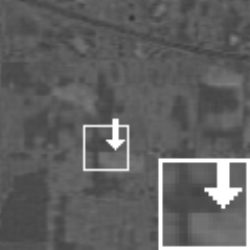}} 
        \end{minipage}
        \begin{minipage}{0.125\hsize}
        	\centerline{\includegraphics[width=\hsize]{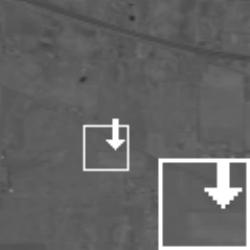}} 
        \end{minipage}
        \begin{minipage}{0.125\hsize}
            \centerline{\includegraphics[width=\hsize]{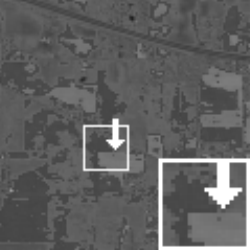}} 
        \end{minipage}
        \begin{minipage}{0.125\hsize}
        	\centerline{\includegraphics[width=\hsize]{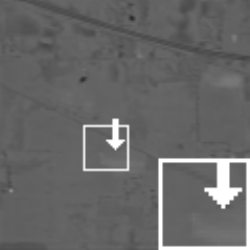}} 
        \end{minipage}
        \begin{minipage}{0.125\hsize}
        	\centerline{\includegraphics[width=\hsize]{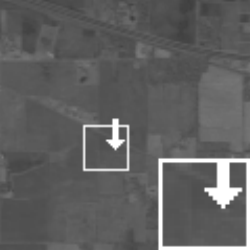}} 
        \end{minipage}
        \begin{minipage}{0.125\hsize}
            \centerline{\includegraphics[width=\hsize]{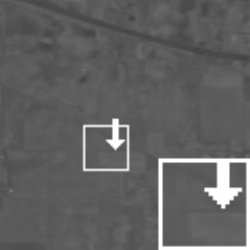}} 
        \end{minipage}
        
        \vspace{1mm}

        \begin{minipage}{0.125\hsize}
			\centerline{\hspace{\hsize}} 
		\end{minipage}
		\begin{minipage}{0.125\hsize}
			\centerline{\small{(i)}}
		\end{minipage}
		\begin{minipage}{0.125\hsize}
			\centerline{\small{(j)}}
		\end{minipage}
		\begin{minipage}{0.125\hsize}
			\centerline{\small{(k)}}
		\end{minipage}
		\begin{minipage}{0.125\hsize}
			\centerline{\small{(l)}}
		\end{minipage}
		\begin{minipage}{0.125\hsize}
			\centerline{\small{(m)}}
		\end{minipage}
        \begin{minipage}{0.125\hsize}
			\centerline{\small{(n)}}
		\end{minipage}
    \end{center}
	
    \vspace{-3mm}
    \caption{Denoising and destriping results for Indian Pines with the 88th band, multiplied by 1.5 for visibility. (b) Observed noisy image. (c) SSTV. (d) HSSTV1. (e) HSSTV2. (f) $\llHTV$. (g) STV. (h) SSST. (i) LRTDTV. (j) FGSLR. (k) TPTV. (l) FastHyMix. (m) QRNN3D. (n) $\SSSTTV$ (ours).}
    \label{fig:result_image_IndianPines}
\end{figure*}
\begin{figure*}[t]
	\begin{center}
		\begin{minipage}{0.125\hsize}
			\centerline{\includegraphics[width=\hsize]{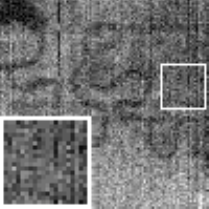}} 
		\end{minipage}
		\begin{minipage}{0.125\hsize}
			\centerline{\includegraphics[width=\hsize]{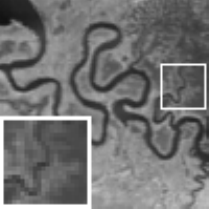}} 
		\end{minipage}
		\begin{minipage}{0.125\hsize}
			\centerline{\includegraphics[width=\hsize]{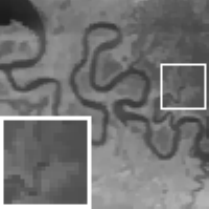}} 
		\end{minipage}
		\begin{minipage}{0.125\hsize}
			\centerline{\includegraphics[width=\hsize]{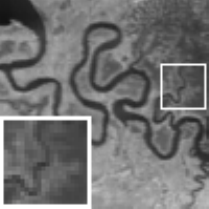}} 
		\end{minipage}
		\begin{minipage}{0.125\hsize}
			\centerline{\includegraphics[width=\hsize]{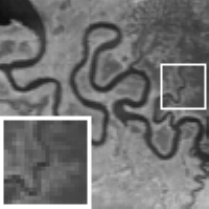}} 
		\end{minipage}
		\begin{minipage}{0.125\hsize}
			\centerline{\includegraphics[width=\hsize]{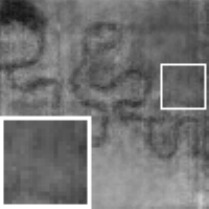}} 
		\end{minipage}
		\begin{minipage}{0.125\hsize}
			\centerline{\includegraphics[width=\hsize]{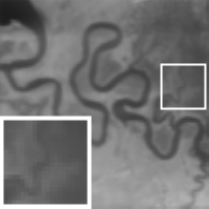}} 
		\end{minipage}
		
		\vspace{1mm}
		
		\begin{minipage}{0.125\hsize}
			\centerline{\small{(b)}}
		\end{minipage}
		\begin{minipage}{0.125\hsize}
			\centerline{\small{(c)}}
		\end{minipage}
		\begin{minipage}{0.125\hsize}
			\centerline{\small{(d)}}
		\end{minipage}
		\begin{minipage}{0.125\hsize}
			\centerline{\small{(e)}}
		\end{minipage}
		\begin{minipage}{0.125\hsize}
			\centerline{\small{(f)}}
		\end{minipage}
		\begin{minipage}{0.125\hsize}
			\centerline{\small{(g)}}
		\end{minipage}
		\begin{minipage}{0.125\hsize}
			\centerline{\small{(h)}}
		\end{minipage}
		
		\vspace{2mm}

		\begin{minipage}{0.125\hsize}
			\centerline{\hspace{\hsize}} 
		\end{minipage}
		\begin{minipage}{0.125\hsize}
			\centerline{\includegraphics[width=\hsize]{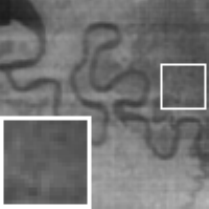}} 
		\end{minipage}
		\begin{minipage}{0.125\hsize}
			\centerline{\includegraphics[width=\hsize]{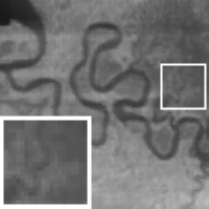}} 
		\end{minipage}
		\begin{minipage}{0.125\hsize}
			\centerline{\includegraphics[width=\hsize]{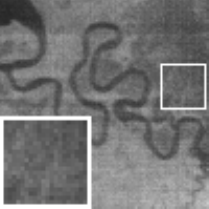}} 
		\end{minipage}
		\begin{minipage}{0.125\hsize}
			\centerline{\includegraphics[width=\hsize]{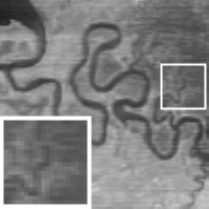}} 
		\end{minipage}
		\begin{minipage}{0.125\hsize}
			\centerline{\includegraphics[width=\hsize]{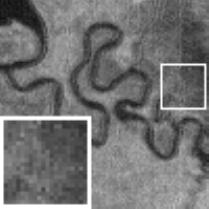}} 
		\end{minipage}
		\begin{minipage}{0.125\hsize}
			\centerline{\includegraphics[width=\hsize]{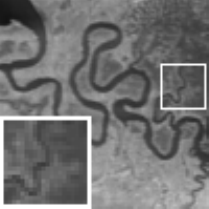}} 
		\end{minipage}
		
		\vspace{1mm}

		\begin{minipage}{0.125\hsize}
			\centerline{\hspace{\hsize}} 
		\end{minipage}
		\begin{minipage}{0.125\hsize}
			\centerline{\small{(i)}}
		\end{minipage}
		\begin{minipage}{0.125\hsize}
			\centerline{\small{(j)}}
		\end{minipage}
		\begin{minipage}{0.125\hsize}
			\centerline{\small{(k)}}
		\end{minipage}
		\begin{minipage}{0.125\hsize}
			\centerline{\small{(l)}}
		\end{minipage}
		\begin{minipage}{0.125\hsize}
			\centerline{\small{(m)}}
		\end{minipage}
		\begin{minipage}{0.125\hsize}
			\centerline{\small{(n)}}
		\end{minipage}
	\end{center}
	
	\vspace{-3mm}
	\caption{Denoising and destriping results for Suwannee with the 197th band. (b) Observed noisy image. (c) SSTV. (d) HSSTV1. (e) HSSTV2. (f) $\llHTV$. (g) STV. (h) SSST. (i) LRTDTV. (j) FGSLR. (k) TPTV. (l) FastHyMix. (m) QRNN3D. (n) $\SSSTTV$ (ours).}
	\label{fig:result_image_Suwannee}
\end{figure*}
We employed the following two datasets:
\subsubsection{Indian Pines}  This HS image was captured using the AVIRIS sensor over the Indian Pines test site in North-western Indiana.
The resolution of the original data is $145 \times 145$ pixels, and each pixel has spectral information with 224 bands ranging from 400 nm to 2500 nm.
After removing several noisy bands and cropping the original data, we obtained the HS image with $120 \times 120$ pixels and 198 bands.

\subsubsection{Suwannee} This HS image was captured using a SpecTIR sensor over the Suwannee River Basin in Florida, USA.
The resolution of the original HS image is $1200 \times 320$ pixels, and each pixel has spectral information with 360 bands ranging from 400 nm to 2500 nm.
We cropped the HS image to $100 \times 100$ pixels and 360 bands.

All the intensities of both HS images were normalized within the range $[0, 1]$.
The block size of $\SSSTTV$ was set to $10 \times 10 \times \NumBand$.
For the radii $\RadiusSparse$, $\RadiusStripe$, and $\RadiusFidel$, we adjusted them to appropriate values after empirically estimating the intensity of the noise in the real HS image.
Specifically, for the Indian Pines, $\RadiusSparse$, $\RadiusStripe$, and $\RadiusFidel$ were set to 200, 100, and 30, respectively, and for Suwannee, they were set to 800, 5000, and 100, respectively.
The stopping criterion of Alg.~1 were set as \eqref{eq:StopCri_simulated}.

Since no reference clean HS image is available, we compare the denoising performance using visual results.
Fig.~\ref{fig:result_image_IndianPines} shows the HS image denoising and destriping results for Indian Pines. 
HSSTV1, STV, and SSST cause over-smoothing in the restored images of $\IndexHSSTVOne$, $\IndexSTV$, and $\IndexSSST$. The restored image by QRNN3D in $\IndexQRNND$ has some edges and texture, but it is overall flat. For LRTDTV and TPTV in $\IndexLRTDTV$ and $\IndexTPTV$, noise remains in the restored images. On the other hand, SSTV, HSSTV2, $\llHTV$, FastHyMix, and $\SSSTTV$ achieve sufficient noise removal while preserving edges and textures. In particular, for the enlarged area, SSTV, HSSTV2, FGSLR, and $\SSSTTV$ restore the edges indicated by the arrows, while $\llHTV$ and FastHyMix lose them. This is due to the characterization of the spatial similarity between adjacent bands of HS images using the second-order spatio-spectral differences.



Fig.~\ref{fig:result_image_Suwannee} shows the HS image denoising and destriping results for Suwannee.
The restored images by STV, LRTDTV, TPTV, and FastHyMix in $\IndexSTV$, $\IndexLRTDTV$, $\IndexTPTV$, and $\IndexFastHyMix$ still retain vertical stripe noise.
For LRTDTV, TPTV, and FastHyMix, this is due to the limitation of stripe noise removal, similar to the results seen in the simulation experiments of Figs.~\ref{fig:result_image_Case4_PU}, ~\ref{fig:result_image_Case5_BV}, and \ref{fig:result_image_Case8_JR}. QRNN3D in $\IndexQRNND$ removes the vertical stripe noise, but residual horizontal noise remains, and the river structure in the enlarged area appears broken.
In the case of STV, its regularization method, which mainly captures the spatial piecewise-smoothness in HS images, is insufficient to separate the HS image and vertically smooth stripe noise.
On the other hand, SSTV, HSSTV1, HSSTV2, $\llHTV$, SSST, and $\SSSTTV$ sufficiently remove stripe noise.
This indicates that the second and third constraints characterizing stripe noise in Eq.~\eqref{prob:S3TTV_denoising} are effective for real stripe noise.
However, the edges of the restored images by HSSTV1 and SSST in $\IndexHSSTVOne$ and $\IndexSSST$ are smoothed in the enlarged areas.
Unlike these methods, SSTV, HSSTV2, $\llHTV$, and the proposed method, $\SSSTTV$, achieve the preservation of the narrow river structure while removing noise.

\subsection{Discussion}
\label{subsec:Discussion}

\subsubsection{Convergence Analysis}
\label{subsubsec:ConvAnal}
\begin{figure*}[t]
    \begin{center}
        \begin{minipage}{0.240\hsize}
            \centerline{\includegraphics[width=\hsize]{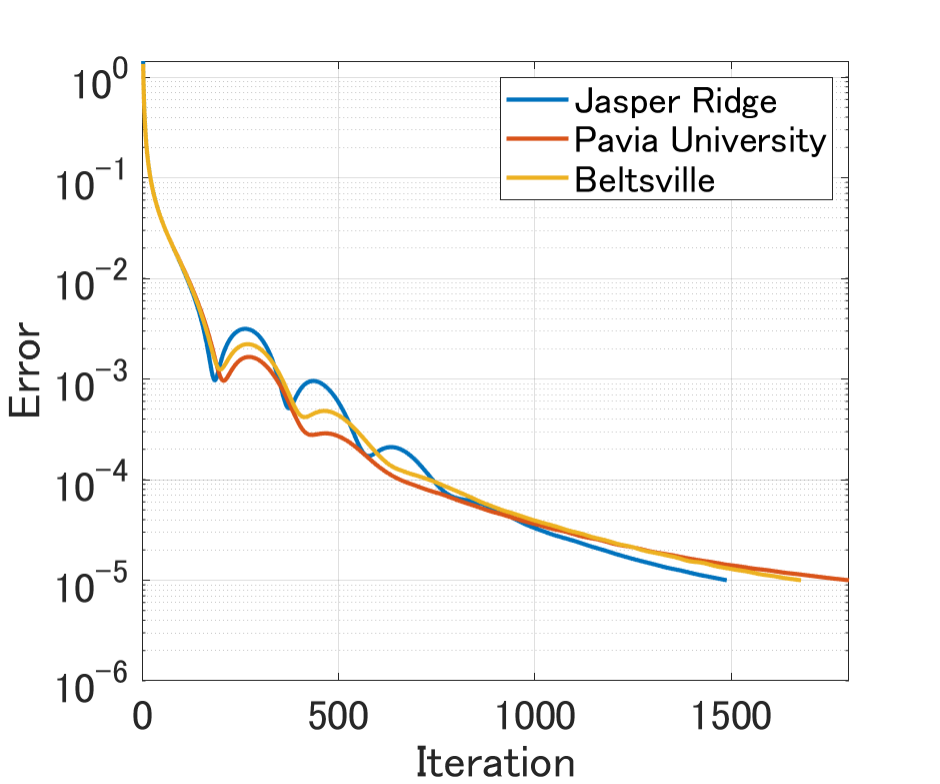}}
        \end{minipage}
        \begin{minipage}{0.240\hsize}
            \centerline{\includegraphics[width=\hsize]{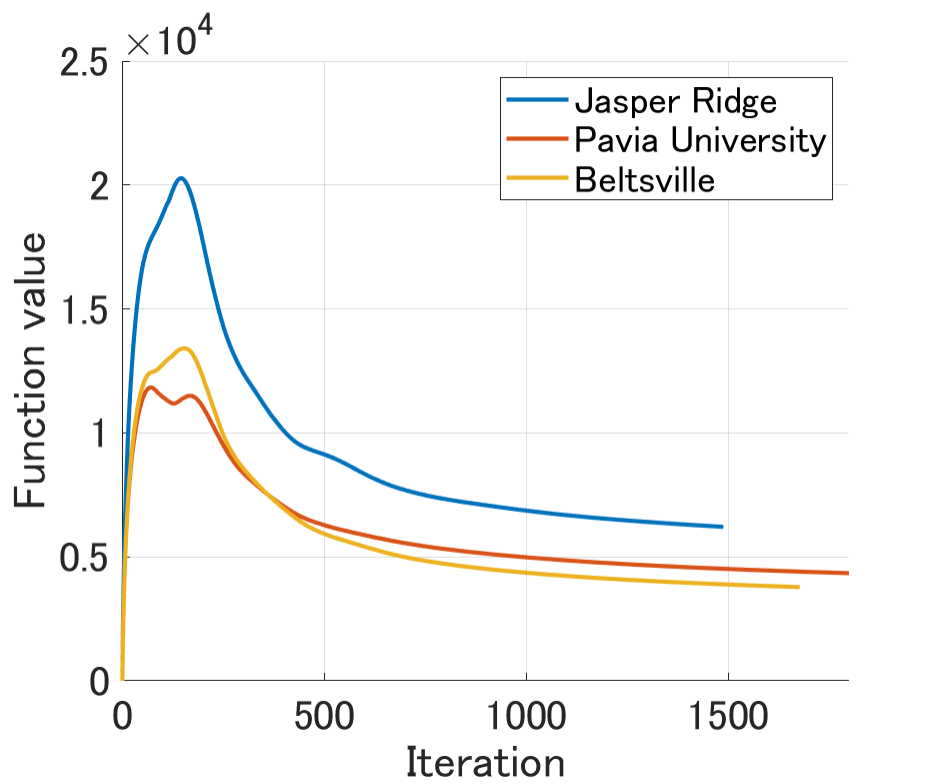}}
        \end{minipage}
        \begin{minipage}{0.240\hsize}
            \centerline{\includegraphics[width=\hsize]{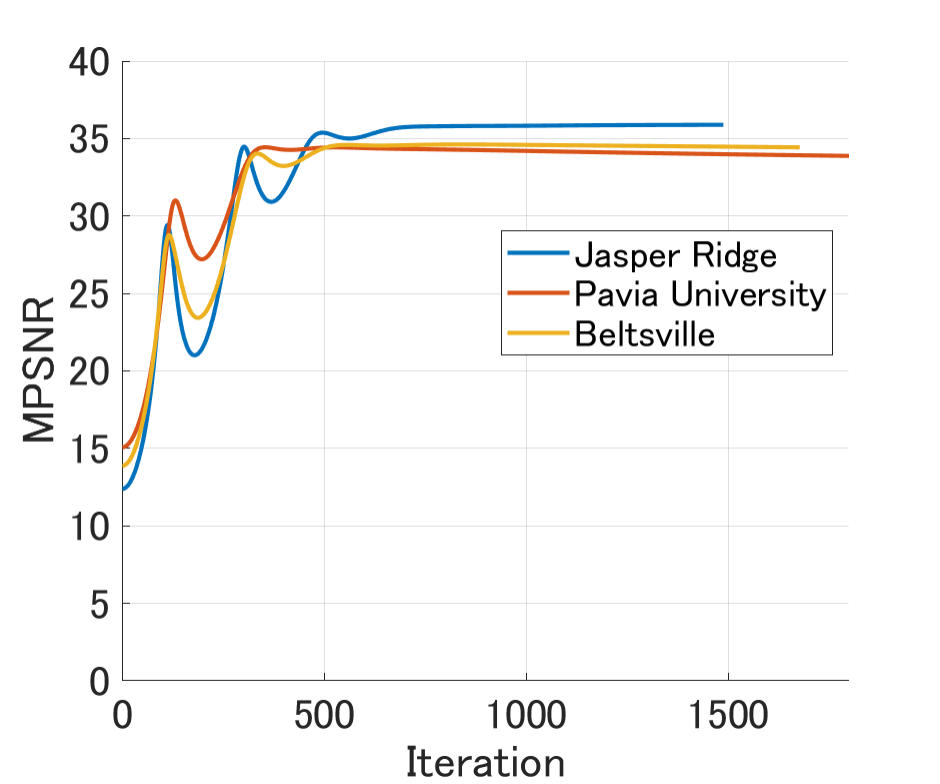}}
        \end{minipage}
        \begin{minipage}{0.240\hsize}
            \centerline{\includegraphics[width=\hsize]{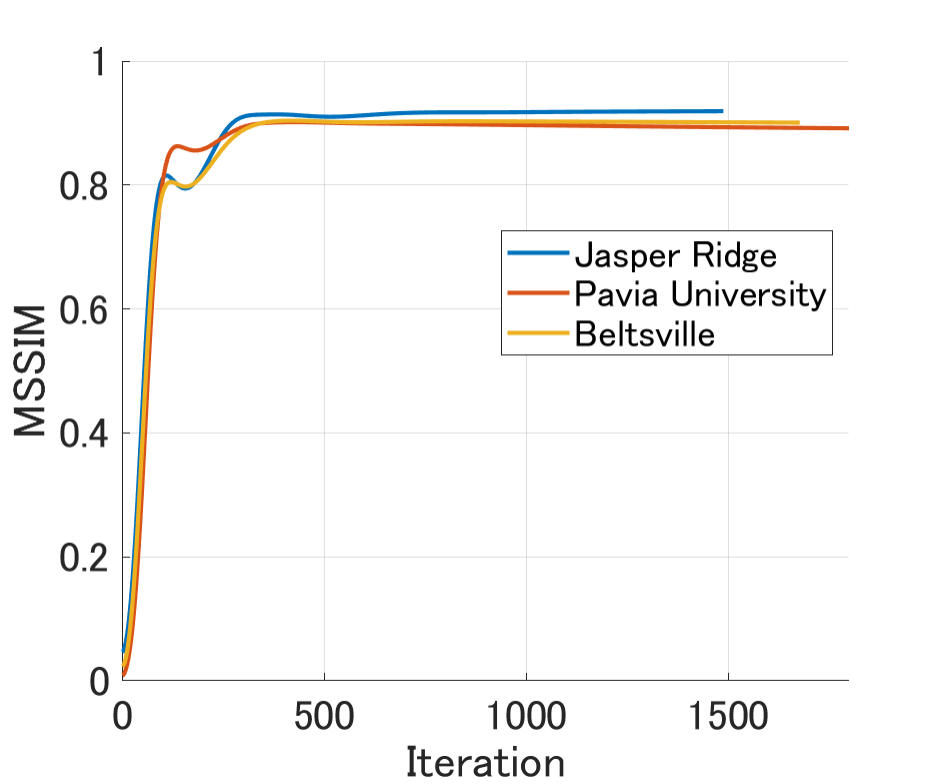}}
        \end{minipage}

        \begin{minipage}{0.240\hsize}
            \centerline{\small{(a)}}
		\end{minipage}
		\begin{minipage}{0.240\hsize}
			\centerline{\small{(b)}}
		\end{minipage}
		\begin{minipage}{0.240\hsize}
			\centerline{\small{(c)}}
		\end{minipage}
		\begin{minipage}{0.240\hsize}
			\centerline{\small{(d)}}
		\end{minipage}
    \end{center}
	
    \vspace{-3mm}
    \caption{Convergence analysis of $\SSSTTV$. (a): The relative error $\| \HSIClean^{(\IndexAlg+1)} - \HSIClean^{(\IndexAlg)} \|_2 / \| \HSIClean^{(\IndexAlg)}\|_{2}$ versus iteration $\IndexAlg$. (b): Objective function value $\SSSTTV(\HSIClean)$ versus iteration $\IndexAlg$. (c): MPSNR versus iteration $\IndexAlg$. (d): MSSIM  versus iteration $\IndexAlg$. }
    \label{fig:Conv_Anal}
\end{figure*}
It is mathematically guaranteed that the variables $\HSIClean, \NoiseSparse, \NoiseStripe$ generated by Alg.~1 converge to a global solution of the proposed convex optimization problem~\cite{Pock2011PPDS}.
Furthermore, we experimentally examined the convergence of our algorithm.
Fig.~\ref{fig:Conv_Anal} plots the relative error of HS images: $\| \HSIClean^{(\IndexAlg+1)} - \HSIClean^{(\IndexAlg)} \|_2 / \| \HSIClean^{(\IndexAlg)}\|_{2}$, the objective function values: $\SSSTTV(\HSIClean)$, the mean peak signal-to-noise ratio (MPSNR), and the mean structural similarity index (MSSIM) for \textit{Jasper Ridge},  \textit{Pavia University}, and \textit{Beltsville} in Case 8.
The relative error of the HS images decreased (Fig.~\ref{fig:Conv_Anal}-(a)).
The objective function values, MPSNRs, and MSSIMs approached certain values (Fig.~\ref{fig:Conv_Anal}-(b), (c), and (d)).
From these results, we can confirm experimentally that P-PDS updates the variables to approach the solution of our constrained convex optimization problem.

\subsubsection{Parameter Analysis}
\label{subsubsec:ParamAnal}
\begin{figure*}[t]
    \begin{center}
        \begin{minipage}{0.240\hsize}
            \centerline{\includegraphics[width=\hsize]{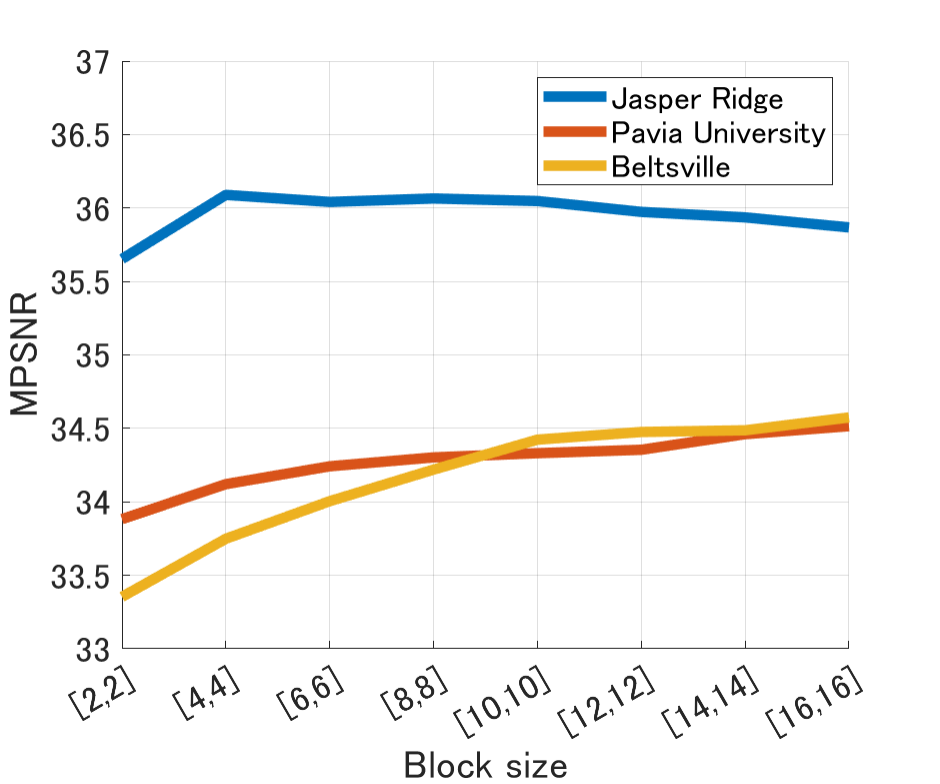}}
        \end{minipage}
        \begin{minipage}{0.240\hsize}
            \centerline{\includegraphics[width=\hsize]{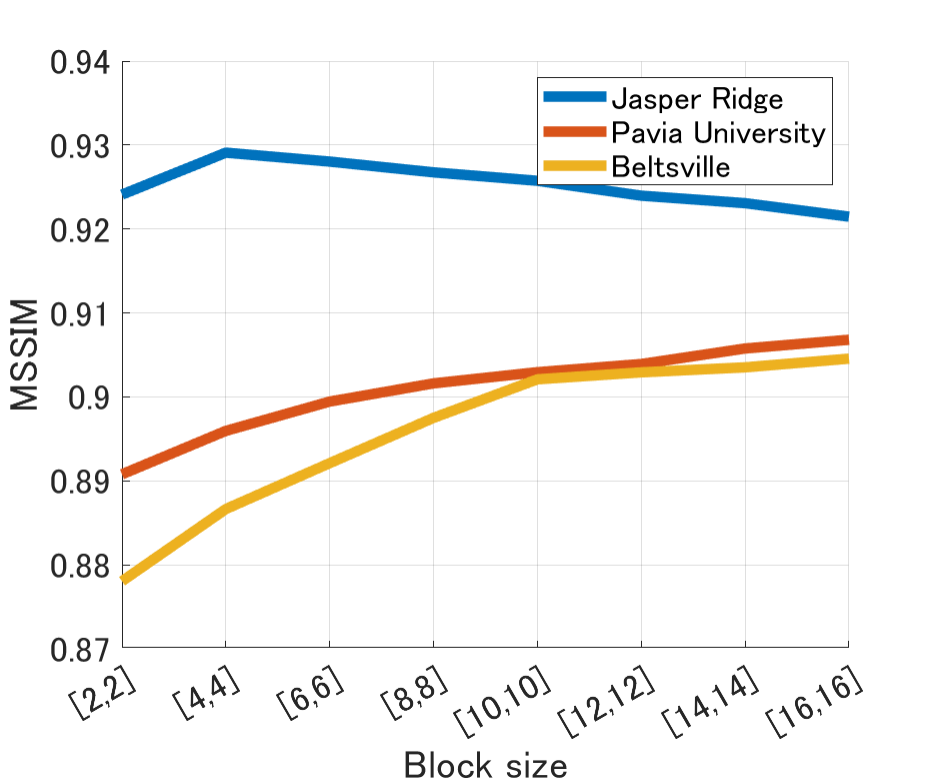}}
        \end{minipage}
        \begin{minipage}{0.240\hsize}
            \centerline{\includegraphics[width=\hsize]{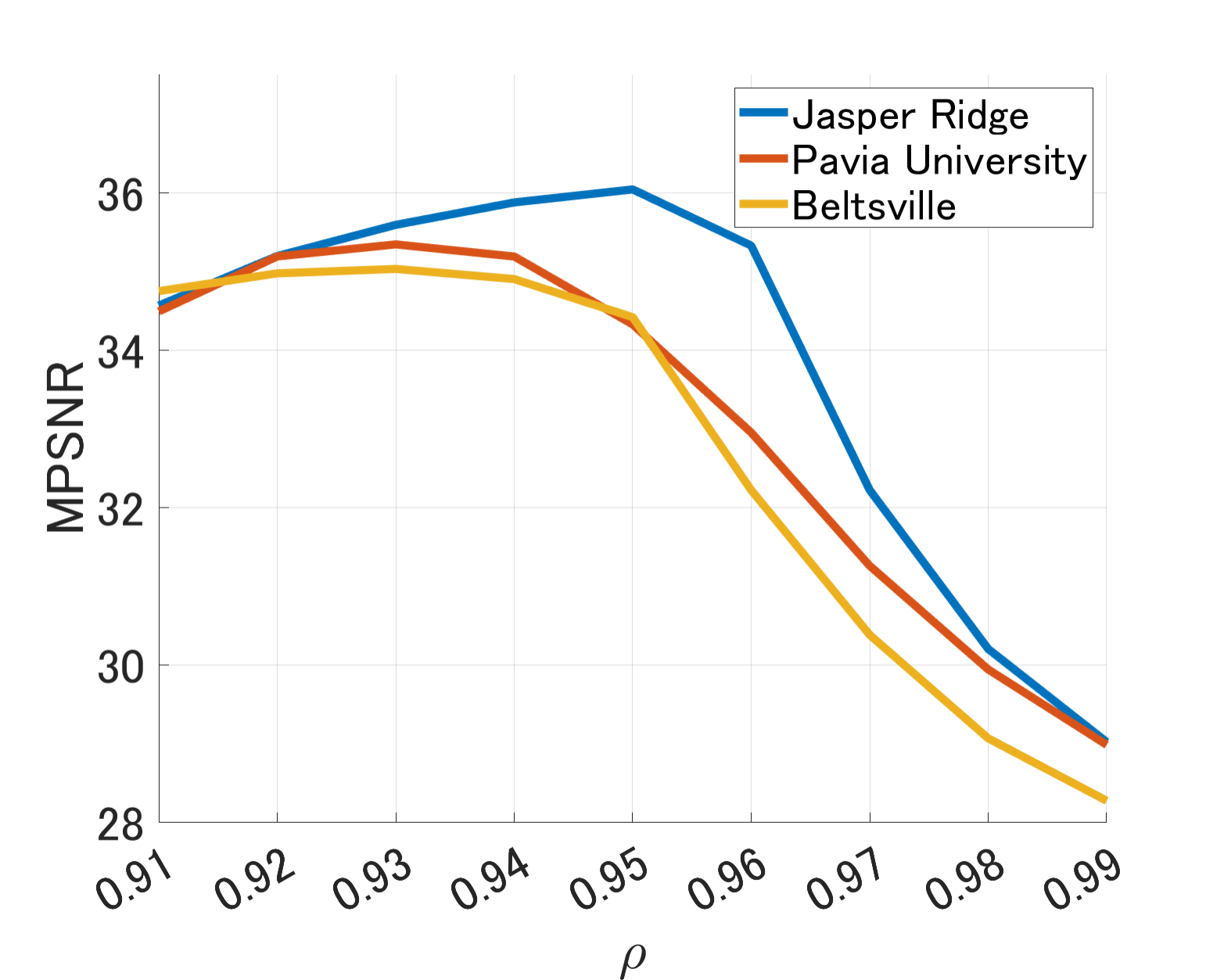}}
        \end{minipage}
        \begin{minipage}{0.240\hsize}
            \centerline{\includegraphics[width=\hsize]{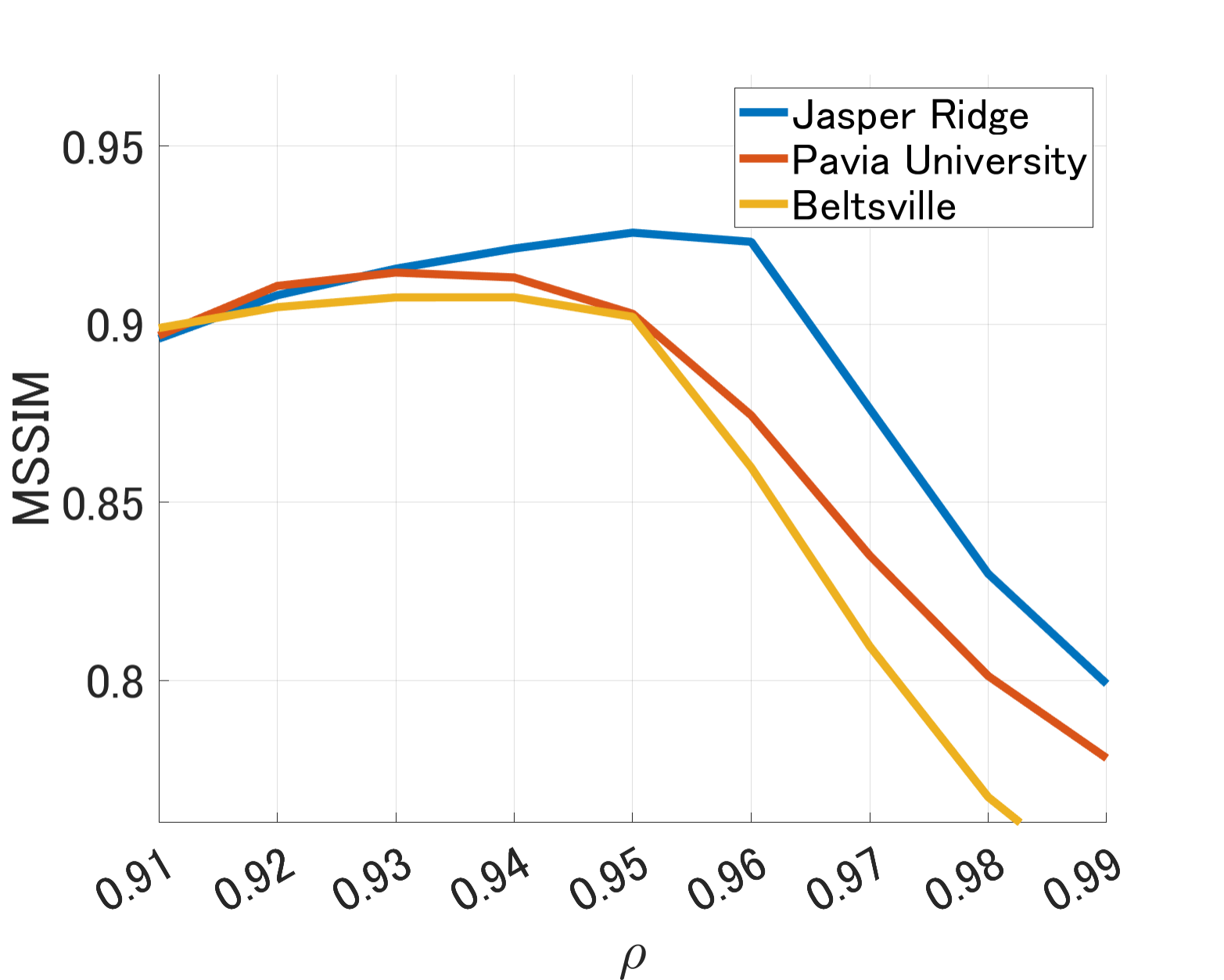}}
        \end{minipage}

        \begin{minipage}{0.240\hsize}
            \centerline{\small{(a)}}
		\end{minipage}
		\begin{minipage}{0.240\hsize}
			\centerline{\small{(b)}}
		\end{minipage}
		\begin{minipage}{0.240\hsize}
			\centerline{\small{(c)}}
		\end{minipage}
		\begin{minipage}{0.240\hsize}
			\centerline{\small{(d)}}
		\end{minipage}
    \end{center}
	
    \vspace{-3mm}
    \caption{Analysis of parameters in the proposed method: block size and $\ParamsRadius$. (a): MPSNR versus block size. (b): MSSIM versus block size. (c): MPSNR versus $\ParamsRadius$. (d): MSSIM versus $\ParamsRadius$. }
    \label{fig:Param_Anal}
\end{figure*}
In the proposed method, there are two key parameters: the block size of the semi-local area and the noise-related parameter $\ParamsRadius$. Fig.~\ref{fig:Param_Anal} shows the relationship between these two parameters and the MPSNRs and MSSIMs. For the block size analysis, $\ParamsRadius$ was fixed at $0.95$, while for the $\ParamsRadius$ analysis, the block size was fixed at $[10, 10]$. Due to hardware limitations, the maximum block size was $[16, 16]$.

In Fig.~\ref{fig:Param_Anal}-(a) and (b), the denoising performance is less sensitive to the block size compared to $\ParamsRadius$. As the block size increases, the performance gradually improves; however, larger block sizes also lead to higher computational costs. Therefore, we recommend using a block size of $[10, 10]$.
In Fig.~\ref{fig:Param_Anal}-(c) and (d), the optimal $\ParamsRadius$ slightly varies depending on the image, but $\ParamsRadius$ in the range of $0.93$ to $0.96$ consistently achieve high performance. In mixed noise conditions, we recommend selecting $\ParamsRadius$ within this range. In this paper, we set the block size to $[10, 10]$ and $\ParamsRadius$ to $0.95$ for all images in the mixed noise condition experiments.

\subsubsection{Running time Analysis}
We measured the actual running times on a Windows 11 computer with an Intel Core i9-13900 1.0 GHz processor, 32GB RAM, and NVIDIA GeForce RTX 4090. For the experiments, all methods except QRNN3D were implemented in MATLAB R2024b. QRNN3D was run in a WSL environment using Python 3.6.13. For SSTV, HSSTV, $\llHTV$, STV, SSST, and $\SSSTTV$, the stopping criterion was set as \eqref{eq:StopCri_simulated}, while for the other methods, the recommended stopping criteria from their respective papers were used.

Table~\ref{tab:Runtime} shows the running times averaged over all noise cases for each image. FastHyMix is the fastest among all methods. The LR-based methods (LRTDTV, FGSLR, and TPTV) are the next fastest. Although the test time of QRNN3D is comparable to these methods, it requires a significant amount of time for pretraining ($4.94 \times 10^{4}$ sec) and fine-tuning ($2.50 \times 10^{3}$ sec). Among the methods based on the same algorithm, the SSTV-based methods i.e., SSTV, HSSTV1, HSSTV2, $\llHTV$ are faster than the STV-based methods i.e., STV, SSST, and $\SSSTTV$. However, the current implementation of the proposed method still has room for computational improvement. The main bottleneck is the calculation of the singular value decomposition used in the proximity operator of the nuclear norm~\ref{eq:prox_nuclear_norm}. This can be accelerated by employing parallel computation or more efficient SVD techniques~\cite{Musco2015Randomized,Struski2024GPUSVD}. In this paper, we adopted a straightforward implementation and prioritized the enhancement of denoising and destriping performance.

\begin{table}[!t]
	\begin{center}
		\caption{Averages of Running Times [sec] in All Noise Cases.}
		\label{tab:Runtime}
			\begin{tabular}{c ccc}
				\toprule
                Methods & Jasper Ridge & Pavia University & Beltsville \\
				\cmidrule(r){1-1}
				\cmidrule(l){2-4}
				SSTV~\cite{Aggarwal2016SSTV}   & $1.92 \times 10^{2}$ & $2.39 \times 10^{2}$ & $2.47 \times 10^{2}$ \\
                HSSTV1~\cite{Takeyama2020HSSTV} & $2.03 \times 10^{2}$ & $2.05 \times 10^{2}$ & $1.81 \times 10^{2}$ \\
				HSSTV2~\cite{Takeyama2020HSSTV} & $6.07 \times 10^{1}$ & $5.97 \times 10^{1}$ & $4.71 \times 10^{1}$ \\
                $\llHTV$~\cite{Wang2021l0l1HTV} & $2.97 \times 10^{2}$ & $2.19 \times 10^{2}$ & $2.13 \times 10^{2}$ \\
                STV~\cite{Lefkimmiatis2015STV} & $1.82 \times 10^{4}$ & $1.83 \times 10^{4}$ & $9.66 \times 10^{3}$ \\
                SSST~\cite{Kurihara2017SSST} & $6.95 \times 10^{4}$ & $1.47 \times 10^{4}$ & $9.01 \times 10^{4}$ \\
                LRTDTV~\cite{Wang2018LRTDTV} & $1.02 \times 10^{1}$ & $7.43 \times 10^{0}$ & $1.15 \times 10^{1}$ \\
                FGSLR~\cite{Chen2022FGSLR} & $2.07 \times 10^{1}$ & $1.71 \times 10^{1}$ & $1.09 \times 10^{1}$ \\
                TPTV~\cite{Chen2023TPTV} & $2.05 \times 10^{1}$ & $1.25 \times 10^{1}$ & $1.30 \times 10^{1}$ \\
                QRNN3D~\cite{Wei2021QRNN3D} & $4.32 \times 10^{2}$ & $4.37 \times 10^{2}$ & $5.41 \times 10^{2}$ \\
                FastHyMix~\cite{Zhuang2023FastHyMix} & $3.43 \times 10^{-2}$ & $5.18 \times 10^{-2}$ & $3.11 \times 10^{-2}$ \\
                $\SSSTTV$ (ours) & $1.37 \times 10^{4}$ & $2.05 \times 10^{4}$ & $1.74 \times 10^{4}$ \\
				\bottomrule
			\end{tabular}
	\end{center}
	\vspace{-3mm}
\end{table}

\subsubsection{Analysis of First-Order and Second-Order Differences for Regularization}
\label{subsubsec:SparsityAnal}
%
%
%

\begin{figure*}[t]
	\begin{center}
		\makebox[0pt][r]{\raisebox{-6mm}{\rotatebox{90}{\shortstack{First-order\\ differences}}} \hspace{2mm}}%
		\begin{minipage}{0.320\hsize}
			\centerline{\includegraphics[width=\hsize]{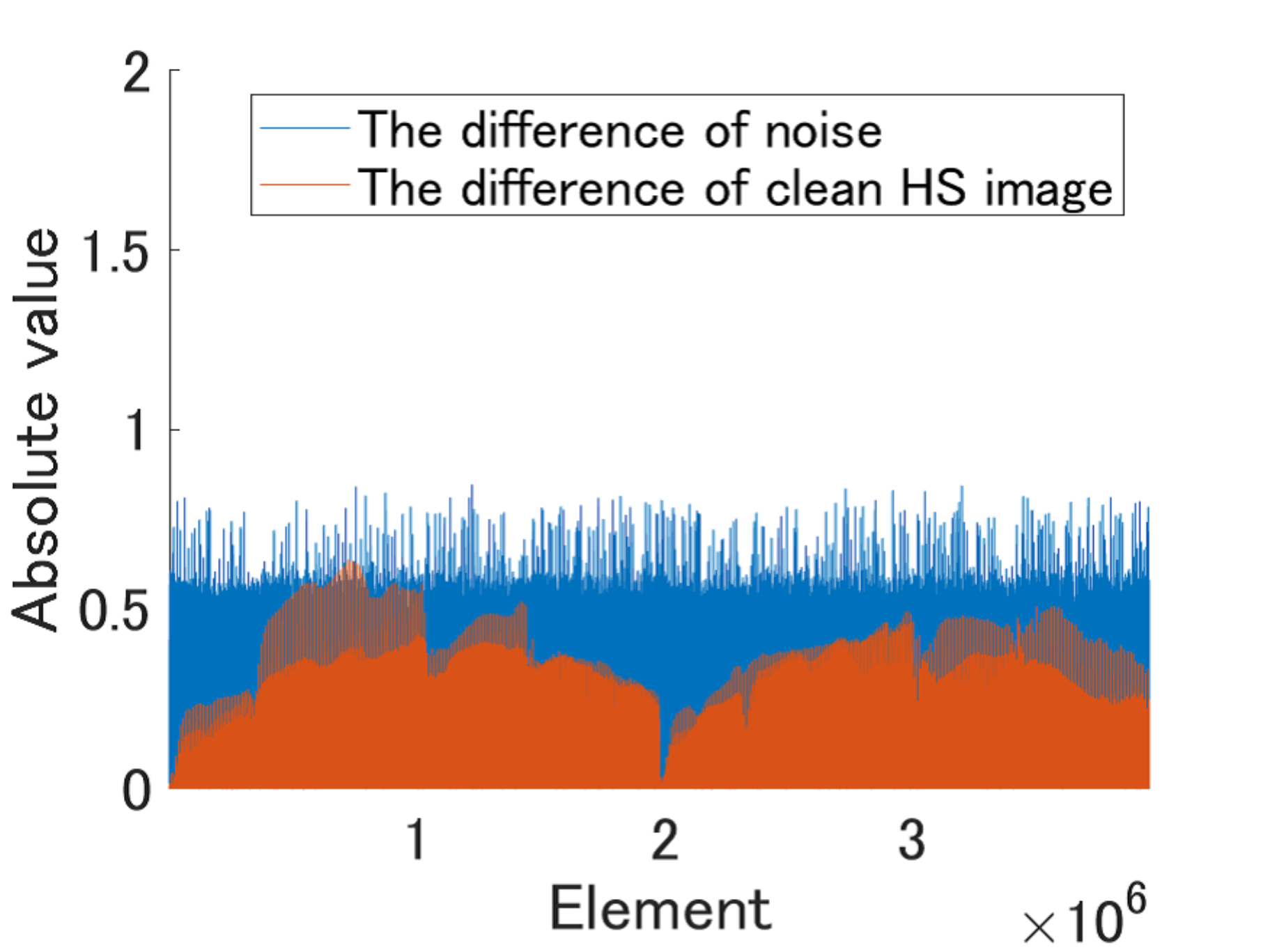}}
		\end{minipage}
		\begin{minipage}{0.320\hsize}
			\centerline{\includegraphics[width=\hsize]{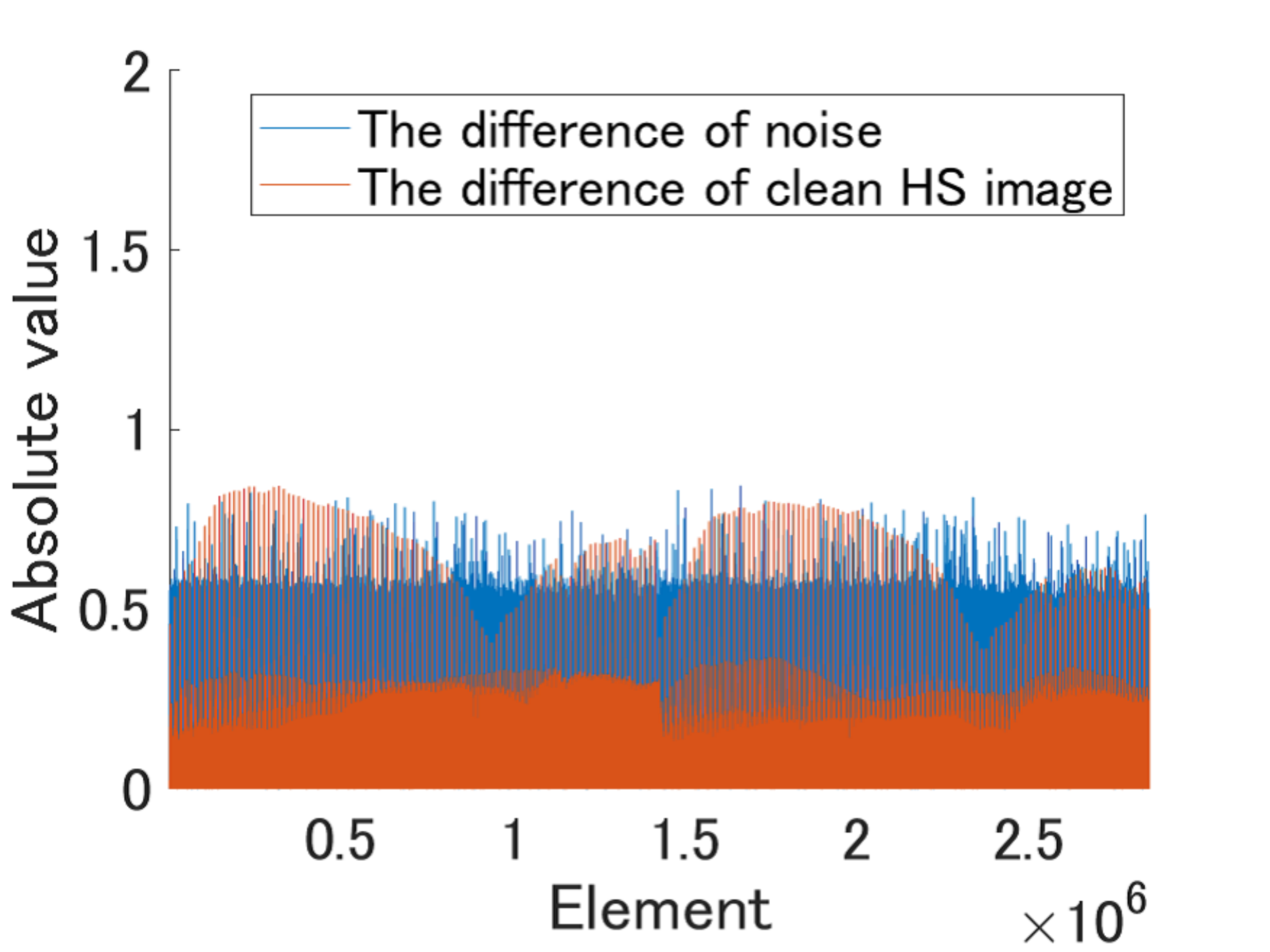}}
		\end{minipage}
		\begin{minipage}{0.320\hsize}
			\centerline{\includegraphics[width=\hsize]{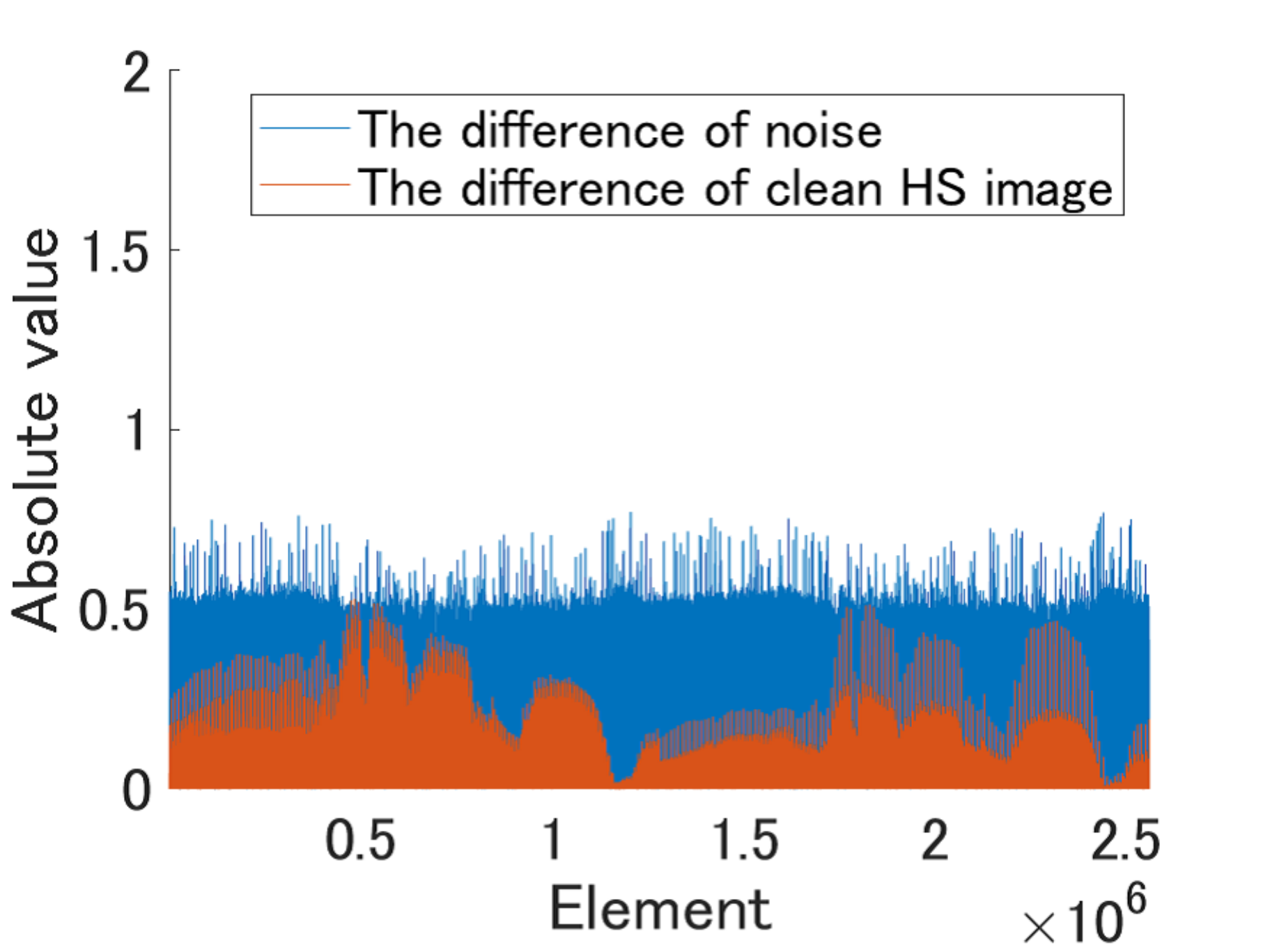}}
		\end{minipage}
		
		\vspace{1mm}
		
		\makebox[0pt][r]{\raisebox{-6mm}{\rotatebox{90}{\shortstack{Second-order\\ differences}}} \hspace{2mm}}%
		\begin{minipage}{0.320\hsize}
			\centerline{\includegraphics[width=\hsize]{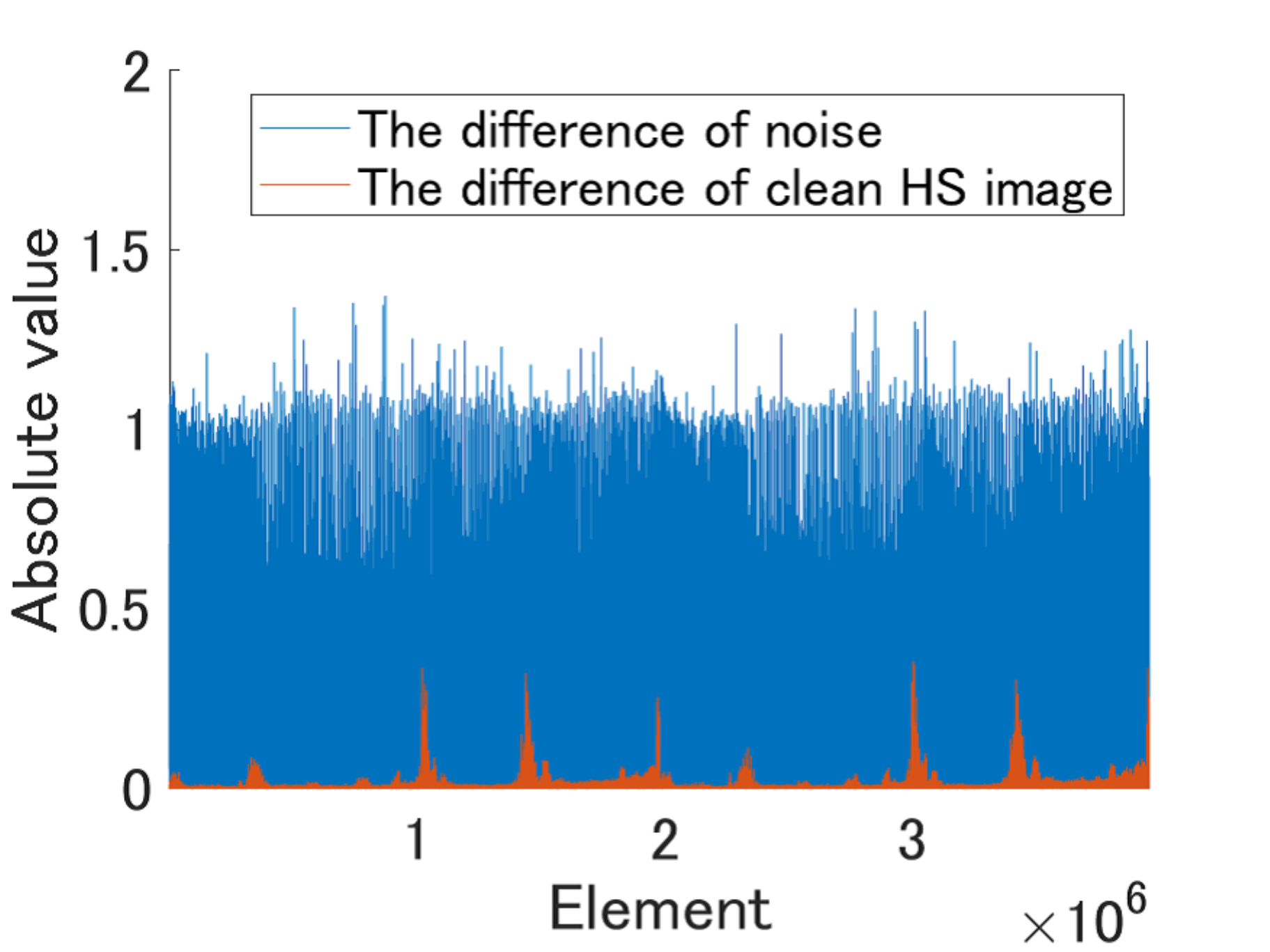}}
		\end{minipage}
		\begin{minipage}{0.320\hsize}
			\centerline{\includegraphics[width=\hsize]{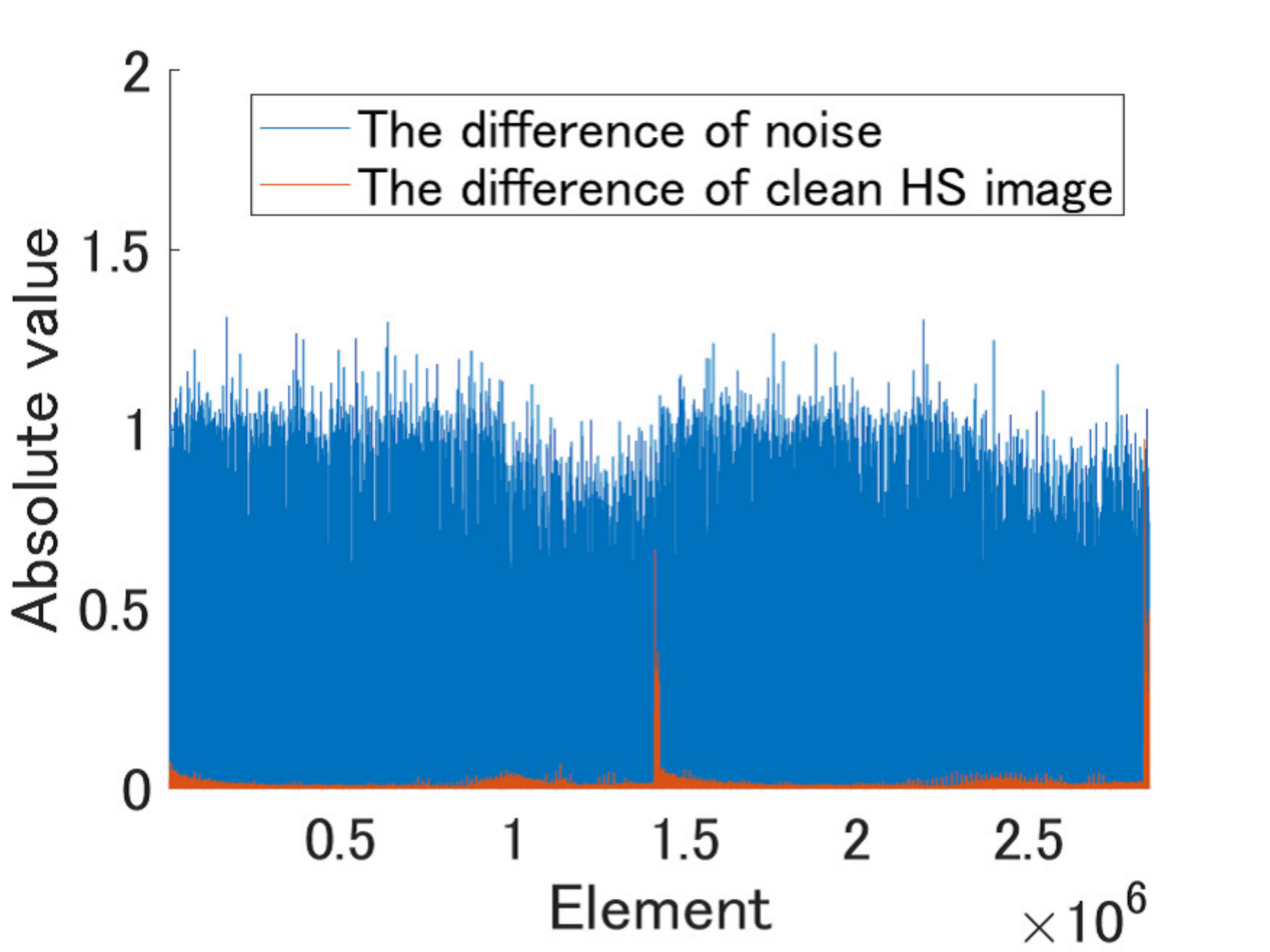}}
		\end{minipage}
		\begin{minipage}{0.320\hsize}
			\centerline{\includegraphics[width=\hsize]{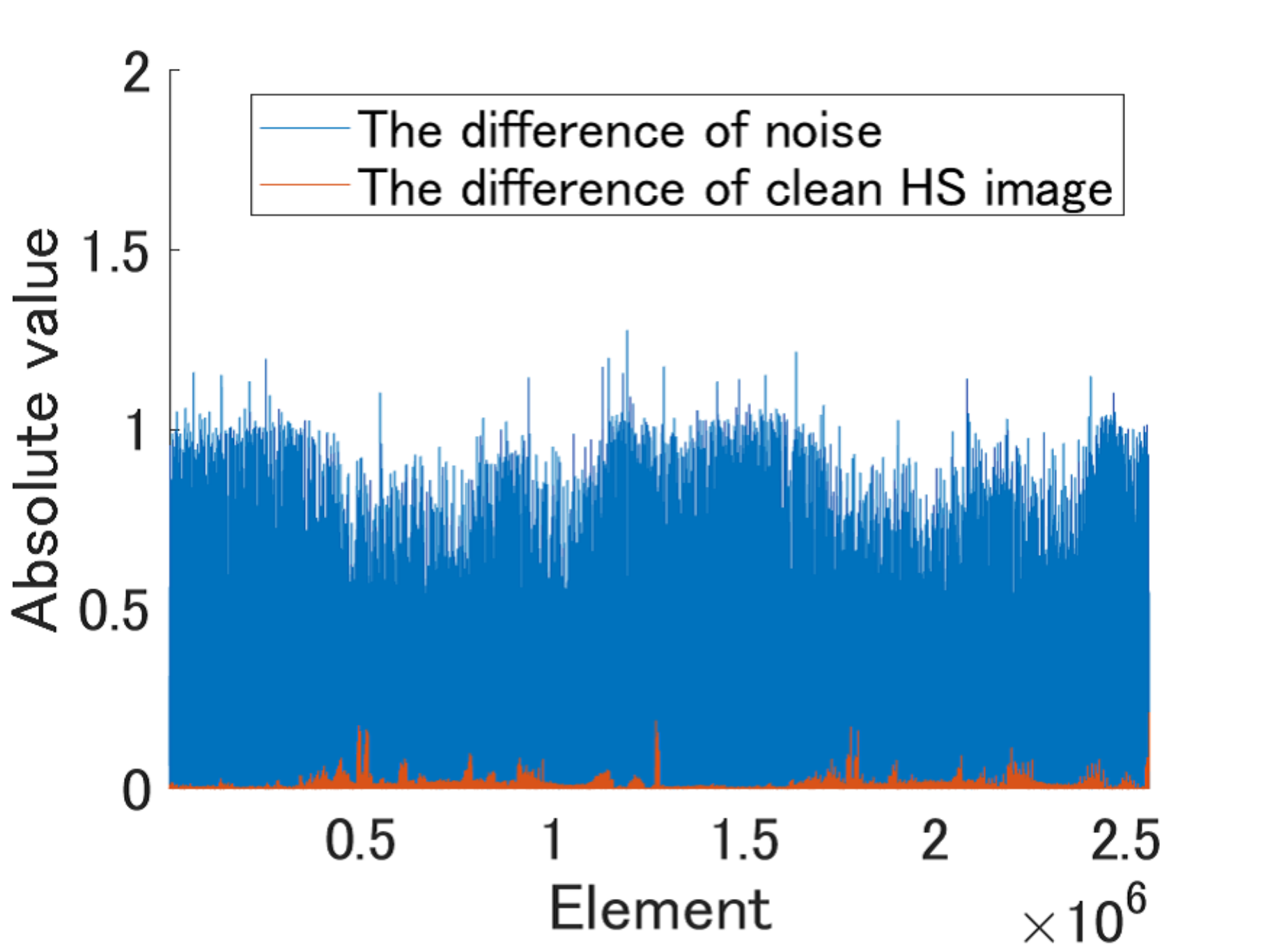}}
		\end{minipage}
		
		\begin{minipage}{0.005\hsize}
			\centerline{\hspace{\hsize}} 
		\end{minipage}
		\begin{minipage}{0.320\hsize}
			\centerline{\small{(a)}}
		\end{minipage}
		\begin{minipage}{0.320\hsize}
			\centerline{\small{(b)}}
		\end{minipage}
		\begin{minipage}{0.320\hsize}
			\centerline{\small{(c)}}
		\end{minipage}
		
	\end{center}
	
	\caption{Absolute values of first-order and second-order differences in each image. (a): Jasper Ridge. (b): Pavia University. (c): Beltsville.}
	
	\label{fig:comparison_sparsity}
\end{figure*}
We experimentally analyze the effectiveness of first-order and second-order differences in our regularization by examining their ability to differentiate noise from clean HS images.
We computed the first-order and second-order differences for clean HS images and noise. The noise condition is Case 8. Fig.~\ref{fig:comparison_sparsity} shows the absolute values of these differences for three types of HS images. The results demonstrate that second-order differences have a higher capability to distinguish noise from clean HS images than first-order differences. This indicates that second-order differences are more appropriate for regularization functions.

\subsection{Summary}
\label{subsec:Expt_summary}
We summarize the insights from the experiments as follows.
\begin{enumerate}
	\item The simulated HS image data experiments demonstrate that $\SSSTTV$ outperforms existing methods in removing mixed noise while preserving edges and textures in HS images with high accuracy. 
	This indicates that $\SSSTTV$ is the most effective in removing mixed noise.
	\item The real HS image data experiments show that $\SSSTTV$ has high performance even when observed HS images are degraded by real noise.
\end{enumerate}


\section{Conclusion}
\label{sec:conclusion}
In this paper, we have proposed a new regularization method, named $\SSSTTV$, for denoising and destriping of HS images. $\SSSTTV$ is defined as the sum of the nuclear norms of matrices consisting of second-order spatio-spectral differences in small spectral blocks, which fully captures the spatial piecewise-smoothness, the spatial similarity between adjacent bands, and the spectral correlation across all bands of HS images. We have formulated the denoising and destriping problem as a constrained convex optimization problem including $\SSSTTV$, and developed the optimization algorithm based on P-PDS. Experiments on HS images with simulated or real noise have demonstrated the superiority of $\SSSTTV$ over existing methods. For future work, we will improve the implementation of the proposed method by accelerating the SVD computations using distributed computing techniques and fast algorithms.

\section*{Appendix}
\subsection*{A. Enhancing Spectral Correlation of differences in HS Images}
To show that improving the spectral correlation of second-order difference enhances the spectral correlation of HS images, we have to show that $\MatHSIClean = \lbrack\HSIClean_{1}^{(\IndexBlock)}, \ldots, \HSIClean_{\NumBand}^{(\IndexBlock)}\rbrack \in \RealSpace{\NumBlockVert \NumBlockHori \times \NumBand}$ is low-rank if $\SmallBlock{\IndexBlock}$ is low-rank. First, if $\SmallBlock{\IndexBlock}$ is low-rank, then $\begin{pmatrix} \lbrack  \DiffOpVert \DiffOpBand \HSIClean \rbrack_{1}^{(\IndexBlock)}, \ldots, \lbrack  \DiffOpVert \DiffOpBand \HSIClean \rbrack_{\NumBand}^{(\IndexBlock)}\end{pmatrix}$ and $\begin{pmatrix} \lbrack  \DiffOpHori \DiffOpBand \HSIClean \rbrack_{1}^{(\IndexBlock)}, \ldots, \lbrack  \DiffOpHori \DiffOpBand \HSIClean \rbrack_{\NumBand}^{(\IndexBlock)}\end{pmatrix}$ are also low-rank. By referring to~\cite{Aggarwal2016SSTV}, we can rewrite them into 
$\DiffOpVert^{\prime} \MatHSIClean \DiffOpBand^{\prime}$ and $\DiffOpHori^{\prime} \MatHSIClean \DiffOpBand^{\prime}$, where $\DiffOpVert^{\prime}$
, $\DiffOpHori^{\prime}$, and $\DiffOpBand^{\prime}$ are the difference operators for the matrix forms of HS images in vertical, horizontal, spectral directions, respectively.
Here, from the the Sylvester’s rank inequality~\cite{Matsaglia1974Equalities}, we can obtain 
\begin{align}
	\rank(\MatHSIClean) \leq 
	& \rank(\DiffOpVert^{\prime} \MatHSIClean \DiffOpBand^{\prime}) + \NumBlockVert \NumBlockHori + \NumBand 
	\nonumber \\
	& - \rank(\DiffOpVert^{\prime}) - \rank(\DiffOpBand^{\prime}),
	\nonumber \\
	\rank(\MatHSIClean) \leq 
	& \rank(\DiffOpHori^{\prime} \MatHSIClean \DiffOpBand^{\prime}) + \NumBlockVert \NumBlockHori + \NumBand 
	\nonumber \\
	& - \rank(\DiffOpHori^{\prime}) - \rank(\DiffOpBand^{\prime}).
\end{align}
Since $\rank(\DiffOpVert^{\prime}) = \NumBlockVert \NumBlockHori - 1$, $\rank(\DiffOpHori^{\prime}) = \NumBlockVert \NumBlockHori - 1$, and $\rank(\DiffOpBand^{\prime}) = \NumBand-1$, we have
\begin{align}
	& \rank(\MatHSIClean) <= \rank(\DiffOpVert^{\prime} \MatHSIClean \DiffOpBand^{\prime}) + 2,
	\nonumber \\
	& \rank(\MatHSIClean) <= \rank(\DiffOpHori^{\prime} \MatHSIClean \DiffOpBand^{\prime}) + 2.
\end{align}
These inequalities indicate that if $\DiffOpVert^{\prime} \MatHSIClean \DiffOpBand^{\prime}$ and $\DiffOpHori^{\prime} \MatHSIClean \DiffOpBand^{\prime}$ are low rank, $\MatHSIClean$ is also low rank.
Therefore, minimizing the nuclear norms of $\SmallBlock{1}, \ldots, \SmallBlock{\NumBlock}$ enhances the spectral correlation of HS images.

\vspace{11pt}


\begin{IEEEbiography}[{\includegraphics[width=1in,height=1.25in,clip,keepaspectratio]{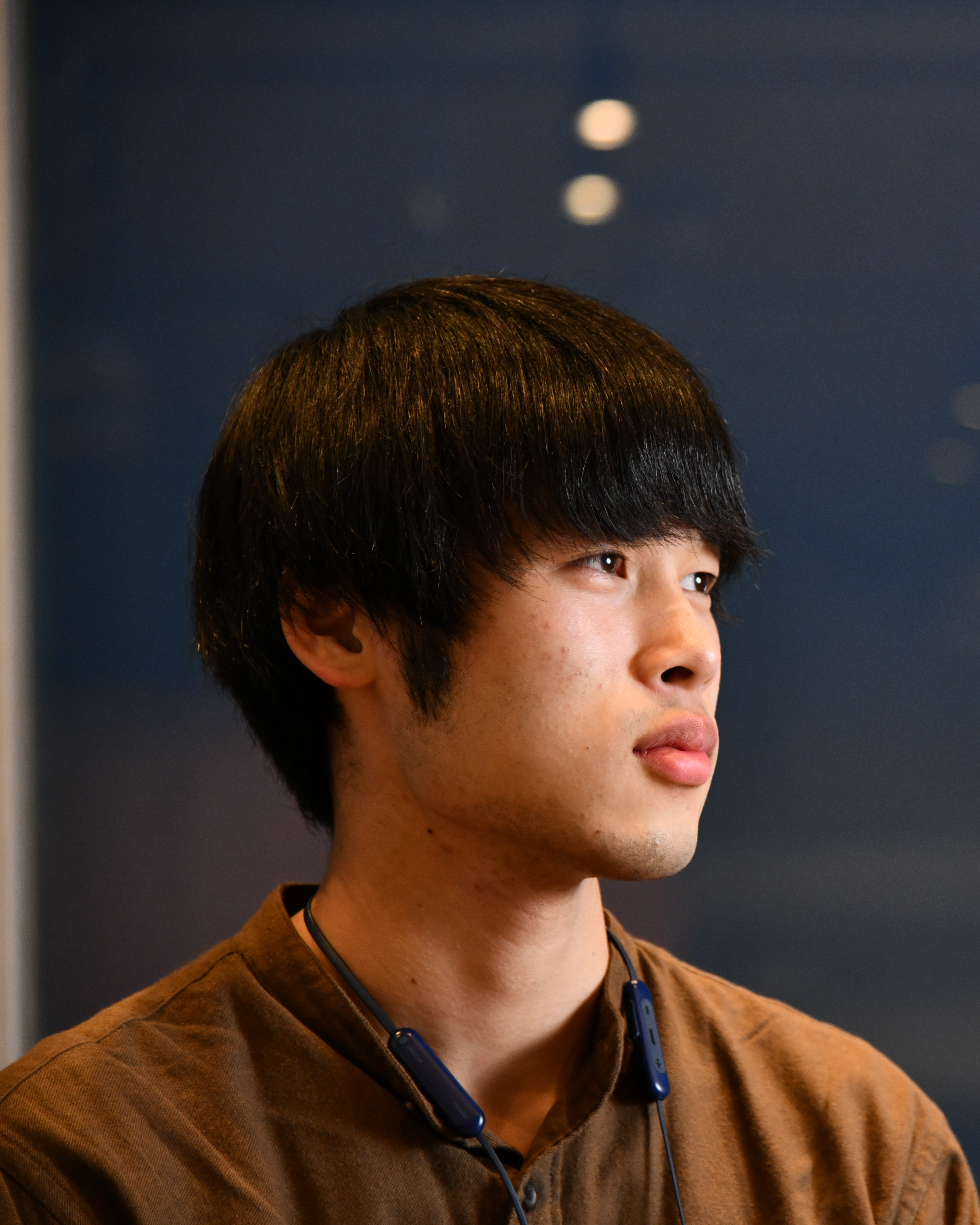}}]{Shingo Takemoto}
    (S’22) received a B.E. degree and M.E. degree in Information and Computer Sciences in 2021 from Sophia University and in 2023 from Institute of Science Tokyo (formerly Tokyo Institute of Technology), respectively. He is currently pursuing a Ph.D. degree at the Department of Computer Science in the Institute of Science Tokyo. Since April 2024, he has been a Research Fellow (DC2) of Japan Society for the Promotion of Science (JSPS). His current research interests are in signal and image processing and optimization theory. Mr. Takemoto received the Student Award from IEEE SPS Tokyo Joint Chapter in 2022.
\end{IEEEbiography}

\begin{IEEEbiography}[{\includegraphics[width=1in,height=1.25in,clip,keepaspectratio]{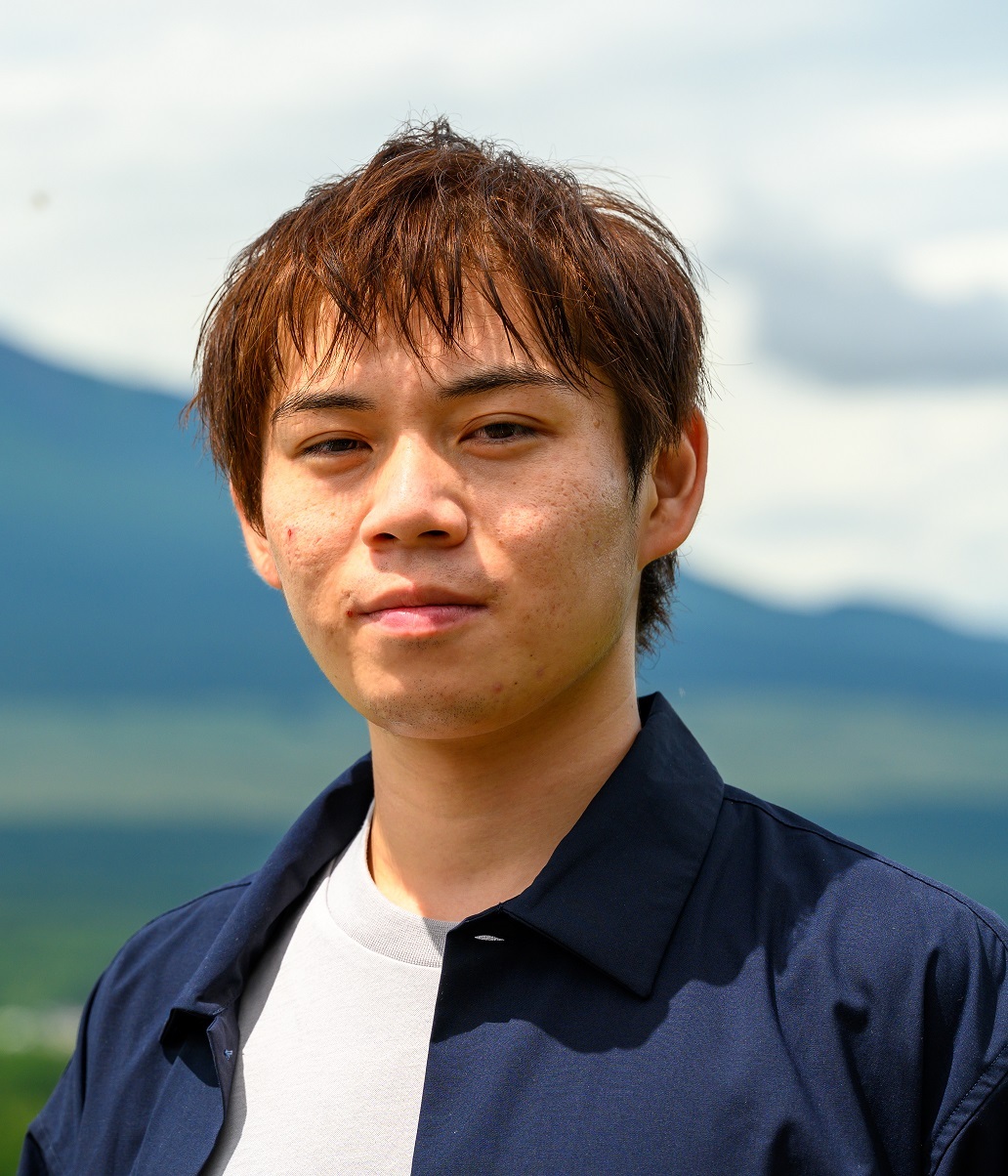}}]{Kazuki Naganuma} (S’21) received a B.E. degree in 2020 from the Kanagawa Institute of Technology and M.E. and Ph.D. degrees in Information and Computer Sciences 2022 and 2024 from the Tokyo Institute of Technology, respectively.
He is an assistant professor at the Institute of Engineering of Tokyo University of Agriculture and Technology.
From April 2023 to March 2025, he was a Research Fellow (DC2) of the Japan Society for the Promotion of Science (JSPS). From October 2023 and April 2025 to present, He is a Research Fellow (PD) of JSPS and a Researcher of ACT-X of the Japan Science and Technology Corporation (JST), Tokyo, Japan.
His current research interests are in signal and image processing and optimization theory.
Dr. Naganuma received the Student Conference Paper Award from IEEE SPS Japan Chapter in 2022, the 38th TELECOM System Technology Student Award from the Telecommunications Advancement Foundation, and the Best Paper Award in APSIPA ASC 2024.
\end{IEEEbiography}

\begin{IEEEbiography}[{\includegraphics[width=1in,height=1.25in,clip,keepaspectratio]{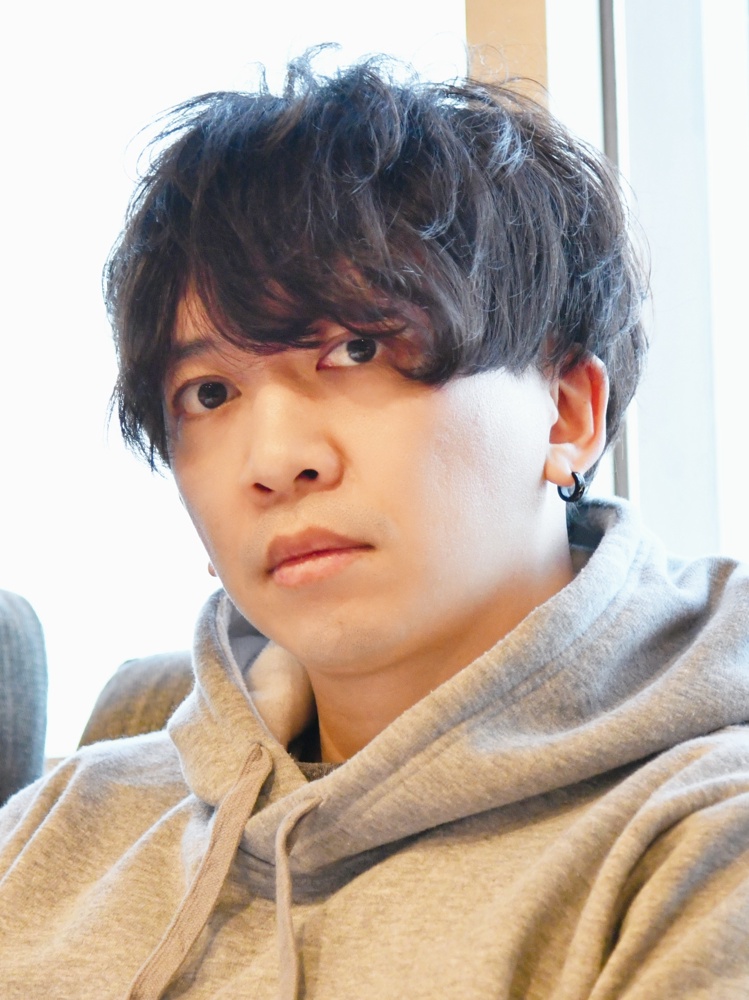}}]{Shunsuke Ono}
(S’11–M’15–SM'23) received a B.E. degree in Computer Science in 2010 and M.E. and Ph.D. degrees in Communications and Computer Engineering in 2012 and 2014 from the Tokyo Institute of Technology, respectively. From 2012 to 2014, he was a Research Fellow (DC1) of the Japan Society for the Promotion of Science (JSPS). He was an Assistant, then an Associate Professor with Tokyo Institute of Technology (TokyoTech), Tokyo, Japan, from 2014 to 2024. From 2016 to 2020, he was a Researcher of Precursory Research for Embryonic Science and Technology (PRESTO), Japan Science and Technology Agency (JST), Tokyo, Japan. Currently, he is an Associate Professor with Institute of Science Tokyo (Science Tokyo), Tokyo, Japan. His research interests include signal processing, image analysis, optimization, remote sensing, and measurement informatics. He has served as an Associate Editor for IEEE TRANSACTIONS ON SIGNAL AND INFORMATION PROCESSING OVER NETWORKS (2019--2024). Dr. Ono was a recipient of the Young Researchers’ Award and the Excellent Paper Award from the IEICE in 2013 and 2014, respectively, the Outstanding Student Journal Paper Award and the Young Author Best Paper Award from the IEEE SPS Japan Chapter in 2014 and 2020, respectively, and the Best Paper Award in APSIPA ASC 2024. He also received the Funai Research Award in 2017, the Ando Incentive Prize in 2021, the MEXT Young Scientists’ Award in 2022, and the IEEE SPS Outstanding Editorial Board Member Award in 2023. 
\end{IEEEbiography}

\vfill


\begin{thebibliography}{10}
\providecommand{\url}[1]{#1}
\csname url@samestyle\endcsname
\providecommand{\newblock}{\relax}
\providecommand{\bibinfo}[2]{#2}
\providecommand{\BIBentrySTDinterwordspacing}{\spaceskip=0pt\relax}
\providecommand{\BIBentryALTinterwordstretchfactor}{4}
\providecommand{\BIBentryALTinterwordspacing}{\spaceskip=\fontdimen2\font plus
\BIBentryALTinterwordstretchfactor\fontdimen3\font minus \fontdimen4\font\relax}
\providecommand{\BIBforeignlanguage}[2]{{%
\expandafter\ifx\csname l@#1\endcsname\relax
\typeout{** WARNING: IEEEtran.bst: No hyphenation pattern has been}%
\typeout{** loaded for the language `#1'. Using the pattern for}%
\typeout{** the default language instead.}%
\else
\language=\csname l@#1\endcsname
\fi
#2}}
\providecommand{\BIBdecl}{\relax}
\BIBdecl

\bibitem{Borengasser2007HSIApplications}
M.~Borengasser, W.~S. Hungate, and R.~Watkins, \emph{Hyperspectral remote sensing: principles and applications}.\hskip 1em plus 0.5em minus 0.4em\relax CRC press, 2007.

\bibitem{Grahn2007Techniques}
H.~Grahn and P.~Geladi, \emph{Techniques and applications of hyperspectral image analysis}.\hskip 1em plus 0.5em minus 0.4em\relax John Wiley \& Sons, 2007.

\bibitem{Thenkabail2016VegetationOverview}
P.~S. Thenkabail, J.~G. Lyon, and A.~Huete, \emph{Hyperspectral remote sensing of vegetation}.\hskip 1em plus 0.5em minus 0.4em\relax CRC press, 2016.

\bibitem{Lu2020AgricultureOverview}
B.~Lu, P.~D. Dao, J.~Liu, Y.~He, and J.~Shang, ``Recent advances of hyperspectral imaging technology and applications in agriculture,'' \emph{Remote Sens.}, vol.~12, no.~16, p. 2659, 2020.

\bibitem{Shen2015DenoisingOverview}
H.~Shen, X.~Li, Q.~Cheng, C.~Zeng, G.~Yang, H.~Li, and L.~Zhang, ``Missing information reconstruction of remote sensing data: A technical review,'' \emph{IEEE Geosci. Remote Sens. Mag.}, vol.~3, no.~3, pp. 61--85, 2015.

\bibitem{Rasti2018DenoisingOverview}
B.~Rasti, P.~Scheunders, P.~Ghamisi, G.~Licciardi, and J.~Chanussot, ``Noise reduction in hyperspectral imagery: Overview and application,'' \emph{Remote Sens.}, vol.~10, no.~3, 2018.

\bibitem{Shen2022DenoisingOverview}
H.~Shen, M.~Jiang, J.~Li, C.~Zhou, Q.~Yuan, and L.~Zhang, ``Coupling model- and data-driven methods for remote sensing image restoration and fusion: Improving physical interpretability,'' \emph{IEEE Geosci. Remote Sens. Mag.}, vol.~10, no.~2, pp. 231--249, 2022.

\bibitem{Bioucas-Dias2012UnmixingOverview}
J.~M. Bioucas-Dias, A.~Plaza, N.~Dobigeon, M.~Parente, Q.~Du, P.~Gader, and J.~Chanussot, ``Hyperspectral unmixing overview: Geometrical, statistical, and sparse regression-based approaches,'' \emph{IEEE J. Sel. Topics Appl. Earth Observ. Remote Sens.}, vol.~5, no.~2, pp. 354--379, 2012.

\bibitem{Ma2014UnmixingOverview}
W.~Ma, J.~M. Bioucas-Dias, T.~Chan, N.~Gillis, P.~Gader, A.~J. Plaza, A.~Ambikapathi, and C.~Chi, ``A signal processing perspective on hyperspectral unmixing: Insights from remote sensing,'' \emph{IEEE Signal Process. Mag.}, vol.~31, no.~1, pp. 67--81, 2014.

\bibitem{Ghamisi2017Classification}
P.~Ghamisi, J.~Plaza, Y.~Chen, J.~Li, and A.~J. Plaza, ``Advanced spectral classifiers for hyperspectral images: A review,'' \emph{IEEE Geosci. Remote Sens. Mag.}, vol.~5, no.~1, pp. 8--32, 2017.

\bibitem{Li2019Classification}
S.~Li, W.~Song, L.~Fang, Y.~Chen, P.~Ghamisi, and J.~A. Benediktsson, ``Deep learning for hyperspectral image classification: An overview,'' \emph{IEEE Trans. Geosci. Remote Sens.}, vol.~57, no.~9, pp. 6690--6709, 2019.

\bibitem{Nicolas2019Classification}
N.~Audebert, B.~L. Saux, and S.~Lefevre, ``Deep learning for classification of hyperspectral data: A comparative review,'' \emph{IEEE Geosci. Remote Sens. Mag.}, vol.~7, no.~2, pp. 159--173, 2019.

\bibitem{Matteoli2014Anomaly}
S.~Matteoli, M.~Diani, and J.~Theiler, ``An overview of background modeling for detection of targets and anomalies in hyperspectral remotely sensed imagery,'' \emph{IEEE J. Sel. Topics Appl. Earth Observ. Remote Sens.}, vol.~7, no.~6, pp. 2317--2336, 2014.

\bibitem{Su2022Anomaly}
H.~Su, Z.~Wu, H.~Zhang, and Q.~Du, ``Hyperspectral anomaly detection: A survey,'' \emph{IEEE Geosci. Remote Sens. Mag.}, vol.~10, no.~1, pp. 64--90, 2022.

\bibitem{Yuan2019HSIDCNN}
Q.~Yuan, Q.~Zhang, J.~Li, H.~Shen, and L.~Zhang, ``Hyperspectral image denoising employing a spatial–spectral deep residual convolutional neural network,'' \emph{IEEE Trans. Geosci. Remote Sens.}, vol.~57, no.~2, pp. 1205--1218, 2019.

\bibitem{Wang2022NL3DCNN}
Z.~Wang, M.~K. Ng, L.~Zhuang, L.~Gao, and B.~Zhang, ``Nonlocal self-similarity-based hyperspectral remote sensing image denoising with 3-d convolutional neural network,'' \emph{IEEE Trans. Geosci. Remote Sens.}, vol.~60, pp. 1--17, 2022.

\bibitem{Wei2021QRNN3D}
K.~Wei, Y.~Fu, and H.~Huang, ``3-\text{D} quasi-recurrent neural network for hyperspectral image denoising,'' \emph{IEEE Trans. Neural Netw. Learn. Syst.}, vol.~32, no.~1, pp. 363--375, 2021.

\bibitem{Li2023SST}
M.~Li, Y.~Fu, and Y.~Zhang, ``Spatial-spectral transformer for hyperspectral image denoising,'' in \emph{AAAI Conf. Artif. Intell. (AAAI)}, vol.~37, no.~1, 2023, pp. 1368--1376.

\bibitem{Zhuang2023FastHyMix}
L.~Zhuang and M.~K. Ng, ``{FastHyMix}: Fast and parameter-free hyperspectral image mixed noise removal,'' \emph{IEEE Trans. Neural Netw. Learn. Syst.}, vol.~34, no.~8, pp. 4702--4716, 2023.

\bibitem{Peng2024RCILD}
J.~Peng, H.~Wang, X.~Cao, Q.~Zhao, J.~Yao, H.~Zhang, and D.~Meng, ``Learnable representative coefficient image denoiser for hyperspectral image,'' \emph{IEEE Trans. Geosci. Remote Sens.}, vol.~62, pp. 1--16, 2024.

\bibitem{Xue2017Robust}
Z.~Xue, P.~Du, J.~Li, and H.~Su, ``Sparse graph regularization for hyperspectral remote sensing image classification,'' \emph{IEEE Trans. Geosci. Remote Sens.}, vol.~55, no.~4, pp. 2351--2366, 2017.

\bibitem{Jiang2022Adaptive}
T.-X. Jiang, L.~Zhuang, T.-Z. Huang, X.-L. Zhao, and J.~M. Bioucas-Dias, ``Adaptive hyperspectral mixed noise removal,'' \emph{IEEE Trans. Geosci. Remote Sens.}, vol.~60, pp. 1--13, 2022.

\bibitem{Naganuma2024Unmixing}
K.~Naganuma and S.~Ono, ``Toward robust hyperspectral unmixing: Mixed noise modeling and image-domain regularization,'' \emph{IEEE J. Sel. Topics Appl. Earth Observ. Remote Sens.}, vol.~17, pp. 8117--8138, 2024.

\bibitem{Qian20123DNLM}
Y.~Qian, Y.~Shen, M.~Ye, and Q.~Wang, ``3-\text{D} nonlocal means filter with noise estimation for hyperspectral imagery denoising,'' in \emph{Proc. IEEE Int. Geosci. Remote Sens. Symp.}, 2012, pp. 1345--1348.

\bibitem{Maggioni2013BM3D}
M.~Maggioni, V.~Katkovnik, K.~Egiazarian, and A.~Foi, ``Nonlocal transform-domain filter for volumetric data denoising and reconstruction,'' \emph{IEEE Trans. Image Process.}, vol.~22, no.~1, pp. 119--133, 2013.

\bibitem{Chen2014BM4D}
G.~Chen, T.~D. Bui, K.~G. Quach, and S.-E. Qian, ``Denoising hyperspectral imagery using principal component analysis and block-matching 4d filtering,'' \emph{Can. J. Remote Sens.}, vol.~40, no.~1, pp. 60--66, 2014.

\bibitem{He2019NGmeet}
W.~He, Q.~Yao, C.~Li, N.~Yokoya, and Q.~Zhao, ``Non-local meets global: An integrated paradigm for hyperspectral denoising,'' in \emph{Proc. IEEE Conf. Comput. Vis. Pattern Recognit. (CVPR)}, 2019, pp. 6861--6870.

\bibitem{Rudin1992TV}
L.~I. Rudin, S.~Osher, and E.~Fatemi, ``Nonlinear total variation based noise removal algorithms,'' \emph{Physica D: Nonlinear Phenomena}, vol.~60, no.~1, pp. 259--268, 1992.

\bibitem{Bresson2008TV}
X.~Bresson and T.~F. Chan, ``Fast dual minimization of the vectorial total variation norm and applications to color image processing,'' \emph{Inverse problems imag.}, vol.~2, no.~4, pp. 455--484, 2008.

\bibitem{Yuan2012HTV}
Q.~Yuan, L.~Zhang, and H.~Shen, ``Hyperspectral image denoising employing a spectral–spatial adaptive total variation model,'' \emph{IEEE Trans. Geosci. Remote Sens.}, vol.~50, no.~10, pp. 3660--3677, 2012.

\bibitem{Chang2015ASSTV}
Y.~Chang, L.~Yan, H.~Fang, and C.~Luo, ``Anisotropic spectral-spatial total variation model for multispectral remote sensing image destriping,'' \emph{IEEE Trans. Image Process.}, vol.~24, no.~6, pp. 1852--1866, 2015.

\bibitem{Aggarwal2016SSTV}
H.~K. Aggarwal and A.~Majumdar, ``Hyperspectral image denoising using spatio-spectral total variation,'' \emph{IEEE Geosci. Remote Sens. Lett.}, vol.~13, no.~3, pp. 442--446, 2016.

\bibitem{Fan2018SSTV-LRTF}
H.~Fan, C.~Li, Y.~Guo, G.~Kuang, and J.~Ma, ``Spatial–spectral total variation regularized low-rank tensor decomposition for hyperspectral image denoising,'' \emph{IEEE Trans. Geosci. Remote Sens.}, vol.~56, no.~10, pp. 6196--6213, 2018.

\bibitem{Ince2019GLSSTV}
T.~Ince, ``Hyperspectral image denoising using group low-rank and spatial-spectral total variation,'' \emph{IEEE Access}, vol.~7, pp. 52\,095--52\,109, 2019.

\bibitem{Takeyama2020HSSTV}
S.~Takeyama, S.~Ono, and I.~Kumazawa, ``A constrained convex optimization approach to hyperspectral image restoration with hybrid spatio-spectral regularization,'' \emph{Remote Sens.}, vol.~12, no.~21, 2020.

\bibitem{Wang2021l0l1HTV}
M.~Wang, Q.~Wang, J.~Chanussot, and D.~Hong, ``$l_0$-$l_1$ hybrid total variation regularization and its applications on hyperspectral image mixed noise removal and compressed sensing,'' \emph{IEEE Trans. Geosci. Remote Sens.}, vol.~59, no.~9, pp. 7695--7710, 2021.

\bibitem{Takemoto2022GSSTV}
S.~Takemoto, K.~Naganuma, and S.~Ono, ``Graph spatio-spectral total variation model for hyperspectral image denoising,'' \emph{IEEE Geosci. Remote Sens. Lett.}, vol.~19, pp. 1--5, 2022.

\bibitem{Ono2017l0Gradient}
S.~Ono, ``$\mathrm{L}_{0}$ gradient projection,'' \emph{IEEE Trans. Image Process.}, vol.~26, no.~4, pp. 1554--1564, 2017.

\bibitem{Zhang2014LRMR}
H.~Zhang, W.~He, L.~Zhang, H.~Shen, and Q.~Yuan, ``Hyperspectral image restoration using low-rank matrix recovery,'' \emph{IEEE Trans. Geosci. Remote Sens.}, vol.~52, no.~8, pp. 4729--4743, 2014.

\bibitem{Chen2022FGSLR}
Y.~Chen, T.~Huang, W.~He, X.~Zhao, H.~Zhang, and J.~Zeng, ``Hyperspectral image denoising using factor group sparsity-regularized nonconvex low-rank approximation,'' \emph{IEEE Trans. Geosci. Remote Sens.}, vol.~60, pp. 1--16, 2022.

\bibitem{He2016LRTV}
W.~He, H.~Zhang, L.~Zhang, and H.~Shen, ``Total-variation-regularized low-rank matrix factorization for hyperspectral image restoration,'' \emph{IEEE Trans. Geosci. Remote Sens.}, vol.~54, no.~1, pp. 178--188, 2016.

\bibitem{Chen2023TPTV}
Y.~Chen, W.~Cao, L.~Pang, J.~Peng, and X.~Cao, ``Hyperspectral image denoising via texture-preserved total variation regularizer,'' \emph{IEEE Trans. Geosci. Remote Sens.}, vol.~61, pp. 1--14, 2023.

\bibitem{Xue2022Tensor1}
J.~Xue, Y.~Zhao, S.~Huang, W.~Liao, J.~C.-W. Chan, and S.~G. Kong, ``Multilayer sparsity-based tensor decomposition for low-rank tensor completion,'' \emph{IEEE Trans. Neural Netw. Learn. Syst.}, vol.~33, no.~11, pp. 6916--6930, 2022.

\bibitem{Xue2022Tensor2}
J.~Xue, Y.~Zhao, Y.~Bu, J.~C.-W. Chan, and S.~G. Kong, ``When laplacian scale mixture meets three-layer transform: A parametric tensor sparsity for tensor completion,'' \emph{IEEE Trans. Cybern.}, vol.~52, no.~12, pp. 13\,887--13\,901, 2022.

\bibitem{Xue2024Tensor}
J.~Xue, Y.~Q. Zhao, T.~Wu, and J.~C.~W. Chan, ``Tensor convolution-like low-rank dictionary for high-dimensional image representation,'' \emph{IEEE Trans. Circuits Syst. Video Technol.}, vol.~34, no.~12, pp. 13\,257--13\,270, 2024.

\bibitem{Wang2018LRTDTV}
Y.~Wang, J.~Peng, Q.~Zhao, Y.~Leung, X.~Zhao, and D.~Meng, ``Hyperspectral image restoration via total variation regularized low-rank tensor decomposition,'' \emph{IEEE J. Sel. Topics Appl. Earth Observ. Remote Sens.}, vol.~11, no.~4, pp. 1227--1243, 2018.

\bibitem{Chen2020LRTDGS}
Y.~Chen, W.~He, N.~Yokoya, and T.-Z. Huang, ``Hyperspectral image restoration using weighted group sparsity-regularized low-rank tensor decomposition,'' \emph{IEEE Trans. Cybern.}, vol.~50, no.~8, pp. 3556--3570, 2020.

\bibitem{Sun2022Tensor}
L.~Sun and C.~He, ``Hyperspectral image mixed denoising using difference continuity-regularized nonlocal tensor subspace low-rank learning,'' \emph{IEEE Geosci. Remote Sens. Lett.}, vol.~19, pp. 1--5, 2022.

\bibitem{Li2024LRTDAHL}
D.~Li, D.~Chu, X.~Guan, W.~He, and H.~Shen, ``Adaptive regularized low-rank tensor decomposition for hyperspectral image denoising and destriping,'' \emph{IEEE Trans. Geosci. Remote Sens.}, vol.~62, pp. 1--17, 2024.

\bibitem{Lefkimmiatis2015STV}
S.~Lefkimmiatis, A.~Roussos, P.~Maragos, and M.~Unser, ``Structure tensor total variation,'' \emph{SIAM J. Imag. Sci.}, vol.~8, no.~2, pp. 1090--1122, 2015.

\bibitem{Wu2017STWNNM}
Z.~Wu, Q.~Wang, J.~Jin, and Y.~Shen, ``Structure tensor total variation-regularized weighted nuclear norm minimization for hyperspectral image mixed denoising,'' \emph{Signal Process.}, vol. 131, pp. 202--219, 2017.

\bibitem{Ono2016ASTV}
S.~Ono, K.~Shirai, and M.~Okuda, ``Vectorial total variation based on arranged structure tensor for multichannel image restoration,'' in \emph{Proc. IEEE Int. Conf. Acoust., Speech, Signal Process. (ICASSP)}, 2016, pp. 4528--4532.

\bibitem{Kurihara2017SSST}
R.~Kurihara, S.~Ono, K.~Shirai, and M.~Okuda, ``Hyperspectral image restoration based on spatio-spectral structure tensor regularization,'' in \emph{Proc. Eur. Signal Process. Conf. (EUSIPCO)}, 2017, pp. 488--492.

\bibitem{Forstner1987Fast}
W.~F{\"o}rstner and E.~G{\"u}lch, ``A fast operator for detection and precise location of distinct points, corners and centres of circular features,'' in \emph{Proc. ISPRS Intercommission Conf. Fast Process. Photogramm. Data}, vol.~6.\hskip 1em plus 0.5em minus 0.4em\relax Interlaken, 1987, pp. 281--305.

\bibitem{Weickert1998Anisotropic}
J.~Weickert, \emph{Anisotropic diffusion in image processing}.\hskip 1em plus 0.5em minus 0.4em\relax Stuttgart, Germany: B. G. Teubner, 1998, vol.~1.

\bibitem{Jahne2005Digital}
B.~J{\"a}hne, \emph{Digital image processing}.\hskip 1em plus 0.5em minus 0.4em\relax Springer Sci. Bus. Media, 2005.

\bibitem{Zhang2022Double}
H.~Zhang, J.~Cai, W.~He, H.~Shen, and L.~Zhang, ``Double low-rank matrix decomposition for hyperspectral image denoising and destriping,'' \emph{IEEE Trans. Geosci. Remote Sens.}, vol.~60, pp. 1--19, 2022.

\bibitem{Afonso2011Constraint}
M.~V. Afonso, J.~M. Bioucas-Dias, and M.~Figueiredo, ``An augmented lagrangian approach to the constrained optimization formulation of imaging inverse problems,'' \emph{IEEE Trans. Image Process.}, vol.~20, no.~3, pp. 681--695, 2011.

\bibitem{Chierchia2015Constraint}
G.~Chierchia, N.~Pustelnik, J.-C. Pesquet, and B.~Pesquet-Popescu, ``Epigraphical projection and proximal tools for solving constrained convex optimization problems,'' \emph{Signal Image Video Process.}, vol.~9, no.~8, pp. 1737--1749, 2015.

\bibitem{Ono2015Constraint}
S.~Ono and I.~Yamada, ``Signal recovery with certain involved convex data-fidelity constraints,'' \emph{IEEE Trans. Signal Process.}, vol.~63, no.~22, pp. 6149--6163, 2015.

\bibitem{Ono2017Constraint}
S.~Ono, ``Primal-dual plug-and-play image restoration,'' \emph{IEEE Signal Process. Lett.}, vol.~24, no.~8, pp. 1108--1112, 2017.

\bibitem{Ono2019Constraint}
S.~Ono, ``Efficient constrained signal reconstruction by randomized epigraphical projection,'' in \emph{Proc. IEEE Int. Conf. Acoust., Speech, Signal Process., (ICASSP)}, 2019, pp. 4993--4997.

\bibitem{Naganuma2022Destriping}
K.~Naganuma and S.~Ono, ``A general destriping framework for remote sensing images using flatness constraint,'' \emph{IEEE Trans. Geosci. Remote Sens.}, vol.~60, pp. 1--16, 2022.

\bibitem{Pock2011PPDS}
T.~Pock and A.~Chambolle, ``Diagonal preconditioning for first order primal-dual algorithms in convex optimization,'' in \emph{Proc. IEEE Int. Conf. Comput. Vis. (ICCV)}, 2011, pp. 1762--1769.

\bibitem{Boyd2011ADMM}
S.~Boyd, N.~Parikh, E.~Chu, B.~Peleato, and J.~Eckstein, ``Distributed optimization and statistical learning via the alternating direction method of multipliers,'' \emph{Found. Trends Mach. Learn.}, vol.~3, no.~1, pp. 1--122, 2011.

\bibitem{Chambolle2011PDS}
A.~Chambolle and T.~Pock, ``A first-order primal-dual algorithm for convex problems with applications to imaging,'' \emph{J. Math. Imag. Vis.}, vol.~40, no.~1, pp. 120--145, 2011.

\bibitem{Condat2013PDS}
L.~Condat, ``A primal--dual splitting method for convex optimization involving lipschitzian, proximable and linear composite terms,'' \emph{J. Optim. Theory Appl.}, vol. 158, no.~2, pp. 460--479, 2013.

\bibitem{Naganuma2023PPDS}
K.~Naganuma and S.~Ono, ``Variable-wise diagonal preconditioning for primal-dual splitting: Design and applications,'' \emph{IEEE Trans. Signal Process.}, vol.~71, pp. 3281--3295, 2023.

\bibitem{Takemoto2023S3TTV}
S.~Takemoto and S.~Ono, ``Enhancing spatio-spectral regularization by structure tensor modeling for hyperspectral image denoising,'' in \emph{Proc. IEEE Int. Conf. Acoust., Speech, Signal Process. (ICASSP)}, 2023, pp. 1--5.

\bibitem{Combettes2013Moreau}
P.~L. Combettes and N.~N. Reyes, ``Moreau’s decomposition in banach spaces,'' \emph{Math. Program.}, vol. 139, no.~1, pp. 103--114, 2013.

\bibitem{Condat2016L1ball}
L.~Condat, ``Fast projection onto the simplex and the $\ell_{1}$ ball,'' \emph{Math. Program.}, vol. 158, no.~1, pp. 575--585, 2016.

\bibitem{Wang2004SSIM}
Z.~Wang, A.~Bovik, H.~Sheikh, and E.~Simoncelli, ``Image quality assessment: from error visibility to structural similarity,'' \emph{IEEE Trans. Image Process.}, vol.~13, no.~4, pp. 600--612, 2004.

\bibitem{Musco2015Randomized}
C.~Musco and C.~Musco, ``Randomized block krylov methods for stronger and faster approximate singular value decomposition,'' \emph{Adv. neural inf. process. syst.}, vol.~28, 2015.

\bibitem{Struski2024GPUSVD}
{\L}.~Struski, P.~Morkisz, P.~Spurek, S.~R. Bernabeu, and T.~Trzci{\'n}ski, ``Efficient gpu implementation of randomized svd and its applications,'' \emph{Expert Syst. Appl.}, vol. 248, p. 123462, 2024.

\bibitem{Matsaglia1974Equalities}
G.~Matsaglia and G.~PH~Styan, ``Equalities and inequalities for ranks of matrices,'' \emph{Linear multilinear Algebra}, vol.~2, no.~3, pp. 269--292, 1974.

\end{thebibliography}
\end{document}